\documentclass[a4paper,11pt]{article}

\usepackage{jheppub}
\usepackage     [utf8]                  {inputenc}
\usepackage     [T1]                    {fontenc}
\usepackage                             {color}
\usepackage                             {amsmath}
\usepackage                             {amssymb}
\usepackage{bbold}
\usepackage                             {graphicx}
\usepackage     [english]               {babel}
\usepackage                             {hyperref}   
\usepackage                             {float}

\usepackage                             {pdflscape}
\usepackage{hyperref}

\usepackage{array}
\newcolumntype{L}[1]{>{\raggedright\let\newline\\\arraybackslash\hspace{0pt}}m{#1}}
\newcolumntype{C}[1]{>{\centering\let\newline\\\arraybackslash\hspace{0pt}}m{#1}}
\newcolumntype{R}[1]{>{\raggedleft\let\newline\\\arraybackslash\hspace{0pt}}m{#1}}

\definecolor{darkblue}{rgb}{0.0,0.0,0.4}
\definecolor{darkgreen}{rgb}{0.0,0.4,0.0}
\hypersetup{
    colorlinks,
    linkcolor=black,
    citecolor=darkgreen,
    urlcolor=darkblue
}

\usepackage{physics}
\usepackage[makeroom]{cancel}
\usepackage{tabularx}

\usepackage{wrapfig}
\usepackage{tikz-feynman}
\allowdisplaybreaks

\bibpunct{[}{]}{,}{n}{}{,}

\newcommand{\diag}{\text{diag}}

\newcommand\nn[0]{\nonumber}					  
\newcommand\ovbb[0]{$0\nu\beta\beta$ }			  

\newcommand{\0}{$0\nu\beta\beta$}

\title{Unraveling the \ovbb Decay Mechanisms}

\author[a,b,c]{Luk\'{a}\v{s} Gr\'{a}f}
\emailAdd{lukas.graf@berkeley.edu}

\author[a]{Manfred Lindner}
\emailAdd{lindner@mpi-hd.mpg.de}

\author[a]{Oliver Scholer}
\emailAdd{scholer@mpi-hd.mpg.de}

\affiliation[a]{Max-Planck-Institut f{\"u}r Kernphysik, Saupfercheckweg 1, 69117 Heidelberg, Germany}
\affiliation[b]{Department of Physics, University of California, Berkeley, CA 94720, USA}
\affiliation[c]{Department of Physics, University of California, San Diego, CA 92093, USA}

\abstract{
We discuss the possibilities of distinguishing among different mechanisms of neutrinoless double beta decay arising in the effective field theory framework. Following the review and detailed investigation of the particular ways of discrimination, we conclude that the 32 different low-energy effective operators can be split into multiple groups that are in principle distinguishable from each other by measurements of the phase-space observables and by comparison of the decay rates obtained using different isotopes. This would require not only a substantial experimental precision but necessarily also a considerable improvement of the current theoretical knowledge of the underlying nuclear physics. Specifically, the limiting aspect in our approach turns out to be the currently unknown or uncertain values of low-energy constants. Besides the study adopting the effective field theory language we also look into several typical UV models.
}

\begin{document}

\maketitle

\tableofcontents

\section{Introduction}
\label{sec:intro}
The unknown origin of neutrino masses, being one of the major puzzles of contemporary particle physics, strongly motivates the quest for lepton number violation in Nature. The prominent way of probing this symmetry is the search for neutrinoless double beta (\0) decay~\cite{Furry:1939qr}, observation of which would imply non-zero Majorana neutrino masses in accordance with the black-box theorem~\citep{PhysRevD.25.2951}. 

Besides the tight connection to neutrino masses as realized in the standard mass mechanism, neutrinoless double beta decay can be triggered in a variety of different ways, and thus potentially involve also other new physics. Generally, one can study higher-dimensional lepton-number-violating operators that can trigger \0 decay~\citep{Pas:1999fc,Pas:2000vn,Deppisch:2012nb,Cirigliano_2017,Cirigliano_2018,Graf:2018ozy,Deppisch:2020ztt}. In fact, while the sole observation of \0 decay would indeed indicate that neutrinos acquire Majorana mass, it remains unclear whether the standard mechanism that gives a contribution proportional to the neutrino mass would be the dominant one. Examples of models beyond the standard model that can induce non-standard contributions to the \0 decay rate include, for instance, the left-right symmetric models~\citep{Pati:1974yy, Mohapatra:1974hk, Mohapatra:1974gc, Senjanovic:1975rk} triggering several distinct mechanisms~\citep{Deppisch:2012nb,Huang_2014}. Sterile neutrinos can also contribute to \0 decay~\citep{,Huang_2014,Giunti:2010wz,Barry:2011wb,Barea:2015zfa,Dekens:2020ttz,Li:2020flq,Bolton:2019pcu}.

There is variety of experiments searching for \0 decay in different double-beta-decaying isotopes~\citep{2020GERDA,2019CUPID,2010NEMO3Zr,2021CUPID-Mo,2016Cd116,2003Te128,2020CUORE130Te,2019EXO,2016KamLandZen,2009NEMO150Nd}. Currently, the best limit on the half-life reaching $2.3\times10^{26}\,$years, is claimed by the KamLAND-Zen collaboration~\citep{KamLAND-Zen:2022tow} studying the decay of \textsuperscript{136}Xe. The most stringent limit on the half-life of \0 decay of \textsuperscript{76}Ge attains $1.8\times10^{26}\,$years, as obtained by the GERDA collaboration~\citep{2020GERDA}. Proposed next generation experiments such as
LEGEND~\citep{2017LEGEND200,legendcollaboration2021legend1000} (\textsuperscript{76}Ge), 
CUPID~\citep{2020CupidProspects} (\textsuperscript{100}Mo),
SNO+~\citep{2021SNOplus} (\textsuperscript{130}Te) and
nEXO~\citep{nexocollaboration2021nexo} (\textsuperscript{136}Xe)
 aim towards testing half-lives of order of $10^{27}-10^{28}\,$years. Some experiments like NEMO-3 are also equipped with the technology to track individual electrons and measure the individual electron energy spectra and the opening angle between the two electrons, which can help to uncover new physics not only in \0 decay, but even in standard double beta decay~\cite{Deppisch:2020mxv}. Recent reviews of the experimental and theoretical efforts in the field of \0 decay can be found, e.g., in Refs.~\citep{2019review_Werner,Agostini:2022zub}.
 
 In this work we focus on different possibilities of experimental discrimination among different mechanisms inducing \0 decay. To do so, we adopt the effective field theory (EFT) framework developed in Refs.~\citep{Cirigliano_2017,Cirigliano_2018}, which is briefly introduced in Section~\ref{sec:EFT}. In subsequent Section~\ref{sec:distinguishing} we study the possible ways of distinguishing among the relevant set of low-energy EFT operators from \0 decay observables. After having discussed the single operator settings, we turn towards more complete models in Section~\ref{sec:specific_models}. Finally, we summarize our findings in Section~\ref{sec:summary}. This work has been carried out utilizing the upcoming \texttt{NuBB} code package~\citep{neutrinotool}.

\section{EFT Approach to \ovbb Decay: The Master Formula}
\label{sec:EFT}
\subsection{The half-life master formula}
As we apply in this work the effective field theory approach introduced by~\citep{Cirigliano_2017,Cirigliano_2018}, let us start by briefly summarizing the most important parts.
Below the scale of electroweak symmetry breaking (EWSB) the \0 decay amplitude can be described in terms of an $SU(3)_C\times U(1)_{Q}$ invariant low-energy effective field theory (LEFT). Including operators up to LEFT dimension 9 the most relevant Lagrangians for \0 are given by~\citep{Cirigliano_2017,Cirigliano_2018}
\begin{align}
\begin{split}
    \mathcal{L}_{\Delta L=2}^{(6)} = \frac{2G_F}{\sqrt{2}}\bigg{[}&C_{\text{VL}}^{(6)}\;\big(\overline{u_L}\gamma^\mu d_L\big)\; \big(\overline{e_R}\gamma_\mu\nu^c_L\big)
    +C_{\text{VR}}^{(6)}\;\big(\overline{u_R}\gamma^\mu d_R\big)\; \big(\overline{e_R}\gamma_\mu\nu^c_L\big)\\
    +&C_{\text{SL}}^{(6)}\;\big(\overline{u_R} d_L\big)\;\big(\overline{e_L}\nu^c_L\big)
    +C_{\text{SR}}^{(6)}\;\big(\overline{u_L} d_R\big)\;\big(\overline{e_L}\nu^c_L\big)\\
    +&C_{\text{T}}^{(6)}\;\big(\overline{u_L}\sigma^{\mu\nu} d_R\big)\; \big(\overline{e_L}\sigma_{\mu\nu}\nu^c_L\big)\bigg{]} + \text{h.c.}
\end{split}
\end{align}
and
\begin{align}
\begin{split}
    \mathcal{L}_{\Delta L=2}^{(7)} = \frac{2G_F}{\sqrt{2}v}\bigg[&C_{\text{VL}}^{(7)}\;\big(\overline{u_L}\gamma^\mu d_L\big)\; \big(\overline{e_L}\overset{\leftrightarrow}{\partial}_\mu\nu^c_L\big) 
    +C_{\text{VR}}^{(7)}\;\big(\overline{u_R}\gamma^\mu d_R\big)\; \big(\overline{e_L}\overset{\leftrightarrow}{\partial}_\mu\nu^c_L\big)\bigg] + \text{h.c.}
\end{split}
\end{align}
for the long-range part, where 
\begin{align}
    \alpha\overset{\leftrightarrow}{\partial}\beta = \alpha(\partial\beta) - (\partial\alpha)\beta.
\end{align}
as well as the dimension 9 short-range Lagrangian
\begin{align}
\begin{split}
    \mathcal{L}_{\Delta L=2}^{(9)} = \frac{1}{v^5}\sum_i\bigg[\bigg(&C_{i,R}^{(9)}\;\big(\overline{e_R}e_R^c\big) + C_{i,L}^{(9)}\;\big(\overline{e_L}e_L^c\big)\bigg)\;\mathcal{O}_i 
    +C_i^{(9)}\;\big(\overline{e}\gamma_\mu\gamma_5e^c\big)\;\mathcal{O}_i^\mu \bigg]
\end{split}
\end{align}
with the scalar $\mathcal{O}_i$ and vector $\mathcal{O}_i^\mu$ four-quark operators~\citep{Graesser_2017,Cirigliano_2018}
\begin{align}
\begin{aligned}
    \mathcal{O}_1 &= \big(\overline{u_L}^\alpha\gamma_\mu d_L^\alpha\big)\;\big(\overline{u_L}^\beta\gamma^\mu d_L^\beta\big)\,,\qquad
    &\mathcal{O}_1{'}&= \big(\overline{u_R}^\alpha\gamma_\mu d_R^\alpha\big)\;\big(\overline{u_R}^\beta\gamma^\mu d_R^\beta\big)\,,\\
    \mathcal{O}_2&=\big(\overline{u_R}^\alpha d_L^\alpha\big)\;\big(\overline{u_R}^\beta d_L^\beta\big)\,,\qquad 
    &\mathcal{O}_2{'}&= \big(\overline{u_L}^\alpha d_R^\alpha\big)\;\big(\overline{u_L}^\beta d_R^\beta\big)\,,\\
    \mathcal{O}_3 &= \big(\overline{u_R}^\alpha d_L^\beta\big)\;\big(\overline{u_R}^\beta d_L^\alpha\big)\,,\qquad
    &\mathcal{O}_3{'}&=\big(\overline{u_L}^\alpha d_R^\beta\big)\;\big(\overline{u_L}^\beta d_R^\alpha\big)\,,\\
    \mathcal{O}_4&=\big(\overline{u_L}^\alpha \gamma_\mu d_L^\alpha\big)\;\big(\overline{u_R}^\beta\gamma^\mu d_R^\beta)\,,\\
    \mathcal{O}_5&=\big(\overline{u_L}^\alpha \gamma_\mu d_L^\beta\big)\;\big(\overline{u_R}^\beta\gamma^\mu d_R^\alpha\big)\,,\\
    \\
    \mathcal{O}_6^\mu&=\big(\overline{u_L}\gamma^\mu d_L\big)\;\big(\overline{u_L}d_R\big)\,,\qquad
    &\mathcal{O}_6^\mu{'}&=\big(\overline{u_R}\gamma^\mu d_R\big)\;\big(\overline{u_R}d_L\big)\,,\\
    \mathcal{O}_7^\mu&=\big(\overline{u_L}t^A\gamma^\mu d_L\big)\;\big(\overline{u_L}t^A d_R\big)\,,\qquad
    &\mathcal{O}_7^\mu{'}&=\big(\overline{u_R}t^A\gamma^\mu d_R\big)\;\big(\overline{u_R}t^A d_L\big)\,,\\
    \mathcal{O}_8^\mu&=\big(\overline{u_L}\gamma^\mu d_L\big)\;\big(\overline{u_R} d_L\big)\,,\qquad
    &\mathcal{O}_8^\mu{'}&=\big(\overline{u_R}\gamma^\mu d_R\big)\;\big(\overline{u_L} d_R\big)\,,\\
    \mathcal{O}_9^\mu&=\big(\overline{u_L}t^A\gamma^\mu d_L\big)\;\big(\overline{u_R}t^A d_L\big)\,,\qquad
    &\mathcal{O}_9^\mu{'}&=\big(\overline{u_R}t^A\gamma^\mu d_R\big)\;\big(\overline{u_L}t^A d_R\big)\,.
\end{aligned}\label{eq:short_range_operators}
\end{align}
Here $\alpha,\beta$ are color-indices and the $t^A$ are the generators of SU(3) in the fundamental representation given by the 8 Gell-Mann matrices $\lambda^A$ as $t^A = \frac{1}{2}\lambda^A,A=1...8$. The operators $\mathcal{O}$ and $\mathcal{O}{'}$ in~\eqref{eq:short_range_operators} are related via parity transformation. Together with the standard mechanism of light Majorana neutrino-exchange, this framework contains 32 different LEFT operators that can trigger \0 decay.

The transition from the quark level to the nuclear level can be achieved employing the chiral effective field theory ($\chi$EFT)~\citep{Prezeau:2003xn}. The expected half-life contributed by the 32 effective operators is then captured by a ``\0 master-formula'' combining the 32 LEFT Wilson coefficients, 6 different phase-space factors (PSFs) given in Table~\ref{tab:PSFs} and nuclear matrix elements (NMEs) summarized in Table~\ref{tab:NMEs}. At the same time, low-energy constants (LECs) that describe the nuclear interactions within $\chi$EFT enter the formula - we summarize these in Table~\ref{tab:LECs}. The \0 half-life is then given in terms of different sub-amplitudes $\mathcal{A}_i$ as
\begin{align}
\begin{split}
    \left(T_{1/2}^{0\nu}\right)^{-1}=&g_A^4\bigg[G_{01}\left(\left|\mathcal{A}_\nu\right|^2+\left|\mathcal{A}_R\right|^2\right)-2\left(G_{01}-G_{04}\right)\Re\left[\mathcal{A}_\nu^*\mathcal{A}_R\right]\\&+4G_{02}\left|\mathcal{A}_E\right|^2
    +2G_{04}\left(\left|\mathcal{A}_{m_e}\right|^2+\Re\left[\mathcal{A}_{m_e}^*\left(\mathcal{A}_\nu+\mathcal{A}_R\right)\right]\right)
    \\&-2G_{03}\Re\left[\left(\mathcal{A}_\nu+\mathcal{A}_R\right)\mathcal{A}_E^*+2\mathcal{A}_{m_e}\mathcal{A}_E^*\right]\\&
    +G_{09}\left|\mathcal{A}_{M}\right|^2+G_{06}\Re\left[\left(\mathcal{A}_\nu-\mathcal{A}_R\right)\mathcal{A}_{M}^*\right]\bigg]
\end{split}
\end{align}
\begin{table}[t]
    \centering
    \begin{tabular}{lrrrrrr}
\hline\hline
 &  $G_{01}$ &  $G_{02}$ &  $G_{03}$ &  $G_{04}$ &  $G_{06}$ &  $G_{09}$ \\ & \\
$^{238}$U  &      6.96 &      3.79 &      2.75 &      5.26 &     14.43 &     17.32 \\
$^{232}$Th &      2.70 &      0.73 &      0.76 &      1.83 &      6.35 &      7.14 \\
$^{198}$Pt &      1.23 &      0.55 &      0.44 &      0.90 &      2.64 &      3.10 \\
$^{160}$Gd &      1.33 &      1.68 &      0.73 &      1.12 &      2.25 &      3.08 \\
$^{154}$Sm &      0.44 &      0.28 &      0.18 &      0.34 &      0.87 &      1.08 \\
$^{150}$Nd &      8.82 &     40.15 &      7.00 &      8.25 &      9.83 &     18.78 \\
$^{148}$Nd &      1.36 &      2.10 &      0.79 &      1.17 &      2.15 &      3.09 \\
$^{136}$Xe &      1.88 &      4.64 &      1.26 &      1.69 &      2.58 &      4.14 \\
$^{134}$Xe &      0.08 &      0.02 &      0.02 &      0.05 &      0.18 &      0.20 \\
$^{130}$Te &      1.81 &      4.68 &      1.22 &      1.63 &      2.43 &      3.96 \\
$^{128}$Te &      0.07 &      0.02 &      0.02 &      0.05 &      0.17 &      0.19 \\
$^{124}$Sn &      1.13 &      2.42 &      0.72 &      1.01 &      1.62 &      2.51 \\
$^{116}$Cd &      2.06 &      6.51 &      1.46 &      1.89 &      2.59 &      4.47 \\
$^{110}$Pd &      0.58 &      0.95 &      0.34 &      0.50 &      0.89 &      1.30 \\
$^{100}$Mo &      1.89 &      6.80 &      1.36 &      1.75 &      2.25 &      4.06 \\
$^{96}$Zr  &      2.42 &     10.43 &      1.81 &      2.26 &      2.68 &      5.15 \\
$^{82}$Se  &      1.15 &      3.96 &      0.80 &      1.06 &      1.37 &      2.47 \\
$^{76}$Ge  &      0.26 &      0.43 &      0.15 &      0.23 &      0.40 &      0.59 \\
\hline\hline
\end{tabular}

    \caption{The different PSFs in terms of $10^{-14}y^{-1}$ used in our calculations.}
    \label{tab:PSFs}
\end{table}
\subsection{Sub-Amplitudes}
The sub-amplitudes $\mathcal{A}_i$ are categorized and defined via their corresponding leptonic currents. They each depend on the Wilson coefficients of different LEFT operators and can be written as
\begin{align}
\begin{split}
\mathcal{ A}_{\nu} =& \frac{m_{\beta\beta} }{m_e} \mathcal M^{(3)}_{\nu}  
+  \frac{m_N}{m_e}  \mathcal M^{(6)}_{\nu}\left(C_{\text{SL}}^{(6)}, C_{\text{SR}}^{(6)}, C_{\text{T}}^{(6)}, C_{\text{VL}}^{(7)}, C_{\text{VR}}^{(7)}\right) 
\\&+    \frac{m^2_N}{m_e v} \mathcal M^{(9)}_{\nu}\left(C_{\text{1L}}^{(9)}, C_{\text{1L}}^{(9)}{'}, C_{\text{2L}}^{(9)}, C_{\text{2L}}^{(9)}{'}, C_{\text{3L}}^{(9)}, C_{\text{3L}}^{(9)}{'}, C_{\text{4L}}^{(9)}, C_{\text{5L}}^{(9)}\right)\,,
\end{split}
\nn\\
\begin{split}
\mathcal A_R =& \frac{m^2_N}{m_e v} \mathcal M^{(9)}_{R}\left(C_{\text{1R}}^{(9)}, C_{\text{1R}}^{(9)}{'}, C_{\text{2R}}^{(9)}, C_{\text{2R}}^{(9)}{'}, C_{\text{3R}}^{(9)}, C_{\text{3R}}^{(9)}{'}, C_{\text{4R}}^{(9)}, C_{\text{5R}}^{(9)}\right)\, , 
\end{split}
\nn\\
\begin{split}
\mathcal A_{E} =&  \mathcal M^{(6)}_{E,L}\left(C_{\text{VL}}^{(6)}\right) +  \, \mathcal M^{(6)}_{E,R}\left(C_{\text{VR}}^{(6)}\right) \, ,  
\end{split}
\nn\\
\begin{split}
\mathcal A_{m_e} =& \mathcal M^{(6)}_{m_e,L}\left(C_{\text{VL}}^{(6)}\right) +  \,\mathcal M^{(6)}_{m_e,R}\left(C_{\text{VR}}^{(6)}\right) \,,
\end{split}
\nn\\
\begin{split}
\mathcal A_{ M} =&  \frac{m_N}{m_e}    \, \mathcal M^{(6)}_{M}\left(C_{\text{VL}}^{(6)}\right) +  \frac{m_N^2}{m_e v} \mathcal M^{(9)}_{M}\left(C_{\text{6}}^{(9)}, C_{\text{6}}^{(9)}{'}, C_{\text{7}}^{(9)}, C_{\text{7}}^{(9)}{'}, C_{\text{8}}^{(9)}, C_{\text{8}}^{(9)}{'}, C_{\text{9}}^{(9)}, C_{\text{9}}^{(9)}{'}\right)\,.
\end{split}\label{eq:A_M}
\end {align}
The matrix elements $\mathcal{M}_i$ depend on the different LECs and Wilson coefficients. We explicitly state the dependency on the different Wilson coefficients within the brackets in~\eqref{eq:A_M}.
\begin{table}[t!]
\setlength{\tabcolsep}{3pt}
\scalebox{0.80}{
    \centering
    \begin{tabular}{c|rrrrrrrrrrrrrrr}
\hline\hline
 &   $M_F$ &  $M_{GT}^{AA}$ &  $M_{GT}^{AP}$ &  $M_{GT}^{PP}$ &  $M_{GT}^{MM}$ &  $M_T^{AA}$  &  $M_T^{AP}$ &  $M_T^{PP}$ &  $M_T^{MM}$ &  $M_{Fsd}$ &  $M^{AA}_{GTsd}$ &  $M^{AP}_{GTsd}$ &  $M^{PP}_{GTsd}$ &  $M^{AP}_{Tsd}$ &  $M^{PP}_{Tsd}$ \\ & \\
 \textsuperscript{76}Ge  & -0.78 &         6.06 &        -0.86 &         0.17 &         0.20 &       0.0 &      0.24 &     -0.06 &      0.04 &    -1.20 &           4.18 &          -1.24 &           0.29 &         -0.77 &          0.23 \\
 \textsuperscript{82}Se  & -0.67 &         4.93 &        -0.71 &         0.14 &         0.17 &       0.0 &      0.24 &     -0.06 &      0.04 &    -1.01 &           3.46 &          -1.03 &           0.25 &         -0.73 &          0.22 \\
 \textsuperscript{96}Zr  & -0.36 &         4.32 &        -0.64 &         0.13 &         0.15 &       0.0 &     -0.21 &      0.05 &     -0.04 &    -0.87 &           3.06 &          -0.89 &           0.21 &          0.64 &         -0.20 \\
 \textsuperscript{100}Mo & -0.51 &         5.55 &        -0.90 &         0.20 &         0.22 &       0.0 &     -0.29 &      0.07 &     -0.05 &    -1.28 &           4.48 &          -1.33 &           0.30 &          0.93 &         -0.28 \\
 \textsuperscript{110}Pd & -0.42 &         4.43 &        -0.76 &         0.17 &         0.18 &       0.0 &     -0.21 &      0.06 &     -0.04 &    -1.07 &           3.72 &          -1.11 &           0.25 &          0.79 &         -0.24 \\
 \textsuperscript{116}Cd & -0.34 &         3.17 &        -0.55 &         0.12 &         0.13 &       0.0 &     -0.12 &      0.04 &     -0.03 &    -0.80 &           2.72 &          -0.81 &           0.18 &          0.49 &         -0.16 \\
 \textsuperscript{124}Sn & -0.57 &         3.37 &        -0.50 &         0.11 &         0.12 &       0.0 &      0.12 &     -0.03 &      0.02 &    -0.82 &           2.56 &          -0.77 &           0.19 &         -0.42 &          0.13 \\
 \textsuperscript{128}Te & -0.72 &         4.32 &        -0.64 &         0.13 &         0.15 &       0.0 &      0.12 &     -0.04 &      0.03 &    -1.03 &           3.24 &          -0.98 &           0.24 &         -0.52 &          0.16 \\
 \textsuperscript{130}Te & -0.65 &         3.89 &        -0.57 &         0.12 &         0.14 &       0.0 &      0.14 &     -0.04 &      0.02 &    -0.94 &           2.95 &          -0.89 &           0.22 &         -0.47 &          0.15 \\
 \textsuperscript{134}Xe & -0.69 &         4.21 &        -0.62 &         0.13 &         0.15 &       0.0 &      0.12 &     -0.04 &      0.03 &    -0.97 &           3.07 &          -0.92 &           0.22 &         -0.48 &          0.15 \\
 \textsuperscript{136}Xe & -0.52 &         3.20 &        -0.45 &         0.09 &         0.11 &       0.0 &      0.12 &     -0.03 &      0.02 &    -0.73 &           2.32 &          -0.69 &           0.17 &         -0.36 &          0.12 \\
 \textsuperscript{148}Nd & -0.36 &         2.52 &        -0.48 &         0.11 &         0.12 &       0.0 &     -0.12 &      0.02 &     -0.02 &    -0.78 &           2.54 &          -0.79 &           0.19 &          0.30 &         -0.09 \\
 \textsuperscript{150}Nd & -0.51 &         3.75 &        -0.76 &         0.17 &         0.19 &       0.0 &     -0.12 &      0.04 &     -0.03 &    -0.74 &           2.46 &          -0.76 &           0.18 &          0.34 &         -0.10 \\
 \textsuperscript{154}Sm & -0.34 &         2.98 &        -0.52 &         0.11 &         0.13 &       0.0 &     -0.12 &      0.03 &     -0.02 &    -0.78 &           2.64 &          -0.79 &           0.19 &          0.39 &         -0.13 \\
 \textsuperscript{160}Gd & -0.42 &         4.22 &        -0.71 &         0.15 &         0.17 &       0.0 &     -0.21 &      0.05 &     -0.03 &    -1.02 &           3.52 &          -1.04 &           0.24 &          0.60 &         -0.19 \\
 \textsuperscript{198}Pt & -0.33 &         2.27 &        -0.50 &         0.11 &         0.12 &       0.0 &     -0.12 &      0.03 &     -0.02 &    -0.78 &           2.57 &          -0.78 &           0.18 &          0.37 &         -0.12 \\
 \textsuperscript{232}Th & -0.44 &         4.17 &        -0.76 &         0.17 &         0.18 &       0.0 &     -0.21 &      0.05 &     -0.04 &    -1.08 &           3.80 &          -1.11 &           0.25 &          0.69 &         -0.22 \\
 \textsuperscript{238}U  & -0.52 &         4.96 &        -0.90 &         0.20 &         0.21 &       0.0 &     -0.21 &      0.06 &     -0.04 &    -1.29 &           4.51 &          -1.32 &           0.30 &          0.82 &         -0.25 \\
 \hline\hline
\end{tabular}}
    \caption{NMEs used in our calculations based on the IBM2 model~\citep{Deppisch_2020}}
    \label{tab:NMEs}
\end{table}
$\mathcal{A}_\nu$ depends on the matrix elements
\begin{align}
\begin{split}
    \mathcal{M}_\nu^{(3)} &= -V_{ud}^2\left(-\frac{1}{g_A^2}M_F+\mathcal{M}_{GT}+\mathcal{M}_T+2\frac{m_\pi^2\mathbf{g_\nu^{NN}}}{g_A^2}M_{F,sd}\right)\label{eq:M(3)nu},
    \\
    \mathcal{M}_\nu^{(6)} &=V_{ud}\left(\frac{B}{m_N}\left(C_{SL}^{(6)}-C_{SR}^{(6)}\right)+\frac{m_\pi^2}{m_N v}\left(C_{VL}^{(7)}-C_{VR}^{(7)}\right)\right)\mathcal{M}_{PS}+V_{ud}C_{\text{T}}^{(6)}\mathcal{M}_{T6},
    \\
    \mathcal{M}_\nu^{(9)} &= -\frac{1}{2m_N^2}C_{\pi\pi L}^{(9)}\left(M_{GT,sd}^{AP}+M_{T,sd}^{AP}\right)-\frac{2m_\pi^2}{g_A^2m_N^2}C_{NNL}^{(9)}M_{F,sd},
\end{split}
\end{align}
where $\mathcal{M}_\nu^{(3)}$ represents the contribution from the standard mass mechanism. In contrast to the traditional approach employing the non-relativistic approximation, the EFT treatment contains also the contribution proportional to $g_\nu^{NN}$, which parametrizes the contact-term contribution originating from the exchange of hard neutrinos~\citep{Cirigliano:2018hja,Cirigliano:2019vdj}.
$\mathcal{A}_R$ is given by
\begin{table}[t!]
    \centering
    \begin{tabular}{l r c l r}
        \hline\hline
        \multicolumn{2}{c}{Known LECs} & & \multicolumn{2}{c}{Unknown LECs} \\
         \cline{1-2} \cline{4-5}
        $g_A$ & $1.271$~ & \qquad & $\left|g_T'\right|$ & $\mathcal{O}(1)$ \\
        $g_S$ & $0.97$~\citep{Bhattacharya_2016} & \qquad &  $\left|g_T{\pi\pi}\right|$ & $\mathcal{O}(1)$ \\
        $g_M$ & $4.7$ &  \qquad & $\left|g_{1,6,7,8,9}^{\pi N}\right|$ & $\mathcal{O}(1)$ \\
        $g_T$ & $0.99$~\citep{Bhattacharya_2016} &  &$\left|g_{VL}^{\pi N}\right|$ & $\mathcal{O}(1)$  \\
        $B$ & $2.7\,\mathrm{GeV}$ & & $\left|g_{T}^{\pi N}\right|$ & $\mathcal{O}(1)$  \\
        $g_1^{\pi\pi}$ & $0.36$~\citep{Nicholson_2018} &  &$\left|g_{1,6,7}^{NN}\right|$ & $\mathcal{O}(1)$  \\
        $g_2^{\pi\pi}$ & $2.0$~\citep{Nicholson_2018} &   &$\left|g_{2,3,4,5}^{NN}\right|$ & $\mathcal{O}(16\pi^2)$  \\
        $g_3^{\pi\pi}$ & $-0.62$~\citep{Nicholson_2018} &   &$\left|g_{VL}^{NN}\right|$ & $\mathcal{O}(1)$  \\
        $g_4^{\pi\pi}$ & $-1.9$~\citep{Nicholson_2018} &   &$\left|g_{T}^{NN}\right|$ & $\mathcal{O}(1)$  \\
        $g_5^{\pi\pi}$ & $-8.0$~\citep{Nicholson_2018} &   &  $\left|g_{VL,VR}^{E,m_e}\right|$ & $\mathcal{O}(1)$ \\
           $g_{\nu}^{NN}$ & $-92.9\,\mathrm{GeV^{-2}} \pm 50\%$~\citep{Cirigliano_2021_1,Cirigliano_2021,Wirth:2021pij} & & \\
        \hline\hline
    \end{tabular}
    \caption[Summary of the low-energy constants necessary to calculate the \0 half-life.]{Summary of the low-energy constants necessary to calculate the \0 half-life for all 32 different operators. The table is taken from~\citep{Cirigliano_2018} and restructured.}
    \label{tab:LECs}
\end{table}
\begin{align}
\begin{split}
    \mathcal{M}_R^{(9)} &= \mathcal{M}_\nu^{(9)}|_{L\rightarrow R},
\end{split}
\end{align}
for $\mathcal{A}_E$ the different contributions are
\begin{align}
\begin{split}
    \mathcal{M}_{E,L}^{(6)} &= -\frac{V_{ud}C_{VL}^{(6)}}{3}\left(\frac{g_V^2}{g_A^2}M_F+\frac{1}{3}\left(2M_{GT}^{AA}+M_T^{AA}\right)+\frac{6\mathbf{g_{V\;L}^E}}{g_A^2}M_{F,sd}\right),
    \\
    \mathcal{M}_{E,R}^{(6)} &= -\frac{V_{ud}C_{VR}^{(6)}}{3}\left(\frac{g_V^2}{g_A^2}M_F-\frac{1}{3}\left(2M_{GT}^{AA}+M_T^{AA}\right)+\frac{6\mathbf{g_{V\;R}^E}}{g_A^2}M_{F,sd}\right),
\end{split}
\end{align}
$\mathcal{A}_{m_e}$ is determined by
\begin{align}
    \begin{split}
    \mathcal{M}_{m_e,L}^{(6)} &= \frac{V_{ud}C_{VL}^{(6)}}{6}\bigg(\frac{g_V^2}{g_A^2}M_F-\frac{1}{3}\left(M_{GT}^{AA}-4M_T^{AA}\right)-3\left(M_{GT}^{AP}+M_{GT}^{PP}+M_T^{AP}+M_T^{PP}\right)\\
    &\qquad\qquad\;\;\;\;\;-\frac{12\mathbf{g_{V\;L}^{m_e}}}{g_A^2}M_{F,sd}\bigg),
\\
    \mathcal{M}_{m_e,R}^{(6)} &= \frac{V_{ud}C_{VR}^{(6)}}{6}\bigg(\frac{g_V^2}{g_A^2}M_F+\frac{1}{3}\left(M_{GT}^{AA}-4M_T^{AA}\right)+3\left(M_{GT}^{AP}+M_{GT}^{PP}+M_T^{AP}+M_T^{PP}\right)\\
    &\qquad\qquad\;\;\;\;\;-\frac{12\mathbf{g_{V\;R}^{m_e}}}{g_A^2}M_{F,sd}\bigg),
    \end{split}
\end{align}
and finally $\mathcal{A}_M$ is given by
\begin{align}
    \begin{split}
    \mathcal{M}_M^{(6)} &= V_{ud}C_{VL}^{(6)}\bigg[2\frac{g_A}{g_M}\left(M_{GT}^{MM}+M_T^{MM}\right)
    \\
    &\qquad\qquad\;\;\;\;\;+\frac{m_\pi^2}{m_N^2}\left(-\frac{2}{g_A^2}\mathbf{g_{VL}^{NN}}M_{F,sd}+\frac{1}{2}\mathbf{g_{VL}^{\pi N}}\left(M_{GT,sd}^{AP}+M_{T,sd}^{AP}\right)\right)\bigg],
    \\
    \mathcal{M}_M^{(9)} &=\frac{m_\pi^2}{m_N^2}\bigg[-\frac{2}{g_A^2}\big(\mathbf{g_6^{NN}}C_V^{(9)}+\mathbf{g_7^{NN}}\Tilde{C}_V^{(9)}\big)M_{F,sd}\\
    &\qquad\qquad+\frac{1}{2}\big(\mathbf{g_V^{\pi N)}}C_V^{(9)} + \mathbf{\Tilde{g}_V^{\pi N}}\Tilde{C}_V^{(9)}\big)\big(M_{GT,sd}^{AP} + M_{T,sd}^{AP}\big)\bigg]\label{eq:M(9)M}.
\end{split}
\end{align}
In the above formulas we have defined the combined NMEs
\begin{align}
    \begin{split}
            \mathcal{M}_{GT} &= M_{GT}^{AA}+M_{GT}^{AP}+M_{GT}^{PP}+M_{GT}^{MM},
    \\
    \mathcal{M}_T &= M_T^{AP}+M_T^{PP}+M_T^{MM},
    \\
    \mathcal{M}_{PS} &= \frac{1}{2}M_{GT}^{AP}+M_{GT}^{PP}+\frac{1}{2}M_T^{AP}+M_T^{PP},
    \\
    \mathcal{M}_{T6} &= 2\frac{\mathbf{g_T'}-\mathbf{g_T^{NN}}}{g_A^2}\frac{m_\pi^2}{m_N^2}M_{F,sd}-\frac{8g_T}{g_M}\left(M_{GT}^{MM}+M_T^{MM}\right)
    \\
    &\quad+\mathbf{g_T^{\pi N}}\frac{m_\pi^2}{4m_N^2}\left(M_{GT,sd}^{AP}+M_{T,sd}^{AP}\right)+\mathbf{g_T^{\pi\pi}}\frac{m_\pi^2}{4m_N^2}\left(M_{GT,sd}^{PP}+M_{T,sd}^{PP}\right).
    \end{split}
\end{align}
The short-range dimension-9 LEFT operators contribute to the $C_{V,\pi\pi L,\pi NL,NNL}^{(9)}$ couplings that appear in the chiral Lagrangian. They are given by
\begin{align}
    \begin{split}
        &C_V^{(9)} = C_6^{(9)}+C_6^{(9)}{'}+C_8^{(9)}+C_8^{(9)}{'}\,,\quad
    \Tilde{C}_V^{(9)} = C_7^{(9)}+C_7^{(9)}{'}+C_9^{(9)}+C_9^{(9)}{'}
    \\
    \begin{split}
        &C_{\pi\pi L}^{(9)} = g_2^{\pi\pi}\left(C_{2L}^{(9)}+C_{2L}^{(9)}{'}\right) + g_3^{\pi\pi}\left(C_{3L}^{(9)}+C_{3L}^{(9)}{'}\right)
        -g_4^{\pi\pi}C_{4L}^{(9)} - g_5^{\pi\pi}C_{5L}^{(9)} - \frac{5}{3}g_1^{\pi\pi}m_\pi^2\left(C_{1L}^{(9)}+C_{1L}^{(9)}{'}\right)
    \end{split}
    \\
    &C_{\pi NL}^{(9)} = \left(\mathbf{g_1^{\pi N}} - \frac{5}{6}g_1^{\pi\pi}\right)\left(C_{1L}^{(9)}+C_{1L}^{(9)}{'}\right)
    \\
    \begin{split}
        &C_{NNL}^{(9)} = \mathbf{g_1^{NN}}\left(C_{1L}^{(9)}+C_{1L}^{(9)}{'}\right) +\mathbf{g_2^{NN}}\left(C_{2L}^{(9)}+C_{2L}^{(9)}{'}\right) 
        + \mathbf{g_3^{NN}}\left(C_{3L}^{(9)}+C_{3L}^{(9)}{'}\right) +\mathbf{g_4^{NN}}C_{4L}^{(9)}+\mathbf{g_5^{NN}}C_{5L}^{(9)}
    \end{split}
    \\
    &C_{\left\{\pi\pi,\pi N,NN\right\}R} = C_{\left\{\pi\pi,\pi N,NN\right\}L}|_{L\rightarrow R}\;.
    \end{split}\label{eq:short-range_couplings}
\end{align}
The two LECs $g_V^{\pi N}$ and $\Tilde{g}_V^{\pi N}$ are defined as
\begin{align}
\begin{split}
    \mathbf{g_V^{\pi N}} = \mathbf{g_6^{\pi N}} + \mathbf{g_8^{\pi N}}\\
    \mathbf{\Tilde{g}_V^{\pi N}} = \mathbf{g_7^{\pi N}} + \mathbf{g_9^{\pi N}}
\end{split}
\end{align}
For the sake of convenience, we have marked all currently unknown LECs including $g_\nu^{NN}$ in \textbf{bold} within the above formulas.

In this work we will study the different LEFT operators at the matching scale of $\Lambda=m_W$ at which one would usually match the BSM model of interest onto LEFT. The running of the operators down to the scale of $\chi$PT at $\Lambda_\chi\simeq 2\,$GeV is described in~\citep{Cirigliano_2018}.

\subsection{Relation to literature}
A different basis to describe \0 decay developed first in~\citep{Pas:1999fc} and~\citep{Pas:2000vn} that is often used in the literature is defined by a set of 29 dimension-6 and dimension-9 lepton number violating LEFT operators given by
\begin{align}
    \mathcal{L}_6 = \frac{G_F}{\sqrt{2}}
    \sum_{i,k}\epsilon_k^i j_i J_k
\end{align}
for the long-range part with $i,k\in\left\{V\pm A,S\pm P,T_L, T_R, \right\}$ and 
\begin{align}
    \begin{split}
    \mathcal{L}_9 = \frac{G_F^2}{2m_N}\sum_{l,m,n}
    \Big[&
    \epsilon_1^{lmn}J_l J_m j_n
    +\epsilon_2^{lmn}J^{\mu\nu}_l {J_{\mu\nu}}_m j_n
    +\epsilon_3^{lmn}J^{\mu}_l {J_{\mu}}_m j_n
    \\&+\epsilon_4^{lmn}J^{\mu}_l {J_{\mu\nu}}_m j^\nu
    +\epsilon_5^{lmn}J^{\mu}_l J_m j_{\mu n}
    \Big]
    \end{split}
\end{align}
for the short-range part with $l,m,n\in\left\{L,R\right\}$. Here, $\epsilon_k^i$ and $\epsilon^{lmn}$ are the Wilson coefficients of the different long- and short-range operators. The quark currents $J$ are given by\footnote{We keep the two different types of indices for the short-range currents to stick with the literature}
\begin{align}
\begin{split}
    J_{S\pm P} = J_{R,L} &= \overline{u}\left(1\pm\gamma_5\right)d\;,\quad
    J_{V\pm A} = J_{R,L}^\mu = \overline{u}\gamma^\mu\left(1\pm\gamma_5\right)d,\\
    J_{T_{R,L}} = J^{\mu\nu}_{R,L} &= \overline{u}\sigma^{\mu\nu}\left(1\pm\gamma_5\right)d,
\end{split}
\end{align}
and the lepton currents $j$ are given by
\begin{align}
\begin{split}
    j_{S\pm P} &= \overline{e}\left(1\pm\gamma_5\right)\nu^c\;,\quad
    j_{V\pm A} = \overline{e}\gamma^\mu\left(1\pm\gamma_5\right)\nu^c\\
    j_{T_{R,L}} &= \overline{e}\sigma^{\mu\nu}\left(1\pm\gamma_5\right)\nu^c\,,\quad
    j_{R,L} = \overline{e}\left(1\pm\gamma_5\right)e^c\\
    j_{R,L}^\mu &= \overline{e}\gamma^\mu\left(1\pm\gamma_5\right)e^c.
\end{split}
\end{align}
This framework does not include the dimension 7 operators of the framework utilized in our approach. While the remaining long-range part of the two descriptions can be related easily, the short-range operators are related to each other via Fierz transformations. One finds that
\begin{align}
    &C_{1 L}^{(9)} = \frac{2v}{m_N}\epsilon^{LLL}_3,\quad C_{1 L}^{(9)}{'} = \frac{2v}{m_N}\epsilon^{RRL}_3,\quad C_{1 R}^{(9)} = \frac{2v}{m_N}\epsilon^{LLR}_3,\quad C_{1 R}^{(9)}{'} = \frac{2v}{m_N}\epsilon^{RRR}_3,\\
    \begin{split}
    &C_{2 L}^{(9)} = \frac{2v}{m_N}\left(\epsilon_1^{LLL}-4\epsilon^{LLL}_2\right),\quad C_{2 L}^{(9)}{'} = \frac{2v}{m_N}\left(\epsilon_1^{RRL}-4\epsilon^{RRL}_2\right),\\
    &C_{2 R}^{(9)} = \frac{2v}{m_N}\left(\epsilon_1^{LLR}-4\epsilon^{LLR}_2\right),\quad C_{2 R}^{(9)}{'} = \frac{2v}{m_N}\left(\epsilon_1^{RRR}-4\epsilon^{RRR}_2\right),
    \end{split}\\
    \begin{split}
    &C_{3 L}^{(9)} = -\frac{16v}{m_N}\epsilon_2^{LLL},\quad C_{3 L}^{(9)}{'} = -\frac{16v}{m_N}\epsilon_2^{RRL},\\
    &C_{3 R}^{(9)} = -\frac{16v}{m_N}\epsilon_2^{LLR},\quad C_{3 R}^{(9)}{'} = -\frac{16v}{m_N}\epsilon_2^{RRR},
    \end{split}\\
    &C_{4 L}^{(9)} = \frac{2v}{m_N}\epsilon_3^{RLL},\quad C_{4 R}^{(9)} = \frac{2v}{m_N}\epsilon_3^{RLR},\\
    &C_{5 L}^{(9)} = -\frac{v}{m_N}\epsilon_1^{RLL},\quad C_{5 R}^{(9)} = -\frac{v}{m_N}\epsilon_1^{RLR},\\
    &C_{6}^{(9)} = \frac{v}{m_N}\left(\epsilon_5^{LRR}+i\frac{5}{3}\epsilon_4^{LRR}\right),\quad C_{6}^{(9)}{'} = \frac{v}{m_N}\left(\epsilon_5^{RLR}+i\frac{5}{3}\epsilon_4^{RLR}\right),\\
    &C_{7}^{(9)} = 4i\frac{v}{m_N}\epsilon_4^{LRR},\quad C_{7}^{(9)}{'} = 4i\frac{v}{m_N}\epsilon_4^{RLR},\\
    &C_{8}^{(9)} = \frac{v}{m_N}\left(\epsilon_5^{LLR} - i\frac{5}{3}\epsilon_4^{LLR}\right),\quad C_{8}^{(9)}{'} = \frac{v}{m_N}\left(\epsilon_5^{RRR} - i\frac{5}{3}\epsilon_4^{RRR}\right),\\
    &C_{9}^{(9)} = -4i\frac{v}{m_N}\epsilon_4^{LLR},\quad C_{9}^{(9)}{'} = -4i\frac{v}{m_N}\epsilon_4^{RRR}.
\end{align}
The main difference between the two set of operators is that the $\epsilon$-basis contains short-range tensor operators instead of color octets.

\section{Distinguishing the Effective Operators}\label{sec:distinguishing}
Neutrinoless double-$\beta$-decay, if observed, would be characterized by several experimental observables, precise determination of which can give us some insight into the underlying BSM physics. Generally, the \0 decay experiments can be able to determine decay rate, single electron energy spectrum and angular correlation between the two emitted electrons. Additional information might be obtained by studying different $\beta\beta$ modes or by employing complementary information from other experiments such as the LHC. In the following, we will discuss possible ways of experimentally distinguishing among the 32 different \0 decay inducing LEFT operators with a focus on the limiting factors of a potential confirmation/exclusion of the existence of any additional non-standard scenario contributing to \0 decay alongside the standard mass mechanism.
\subsection{Phase-Space Observables}
While most experimental collaborations only attempt to measure the half-life of \0 decay, some experiments like NEMO-3~\citep{Arnold_2017}, or its future successor SuperNEMO~\citep{Arnold2010}, are designed to also measure the single electron energy spectrum and the angular correlation of the two outgoing electrons. These are associated with different electron currents and within the simplest approximation they can be calculated analytically. More exact solutions require numeric calculations of the exact electron wave functions~\citep{Kotila_2012}. The different PSFs $G_{0k}$ can be written in the form~\citep{_tef_nik_2015}
\begin{figure}[t!]
    \centering
    \includegraphics[scale=0.75]{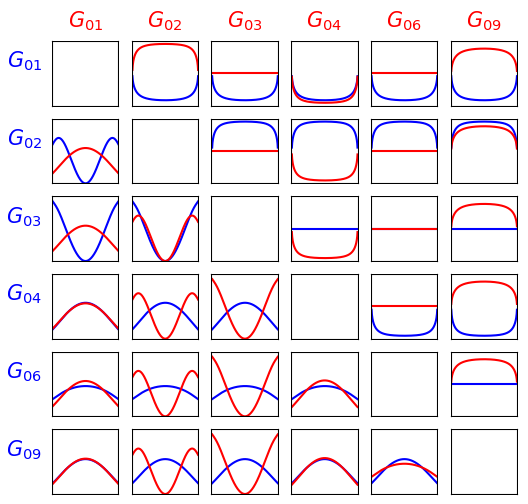}
    \caption[Comparison of the 6 PSFs and the corresponding normalized single electron spectra and angular correlation coefficients]{Comparison of the normalized single electron spectra (lower left) and angular correlation coefficients (upper right) in \textsuperscript{136}Xe that result from the 6 PSFs which appear in the \0 half-life ``master-formula''. Red curves correspond to the red-labelled PSFs on horizontal axis, while blue curves represent the PSFs denoted in blue on the vertical axis. The x-axis covers the range $\Tilde{\epsilon}\in\left[0-1\right]$.}
    \label{fig:6_psf_observables}
\end{figure}
\begin{align}
\begin{split}
    G_{0k} = \frac{(G_FV_{ud})^4m_e^2}{64\pi^5\ln{2}R^2}&\int \delta\bigg(\epsilon_1 + \epsilon_2 +E_f-E_i\bigg) \\
    &\times \bigg(h_{0k}(\epsilon_1, \epsilon_2, R)\cos{\theta} + g_{0k}(\epsilon_1, \epsilon_2, R)\bigg) \\
    &\times p_1 p_2 \epsilon_1 \epsilon_2 \,\text{d}\epsilon_1\,\text{d}\epsilon_2\,\text{d}(\cos{\theta}),\label{eq:PSF_definition}
\end{split}
\end{align}
where $p_{1,2}$ and $\epsilon_{1,2}$ are the momentum and energy of the first and second released electron, $R$ is the radius of the final-state nucleus and $E_{i,f}$ denotes the energy of the initial- or final-state nucleus, respectively. Here, we denote the part of the differential phase-space factor independent of the angle between the two outgoing electrons as $g_{0k}$, while $h_{0k}$ is the angular correlation part proportional to the cosine of the opening angle $\theta$. Additionally, $G_{04, 06, 09}$ have to be rescaled to comply with the definitions in~\citep{Cirigliano_2017, Cirigliano_2018} as
\begin{align}
\begin{split}
    &G_{04} \rightarrow \frac{9}{2}G_{04},\\
    &G_{06} \rightarrow \frac{m_eR}{2}G_{06},\\
    &G_{09} \rightarrow \left(\frac{m_eR}{2}\right)^2G_{09}.
\end{split}
\end{align}
The relations between the electron wave functions and the functions $h_{0k}$ and $g_{0k}$ are given in~\citep{_tef_nik_2015} to which we will refer here. We apply their simplest approximation scheme `A' assuming a uniform charge distribution in the nucleus.
Using Eq.~\eqref{eq:PSF_definition} one can write the angular correlation coefficient $a_1/a_0$ which is defined via
\begin{align}
    \frac{\text{d}\Gamma}{\mathrm{d}\cos{\theta}\mathrm{d}\Tilde{\epsilon}_1} = a_0\left(1+\frac{a_1}{a_0} \cos{\theta}\right)
\end{align}
with
\begin{align}
    \Tilde{\epsilon}_i = \frac{\epsilon_i - m_e}{Q_{\beta\beta}} \in \left[0,1\right]
\end{align}
as
\begin{align}
    \frac{a_1}{a_0}(\Tilde{\epsilon}) = \frac{\sum_i |M_i|^2 h_{0i}\left(\epsilon, \Delta M_{\text{Nuclei}} - \epsilon, R\right)}{\sum_j |M_j|^2 g_{0j}\left(\epsilon, \Delta M_{\text{Nuclei}} - \epsilon, R\right)} \,.\label{eq:angular_correlation_coefficient}
\end{align}
Here, $\Delta M_{\text{Nuclei}}$ is the mass difference between the mother and daughter nuclei and $Q_{\beta\beta}$ denotes the Q-value of the decay. The potential of utilizing the angular correlation of the outgoing electrons for discrimination between different mechanisms of \0 has been discussed e.g.\ in~\citep{Ali:2007ec}. Similarly, the single electron spectra are given by
\begin{align}
\begin{split}
    \frac{\text{d}\Gamma}{\text{d}\epsilon_1} = &\frac{(G_FV_{ud})^4m_e^2}{64\pi^5\ln{2}R^2}
    \bigg(\sum_i |M_i|^2 g_{0i}\left(\epsilon, \Delta M_{\text{Nuclei}} - \epsilon, R\right)\bigg)
     p_1p_2\epsilon\big(\Delta M_{\text{Nuclei}} - \epsilon\big)\,.
\end{split}
\end{align}
\noindent
Consequently, approximating the electron wave functions, we can easily calculate the expected angular correlation factor and single electron spectra for each of the 32 LEFT operators. The normalized single electron spectra as well as the angular correlations corresponding to each of the 6 distinct PSFs are shown in Figure~\ref{fig:6_psf_observables}. As we can see, using these observables the operators associated with distinct PSFs are in principle distinguishable from each other, provided substantial experimental accuracy is reached.
\begin{figure}[t!]
    \centering
    \includegraphics[scale=0.75]{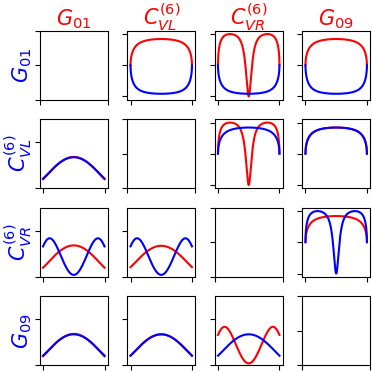}
    \caption[Normalized single electron spectra and angular correlation coefficient for each of the 4 distinguishable groups of operators.]{Normalized single electron spectra (lower left) and angular correlation coefficients (upper right) for each of the 4 distinguishable groups of operators. The shapes are shown for \textsuperscript{136}Xe assuming the NDA values for the currently unknown LECs. However, the particular choice does not result in a significant difference in the general shape of the plots. 
    Red curves correspond to the red-labelled operator group on the horizontal axis, while blue curves represent the operator group denoted in blue on the vertical axis. The x-axis covers the range $\Tilde{\epsilon}\in\left[0-1\right]$.}
    \label{fig:4_PSF_groups}
\end{figure}

However, distinguishing among different \0 mechanisms purely based on the phase-space observables has its obvious limitations. In fact, while $G_{06}$ is only induced in presence of multiple operators the dimension-6 vector operators both trigger several of the remaining PSFs. Taking this into account, we can identify 4 different groups of operators that are in principle distinguishable using the leptonic PSF observables, namely: $C_{VL}^{(6)}, C_{VR}^{(6)}$, the operators corresponding to $G_{01}$ and the ones corresponding to $G_{09}$. The PSF observables that result from each of these 4 groups are shown in Figure~\ref{fig:4_PSF_groups}. Here, we can see that the 
left-handed vector current operator $C_{VL}^{(6)}$ and the operators corresponding to $G_{09}$, while corresponding to distinct PSFs, are practically indistinguishable since the $C_{VL}^{(6)}$ phase-space turns out to be dominated by the contribution from $G_{09}$. The remaining groups are distinguishable from each other using at least one of the considered observables.

\begin{figure}[t!]
    \centering
    \includegraphics[width=0.75\textwidth]{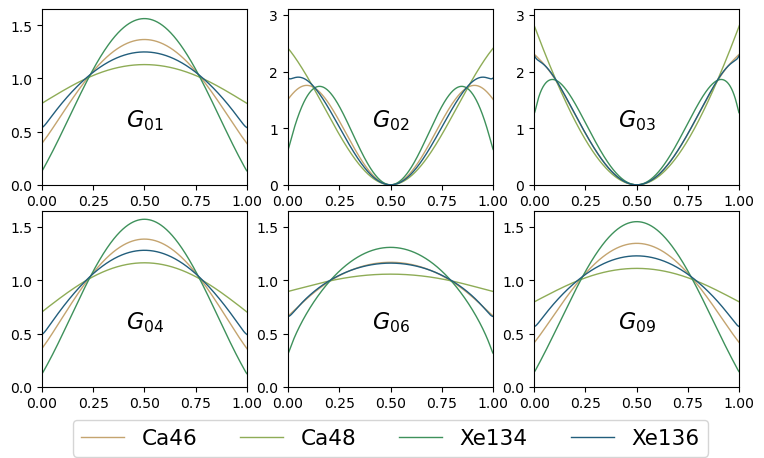}
    \caption[Single electron spectra for 4 different naturally occurring \0 isotopes]{
    The single electron spectra for 4 different naturally occurring $0\nu\beta^-\beta^-$ isotopes are shown. While the exact quantitative curves depend on the choice of the isotope, their shape is mostly independent of this choice. As before, the x-axis shows the normalized electron energy $\Tilde{\epsilon}$.}
    \label{fig:PSF_comparison_all_isotopes_spectra}
\end{figure}

\begin{figure}[t!]
    \centering
    \includegraphics[width=0.75\textwidth]{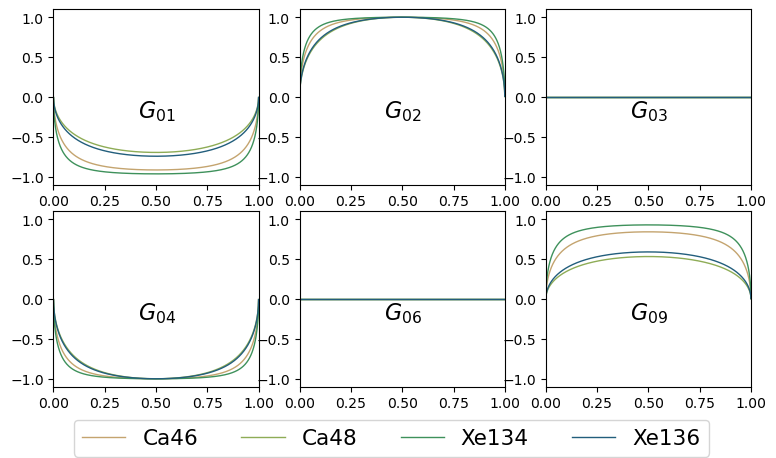}
    \caption[Angular correlation for 4 different naturally occurring \0 isotopes]{Comparison of the angular correlation coefficients in 4 different isotopes as done for the single electron energy spectra in Fig.~\ref{fig:PSF_comparison_all_isotopes_spectra}.}
    \label{fig:PSF_comparison_all_isotopes_angular}
\end{figure}

Note that while the electron wave functions depend on the charge of the daughter nucleus as well as on the decay energy, the general shape of the induced observables is not very dependent on the choice of the decaying isotope. In Figure~\ref{fig:PSF_comparison_all_isotopes_spectra} we show the single electron spectra and in Figure~\ref{fig:PSF_comparison_all_isotopes_angular} the angular correlation coefficients corresponding to the 6 different PSFs in 4 different naturally occurring $0\nu\beta^-\beta^-$ isotopes.
\subsection{Decay Rate Ratios}
The remaining \0 observable is the decay rate $\Gamma$ itself. While the phase-space can be used to distinguish operators with different leptonic currents, information about the decay rates in various isotopes can be also applied to operators with distinct hadronic structures, as these give rise to different NMEs. The isotope dependence of the existing calculations of NMEs can be inferred from Table~\ref{tab:NMEs}. Therefore, one can study the half-life ratios
\begin{align}
    R^{\mathcal{O}_i} (^\text{A}\text{X}) \equiv \frac{T_{1/2}^{\mathcal{O}_i}(^\text{A}\text{X})}{T_{1/2}^{\mathcal{O}_i}(^{76}\text{Ge})} =  \frac{\sum_j|\mathcal{M}_j^{\mathcal{O}_i}(^{76}\text{Ge})|^2 G_j^{\mathcal{O}_i}(^{76}\text{Ge})}{\sum_k|\mathcal{M}_k^{\mathcal{O}_i}(^\text{A}\text{X})|^2 G_k^{\mathcal{O}_i}(^\text{A}\text{X})}\label{eq:Ri}
\end{align}
where $T_{1/2}^{\mathcal{O}_i}(^AX)$ is the half-life induced by the operator $O_i$ in the isotope $^AX$. The sums $\sum_{j,k}$ are taken over all different PSFs generated by the operator $\mathcal{O}_i$ and become relevant only for $C_{VL,VR}^{(6)}$ (see Eqs.~\eqref{eq:CVL(6)} and~\eqref{eq:CVR(6)}). Studying the half-life ratio allows for elimination of the unknown particle physics couplings, as was first discussed in~\citep{Deppisch_2007} and shortly after also in~\citep{Gehman:2007qg}. Here, we take \textsuperscript{76}Ge for the reference isotope. To be able to quantify how well one can distinguish two different operators $\mathcal{O}_{i,j}$ from each other we can take the ratio
\begin{align}
    R_{ij}(^AX) = \frac{R^{\mathcal{O}_i}(^AX)}{R^{\mathcal{O}_j}(^AX)}\label{eq:Rij}.
\end{align}
Specifically, the ratios $R_{im_{\beta\beta}}$ relating the non-standard mechanisms with the standard mass mechanism will be of interest to compare the effect of different higher-dimensional operators and possibly identify the existence of additional exotic contributions to the \0 rate in experiments. Obviously, two operators $\mathcal{O}_{i,j}$ would be indistinguishable via this method if the resulting ratio would equal unity, i.e., if $R_{ij}=1$. Vice versa, they would be perfectly distinguishable for either $R_{ij}\rightarrow\infty$ or $R_{ij}=0$, that is, for $|\log_{10}(R_{ij})|\rightarrow\infty.$

Studying the decay rate ratios has several benefits. First of all, in case only one Wilson coefficient contributes at a time, it drops out. Therefore, the ratio corresponding to a certain operator and its Wilson coefficient is a constant that depends only on the corresponding NMEs, LECs and PSFs. If more Wilson coefficients contribute at the same time, then only the overall magnitude can be factored out. In this case, the relations between different coefficients can, of course, affect the resulting ratios. However, one can still utilize this method to study specific models and see if they are distinguishable from the standard mass mechanism. We will discuss this possibility in section~\ref{sec:specific_models}. 
Additionally, when taking ratios of the half-lives, one can expect that the impact of correlated systematic relative errors on the NMEs decreases as they should (at least partially) cancel. In~\citep{Lisi:2015yma} it was shown that for the NME calculations using QRPA uncertainties arising from unknown $g_A$ quenching and nucleon-nucleon potentials are correlated among different isotopes. Half-life measurements in different isotopes as a tool to discriminate among different mechanisms of \0  decay have also been employed previously in~\citep{Fogli:2009py,Simkovic:2010ka,Meroni:2012qf,Horoi:2018fls}.

Applying this approach to the master-formula framework one can identify 12 different groups of operators that can in principle be distinguished from each other. These groups are summarized in Table~\ref{tab:NME_operator_groups}.
However, the distinguishability of the short-range operators strongly depends on the currently unknown LECs. Taking the most of the unknown LECs to be zero while keeping $g_{6,7}^{NN}=g_V^{\pi N}=\Tilde{g}_V^{\pi N}=1$ (so that the contribution from the short-range vector operators is not omitted) makes it impossible to distinguish the short-range scalar operators $C_{S2-S5}^{(9)}$ as well as the short-range vector operator groups $C^{(9)}_V$ and $\Tilde{C}^{(9)}_V$.

\begin{table}[t]
    \centering
    \begin{tabular}{cccccccccccc}
        \hline\hline
        $m_{\beta\beta}$ & $C_{VL}^{(6)}$ & $C_{VR}^{(6)}$ & $C_{T}^{(6)}$ & $C_{S,V}^{(6,7)}$ & $C_{S1}^{(9)}$ & $C_{S2}^{(9)}$ & $C_{S3}^{(9)}$ & $C_{S4}^{(9)}$ & $C_{S5}^{(9)}$ & $C_{V}^{(9)}$ & $\Tilde{C}_{V}^{(9)}$\\\\
        $m_{\beta\beta}$ & $C_{VL}^{(6)}$ & $C_{VR}^{(6)}$ & $C_{T}^{(6)}$ & $C_{SL}^{(6)}$ & $C_{1L}^{(9)}$ & $C_{2L}^{(9)}$ & $C_{3L}^{(9)}$ & $C_{4L}^{(9)}$ & $C_{5L}^{(9)}$ & $C_{6}^{(9)}$ & $C_{7}^{(9)}$\\
        - & - & - & - & $C_{SR}^{(6)}$ & $C_{1R}^{(9)}$ & $C_{2R}^{(9)}$ & $C_{3R}^{(9)}$ & $C_{4R}^{(9)}$ & $C_{5R}^{(9)}$ & $C_{6}^{(9)}{'}$ & $C_{7}^{(9)}{'}$\\
        - & - & - & - & $C_{VL}^{(7)}$ & $C_{1L}^{(9)}{'}$ & $C_{2L}^{(9)}{'}$ & $C_{3L}^{(9)}{'}$ & - & - & $C_{8}^{(9)}$ & $C_{9}^{(9)}$\\
        - & - & - & - & $C_{VR}^{(7)}$ & $C_{1R}^{(9)}{'}$ & $C_{2R}^{(9)}{'}$ & $C_{3R}^{(9)}{'}$ & - & - & $C_{8}^{(9)}{'}$ & $C_{9}^{(9)}{'}$\\
        \hline\hline
    \end{tabular}
    \caption[Operator groups that can be distinguished via taking decay rate ratios.]{Operator groups that can possibly be distinguished via taking decay rate ratios. The choice of the groups depends on the knowledge of the LECs. If we set the unknown LECs to zero, the short-range scalar operator groups $C_{S2-S5}^{(9)}$ become indistinguishable as well as the short-range vector operator groups $\Tilde{C}_V^{(9)}$ and $\Tilde{C}_V^{(9)}$. Improved knowledge of the LECs, assuming no fine tuning, would allow to distinguish among these operator groups.}
    \label{tab:NME_operator_groups}
\end{table}
\subsubsection{Sensitivity on the unknown LECs}
\begin{figure}[t!]
    \centering
    \includegraphics[width=\textwidth]{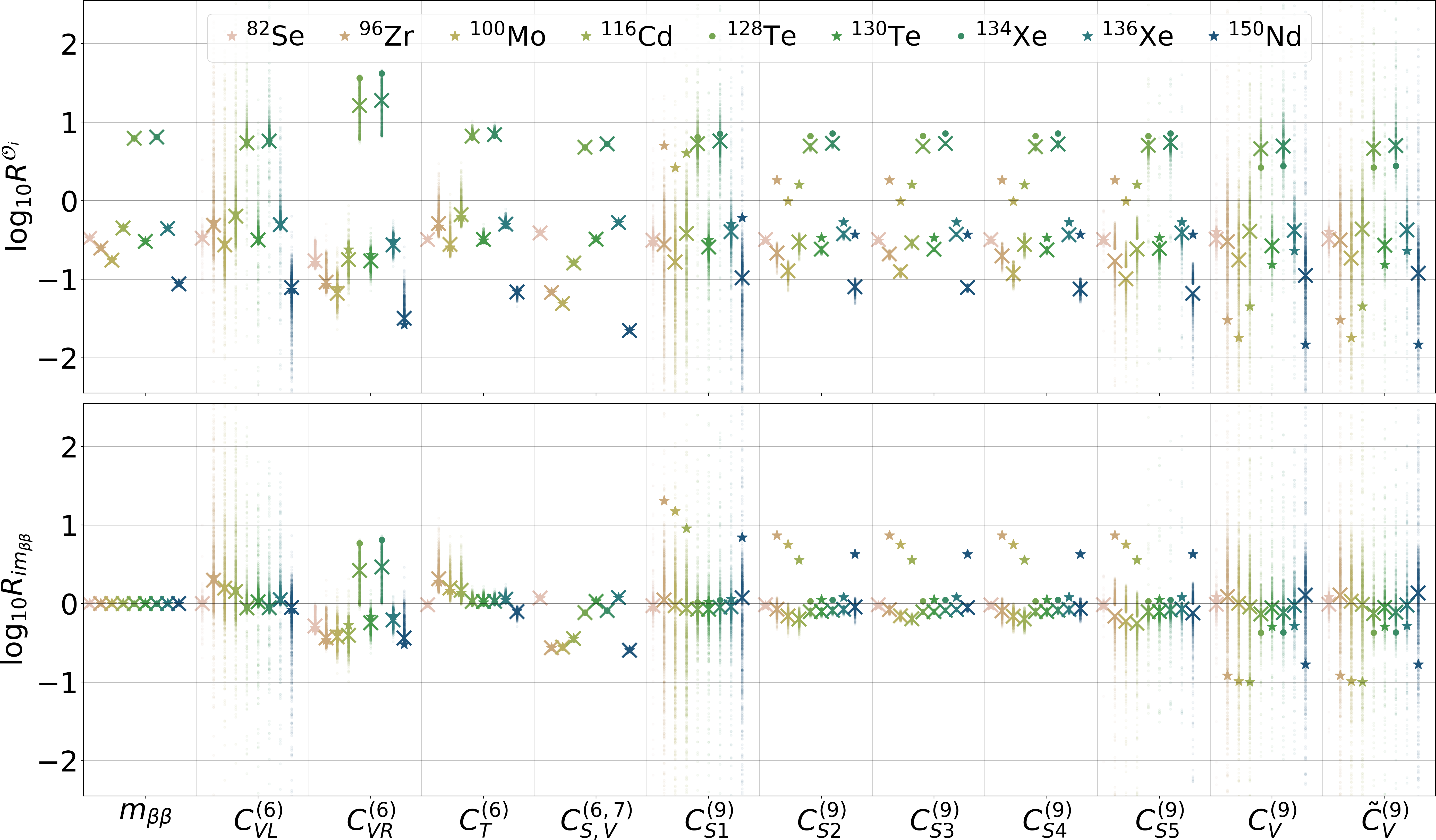}
    \caption{The decay rate ratios $R^{O_i}$ (upper plot) and $R_{i m_{\beta\beta}}$ (lower plot) for the different operator groups are shown. The larger markers represent the choice of vanishing unknown LECs with $g_{6,7}^{NN}=g_V^{\pi N}=\Tilde{g}_V^{\pi N}=1$. Isotopes with a PSF $G_0>10^{-14}\,\mathrm{y^{-1}}$ are represented by stars while isotopes with smaller PSFs are represented by round markers. The additional points represent variations of the different unknown LECs $g_i$ randomly chosen from $\left[-\sqrt{10}, -1/\sqrt{10}\right]\times |g_i|$ and $\left[+1/\sqrt{10}, +\sqrt{10}\right]\times |g_i|$ except for $g_\nu^{NN}$ which is varied in a range of $\pm 50\%$. The crosses represent the central values of the variation i.e. the median values. The reference isotope is chosen to be \textsuperscript{76}Ge. Note that the variation of $g_\nu^{NN}$ does induce a small variation of $R^{m_{\beta\beta}}$ which is, however, not visible in the above plot.}
    \label{fig:ratios_ROi}
\end{figure}

\begin{figure}[t!]
    \centering
    \includegraphics[width=\textwidth]{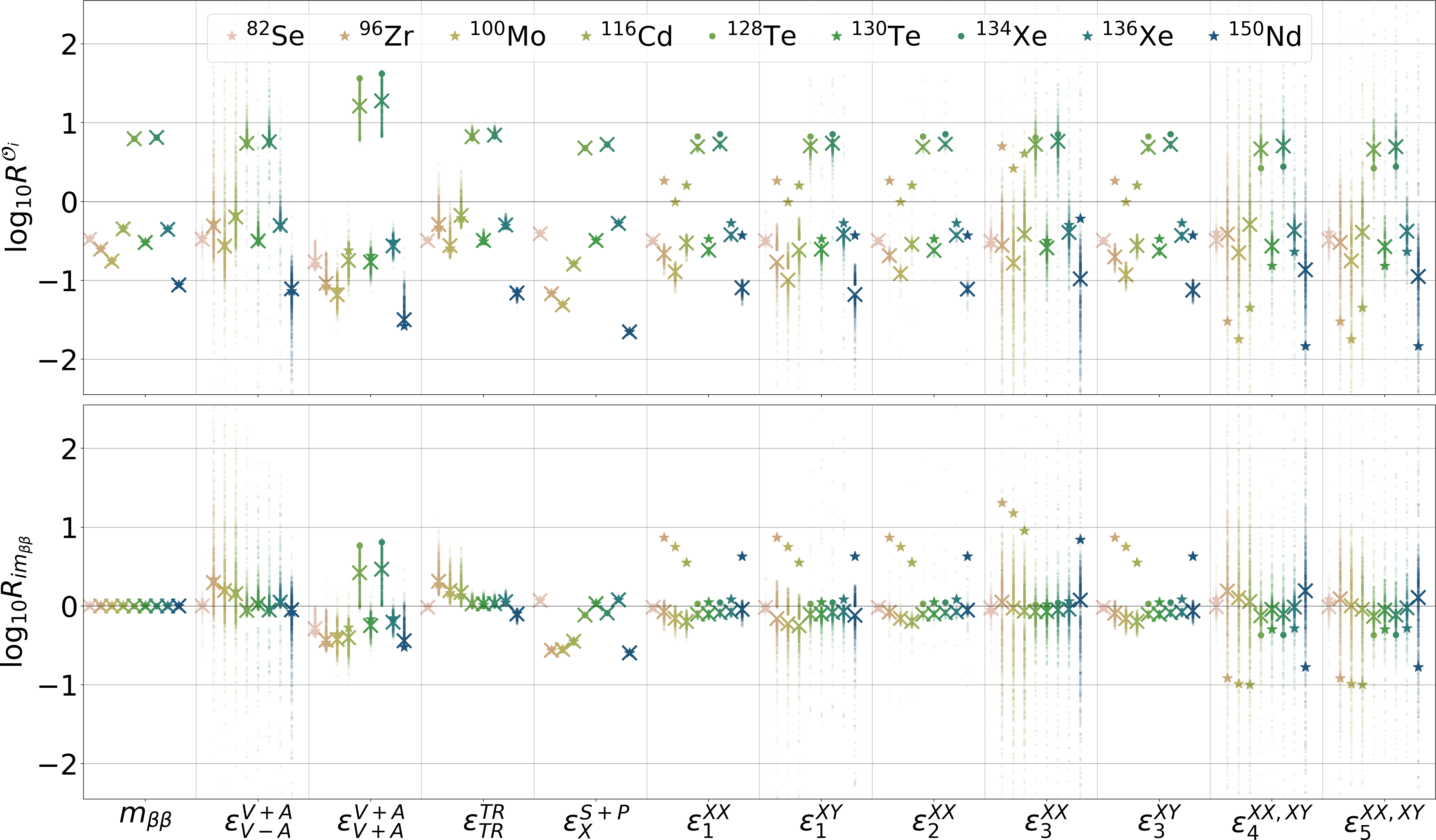}
    \caption{The decay rate ratios $R^{O_i}$ (upper plot) and $R_{i m_{\beta\beta}}$ (lower plot) for the different operator groups in the $\epsilon$-basis similar to Figure~\ref{fig:ratios_ROi} are shown.}
    \label{fig:ratios_ROi_epsilon}
\end{figure}

In Figure~\ref{fig:ratios_ROi} we present the expected ratios $R^{\mathcal{O}_{i}}$ as well as the normalized ratios $R_{im_{\beta\beta}}$ defined in \eqref{eq:Ri} and \eqref{eq:Rij} for the above choice of LECs which will be our benchmark scenario. The ratios for the \textit{$\epsilon$-basis} are shown in Figure~\ref{fig:ratios_ROi_epsilon}. Additionally, to study the uncertainties arising from the unknown LECs, the plots include 1000 points per operator group that each represent variations of the unknown LECs $g_i$ within the ranges $\left[-\sqrt{10}, -1/\sqrt{10}\right]\times |g_i|$ and $\left[1/\sqrt{10}, \sqrt{10}\right]\times |g_i|$
, i.e. we vary the LECs within the range of values given by their expected order of magnitude shown in Table~\ref{tab:LECs}. 
For $g_{\nu}^{NN}$ which generates a short-range component into the standard mass-mechanism we take a variation of $\pm 50\%$. The central values of the variation i.e. the median values are marked by crosses. 

\begin{figure}[t!]
    \centering
    \includegraphics[width = 0.7\textwidth]{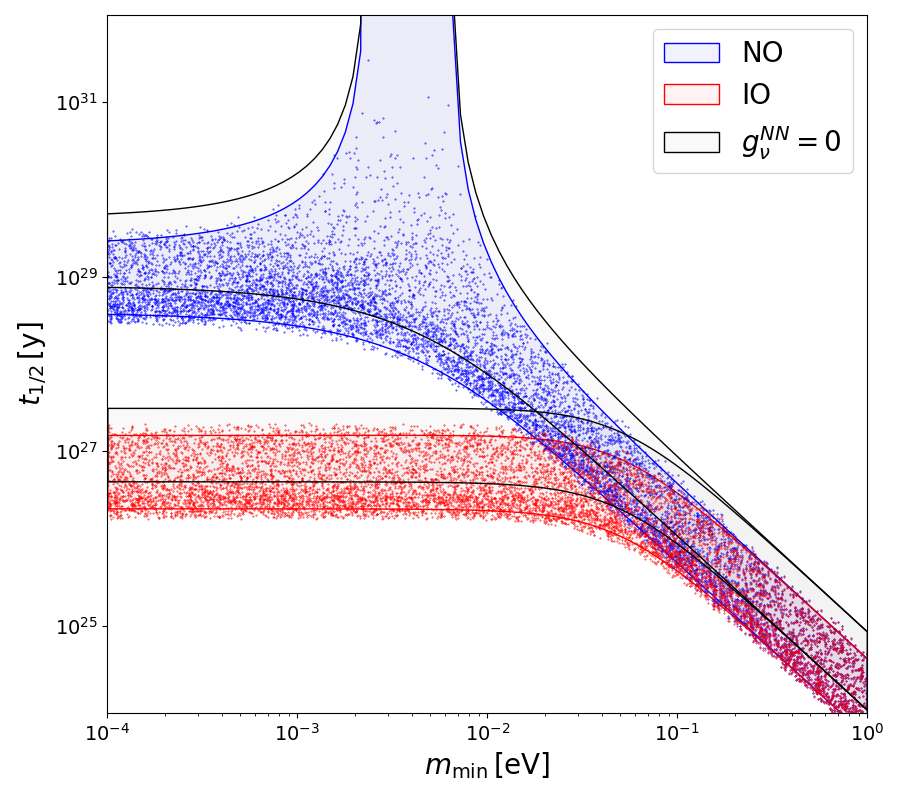}
    \caption{Here we show the half-life for \textsuperscript{76}Ge in dependence on the minimal neutrino mass $m_\mathrm{min}$ for both normal (NO) and inverted (IO) neutrino mass ordering. The scatter points were obtained by marginalizing over $g_\nu^{NN}$ and the unknown Majorana phases. The blue and red contours show the possible half-life ranges when only the phases are varied, while $g_\nu^{NN}$ is fixed. Additionally, the black contours correspond to $g_\nu^{NN}=0$.}
    \label{fig:lobster_mass}
\end{figure}

From the upper panel of Figure~\ref{fig:ratios_ROi} one can infer that the half-life ratios $R^{m_{\beta\beta}}$ corresponding to the standard mass mechanism are not very sensitive to $g_\nu^{NN}$ (they are actually too small to be visible). In Figure~\ref{fig:lobster_mass} we explicitly show the impact of varying the $g_\nu^{NN}$ LEC on the expected half-life in \textsuperscript{76}Ge for the standard mechanism. Again, compared to the impact of the unknown Majorana phases the effect of $g_\nu^{NN}$ is minor. However, it is important to note that the impact of $g_\nu^{NN}$ on the overall magnitude of the half-life cannot be ignored as easily. For comparison, we also present the case where $g_\nu^{NN}=0$ in Figure~\ref{fig:lobster_mass}. 

For the remaining non-standard operators, however, we can see from Figure~\ref{fig:ratios_ROi} that the values of the currently unknown LECs can have quite a significant impact on the expected ratios. Oftentimes, especially for the short-range $C_{i}^{(9)}$ groups, the central values are significantly offset from our benchmark scenario with most unknown LECs turned off. Hence, for these operators the appearance of the unknown LECs has a significant impact on the corresponding \0-decay rate. Although for some operator groups, such as $C_{2S-5S}^{(9)}$, the spread of the values of the ratios obtained by varying the unknown LECs is relatively small, for other groups like the short-range vector contributions $C_V^{(9)}, \Tilde{C}_V^{(9)}$ the variation of the unknown LECs results in a significant stretch around the central values. For these ratios the precise numerical value of the unknown LECs is of particular importance. The different sensitivities of the short-range scalar and vector operators arise from the fact that for the scalar operators some of the relevant LECs, namely those encoding pion-pion interactions $g_i^{\pi\pi}$, are known, while for the short-range vector operators all relevant LECs are unknown. Since we do not fix the sign of the unknown LECs (except $g_\nu^{NN}$) there can be a gap within the LEC-varied ratios resulting in two visible central values for the operator groups, for which the ratios are sensitive to the sign of the LECs. The lower part of Figure~\ref{fig:ratios_ROi} which displays the normalized $R_{im_{\beta\beta}}$ shows that the central values of the LEC-varied ratios are closer to $0$ than the benchmark scenario. Therefore, the inclusion of the unknown LECs tends to impair the distinguishability from the standard mechanism.

The above discussion clearly shows the importance of determining the yet unknown LECs involved in the calculation. This can be achieved, for example, by lattice QCD calculations~\citep{Nicholson:2018mwc,Detmold:2020jqv,Tuo:2019bue}.
\subsubsection{Distinguishing different operators}

\begin{figure}[t!]
    \centering
    \includegraphics[width = 1\textwidth]{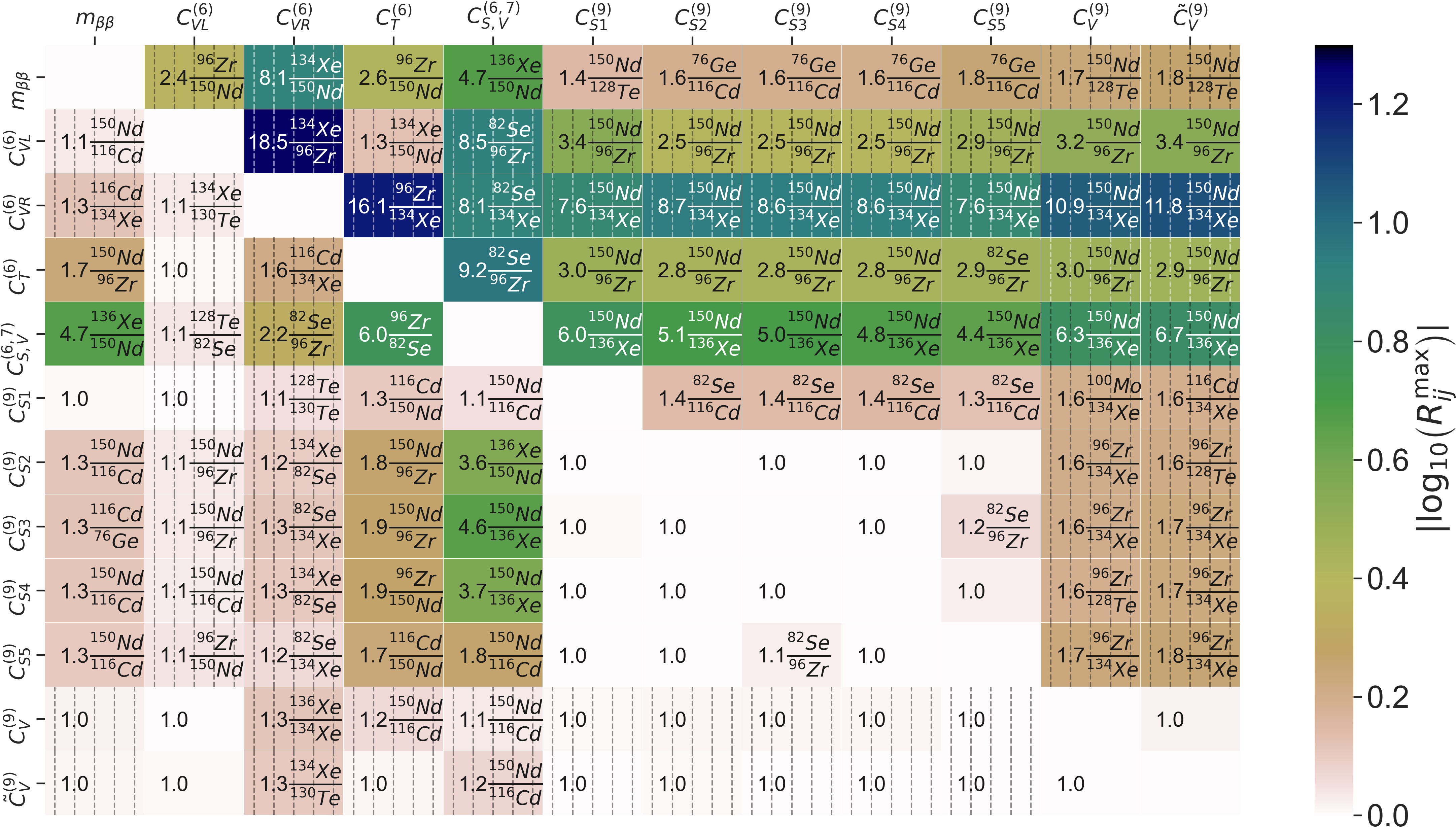}
    \caption{The maximal ratios $R_{ij}^\mathrm{max}$ for all operator combinations $i,j$. The exact values and the corresponding isotopes are displayed in each tile. Additionally, operator combinations that result in different phase-space observables are marked by dashed-line shading. In the upper right half of the plot we show the ratios considering the central values from the variation of the LECs. In the lower left half we show the worst-case scenario considering the values of ratios $R_{ij}$ that are closest to 1 within the range obtained by the variation of the LECs.}
    \label{fig:tile_plot_exp_isotopes}
\end{figure}

When studying the \0 decay rate the basic question to ask is whether and how well could a non-standard contribution be distinguished from the standard light-neutrino-exchange. Employing the half-life measurements in different isotopes, one can try to identify those that are most suitable for discrimination between the mass mechanism and an exotic \0 decay contribution triggered by a particular higher-dimensional operator. In the first row of Figure~\ref{fig:tile_plot_exp_isotopes} we show the maximal ratios $R_{im_{\beta\beta}}^{\mathrm{max}}$ and the corresponding pair of isotopes obtained for each operator group. Here, we consider a ``representative" scenario by studying the central values defined as the median ratio $R_{im_{\beta\beta}}$ of the range of values obtained from the variation of the LECs. At the same time, we identify the ``worst-case scenario" ratio defined as the value within the range that is closest to unity, see the first column of Figure~\ref{fig:tile_plot_exp_isotopes}. In this context we consider only isotopes with existing experimental limits on the half-life, namely, the following: 
\textsuperscript{76}Ge~\citep{2020GERDA}, \textsuperscript{82}Se~\citep{2019CUPID}, \textsuperscript{96}Zr~\citep{2010NEMO3Zr}, \textsuperscript{100}Mo~\citep{2021CUPID-Mo}, \textsuperscript{116}Cd~\citep{2016Cd116}, \textsuperscript{128}Te~\citep{2003Te128}, \textsuperscript{130}Te~\citep{2020CUORE130Te}, \textsuperscript{134}Xe~\citep{2017EXO}, \textsuperscript{136}Xe~\citep{2016KamLandZen} and \textsuperscript{150}Nd~\citep{2009NEMO150Nd}. 
Figure~\ref{fig:tile_plot_exp_isotopes} also presents all the other ratios $R_{ij}^\mathrm{max}$ quantifying the mutual distinguishability of all the operator groups with the values corresponding to the representative scenario above the diagonal and the worst-case scenario below the diagonal. In addition, the dashed lines in Figure~\ref{fig:tile_plot_exp_isotopes} mark the pairs of operators that could be discriminated using the phase-space observables.

Considering the central values, the non-standard long-range operators $ C_{VR}^{(6)}$ and $C_{S,V}^{(6,7)}$ give the most distinct half-life ratios compared to the standard mass mechanism while the remaining long-range operator groups $C_{VL}^{(6)}$ and $C_{T}^{(6)}$ also result in sizable $R_{im_{\beta\beta}}^\mathrm{max}>2$. Additionally, both $C_{VL}^{(6)}$ and $C_{VR}^{(6)}$ could be identified by measuring the angular correlation of the emitted electrons. The non-standard short-range operators generally tend to have lower values of the ratios $R_{im_{\beta\beta}}^\mathrm{max}<2$; thus, there is less potential for identifying their contribution by experiments. However, the short-range vector operators $C_V^{(9)}$ and $\Tilde{C}_V^{(9)}$ are associated with a different angular correlation than $m_{\beta\beta}$. On the contrary, the contributions from scalar short-range operators $C_{S2-S5}^{(9)}$ would be hardest to discriminate, as they do not manifest any significant isotope dependence on $R_{im_{\beta\beta}}$ and do not differ in the phase-space observables, either.

In the worst-case scenario, the operators in the group $C_{S,V}^{(6,7)}$, i.e., lepton number violating long-range scalar and vector interactions, are the only operators that result in $R_{i m_{\beta\beta}}^\mathrm{max}>2$. 
In fact, this is the only operator group that is not affected by any unknown LECs. Besides $C_T^{(6)}$ all remaining operators in this setting have expected ratios
$R_{i m_{\beta\beta}}^\mathrm{max}\leq 1.3$, which would require very precise measurements and accurate knowledge of the theoretically calculated half-lives to be able to claim a detection of any of these non-standard contributions. The contributions from $C_{VL}^{(6)}, C_V^{(9)}$ and 
$\Tilde{C}_V^{(9)}$ could be identified only based on measurements of the angular correlation and the scalar short-range operators, 
$C_{S1}^{(9)}$, 
would be completely indistinguishable from the standard mechanism.
In Appendix~\ref{app:all_isotopes} we show for completeness the same results employing the full set of isotopes, for which there exist numerical values of NMEs computed using IBM2.

To be able to pinpoint the specific non-standard operator group $\mathcal{O}_j$ contributing to \0 decay one needs to consider half-life ratios $R_{ij}$ for all different isotopes. Considering the central values, the best candidate to be clearly identified turns out to be the right-handed vector current $C_{VR}^{(6)}$ for which all the ratios $R_{iC_{VR}^{(6)}}^{\mathrm{max}}$ are large i.e. $\gtrsim 7.6$.


\subsubsection{The impact of nuclear uncertainties}
The uncertainty induced by the nuclear part of the decay rate calculation i.e. the NMEs and LECs highly impacts and limits the above approach of distinguishing among different \0 mechanisms. The approach of comparing theoretically predicted ratios with experimentally measured ratios raises the question how well these theoretical uncertainties must be under control.

To study the impact of nuclear uncertainties, we can use the general formula for the half-life parameterized in terms of a Wilson coefficient $C$, the phase-space factor $G$ and an \emph{effective} NME which we label $M_\mathrm{eff}$,
\begin{align}
    T_{1/2}^{-1} = \left|C\right|^2 G \left|M_\mathrm{eff}\right|^2.
\end{align}
Here, $M_\mathrm{eff}$ is, generally, a weighted sum of combinations of different LECs and NMEs (see App.~\ref{app:all_operators} for the explicit half-life equations of each single operator).
\begin{figure}[t!]
    \centering
    \includegraphics[width = \textwidth]{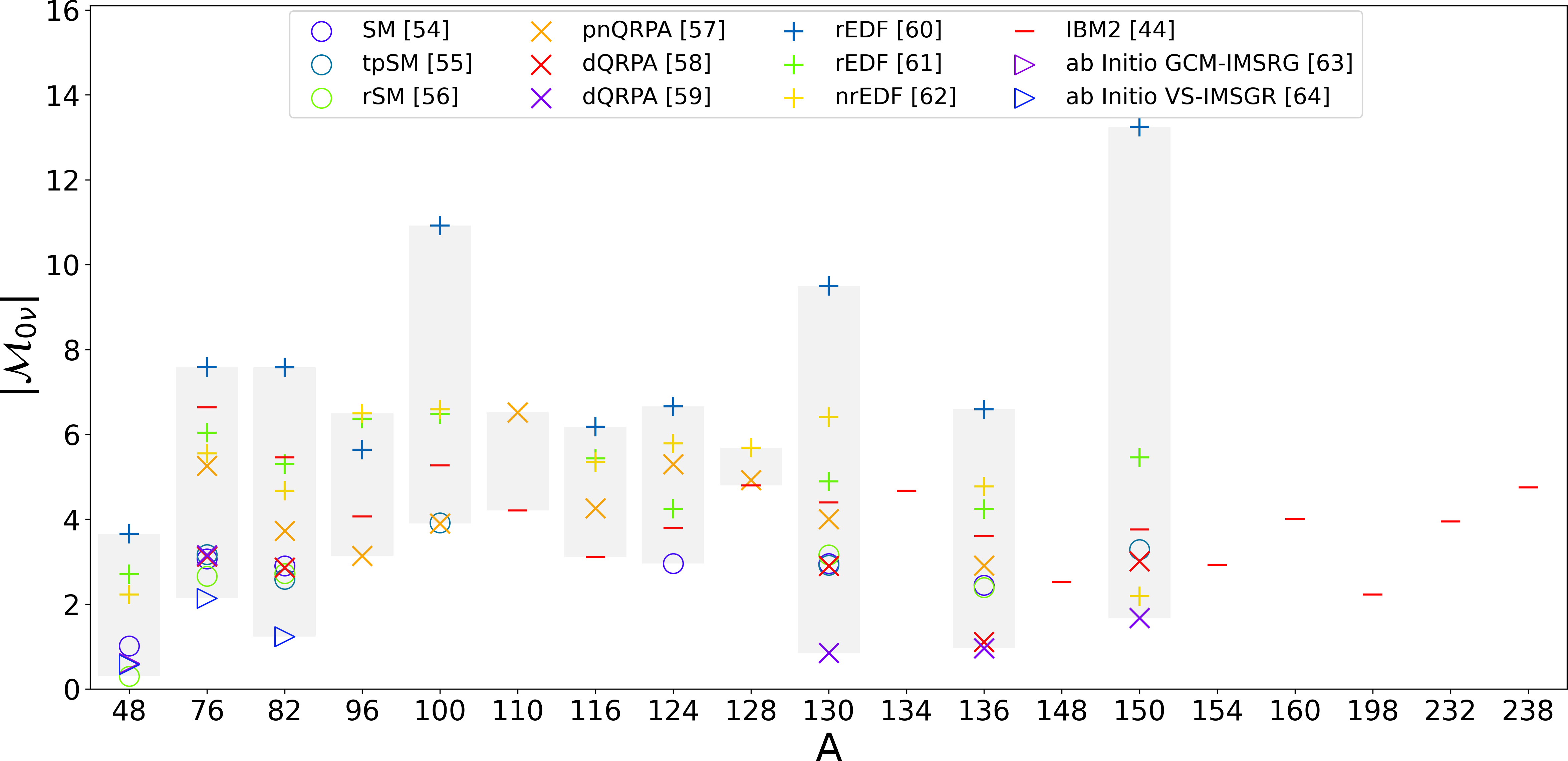}
    \caption{Comparison of the standard NME $\mathcal{M}_{0\nu} = -\frac{1}{g_A^2}M_F + M_{GT} + M_T$ resulting from different calculation methods. Explicitly, we show NMEs obtained from  the interacting shell model (SM)~\citep{menendez2017neutrinoless} and subsequent variants like the triaxial projected shell model (tpSM)~\citep{2021tpSM_NMEs} or realistic shell model (rSM)~\citep{2020rSM_NMEs}, the proton-neutron quasiparticle random phase approximation (pnQRPA)~\citep{PhysRevC.91.024613}, the deformed QRPA (dQRPA)~\citep{2018dQRPAFang,PhysRevC.87.064302}, the relativistic energy density functional method (rEDF) or covariant density functional theory (CDFT)~\citep{PhysRevC.91.024316,2017_rEDF_Song}, the non-relativistic energy density functional method (nrEDF)~\citep{PhysRevLett.111.142501}, the interacting boson model (IBM2)~\citep{Deppisch_2020} and recently introduced ab initio approaches calculating NMEs from basic principles of $\chi$PT~\citep{2020_ab_initio,2021_ab_initio}. The grey bands mark the range of values covered by the different methods.}
    \label{fig:NME_comparison}
\end{figure}

If we consider the theoretical uncertainty of the half-life to be dominated by the uncertainty of $M_\mathrm{eff}$, we can determine the necessary theoretical accuracy of the nuclear physics. To estimate this, we assume $M_\mathrm{eff}$ to be independent of the choice of the isotope, i.e.,
\begin{align}
    \frac{\Delta M_{\mathrm{eff}}}{M_{\mathrm{eff}}}(^AZ) = \frac{\Delta M_{\mathrm{eff}}}{M_{\mathrm{eff}}} = \mathrm{const.}.
\end{align}
Then the necessary theoretical accuracy can be determined from the simple condition that the expected ratios should be distinguishable from unity within the theoretical uncertainty,
\begin{align}
    \Delta R_{ij}\overset{!}{<} |R_{ij}-1|\;.
\end{align}
Hence, the necessary theoretical accuracy for $M_\mathrm{eff}$ reads
\begin{align}
    \frac{\Delta M_{\mathrm{eff}}}{M_{\mathrm{eff}}} \overset{!}{<} \frac{1}{4}\frac{|R_{ij}-1|}{R_{ij}}.
\end{align}
Again, taking the central values as a baseline, the theoretical uncertainty on the overall nuclear part (that is from both LECs and NMEs) would need to be brought down to $\frac{\Delta M_\mathrm{eff}}{M_\mathrm{eff}}\sim7\%$ to be able to identify all possible non-standard contributions assuming single operator dominance.
The exotic contribution easiest to identify using the half-life ratios would be the right-handed vector current $C_{VR}^{(6)}$ requiring an accuracy of $\frac{\Delta M_\mathrm{eff}}{M_\mathrm{eff}}\sim 22\%$. We want to emphasize, again, that this way of estimating nuclear uncertainties assumes that our calculations of the central ratios are a reasonable reflection of reality.

In Figure~\ref{fig:NME_comparison} we show the NME for the standard light-neutrino-exchange mechanism computed employing a variety of different numerical approaches. One can clearly see a significant variation of about a factor of $\sim3$ with some additional outliers corresponding to the rEDF (CDFT) approach. Given the distinct nature of individual nuclear structure computations the spread of the presented values clearly cannot be interpreted as theoretical uncertainty of the NMEs. 

Reaching the estimated required accuracy on both LECs and NMEs seems to be rather challenging considering the current status of the relevant nuclear physics calculations.
However, the recent advances in \emph{ab initio} approaches to the computation of \0 decay NMEs seem to pave the path towards more reliable numerical values and clearer understanding of the theoretical uncertainties involved.
\subsection{Other $\mathbf{0\nu\beta\beta}$ Modes}
Besides the usual $0\nu\beta^-\beta^-$-decay mode one could also make use of searches for neutrinoless modes of other $\beta\beta$ processes. In general there are four of these,
\begin{enumerate}
    \item \textbf{$\mathbf{\beta^-\beta^-}$:}
    \begin{align}
        (A,Z) &\rightarrow(A, Z+2) + 2e^- (+2\overline{\nu}_e)\\
        \Delta m &\overset{!}{>} 0\label{eq:threshold_bmbm}
    \end{align}
    \item \textbf{$\mathbf{\beta^+\beta^+}$:}
    \begin{align}
        (A,Z) &\rightarrow(A, Z-2) + 2e^+ (+2\nu_e)\\
        \Delta m &\overset{!}{>} 4m_e\label{eq:threshold_bpbp}
    \end{align}
    \item \textbf{$\text{EC}\mathbf{\beta^+}$:}
    \begin{align}
        (A,Z)+e^- &\rightarrow(A, Z-2) + e^+ (+2\nu_e)\\
        \Delta m &\overset{!}{>}2m_e\label{eq:threshold_ecbp}
    \end{align}
    \item \textbf{ECEC:}
    \begin{align}
        (A,Z)+2e^-&\rightarrow(A, Z-2) (+2\nu_e)\\
        \Delta m &\overset{!}{>}0\qquad\text{2$\nu$ mode}\label{eq:threshold_ecec}\\
        \Delta m &\overset{!}{=}0\qquad\text{0$\nu$ mode}.
    \end{align}
\end{enumerate}
Here, $\mathrm{EC}$ denotes electron capture. In principle, one could study all of these processes, as any of them would help to distinguish different mechanisms using the decay rate ratios considering nuclear uncertainties to be under control. However, all the other $\beta\beta$  processes listed in Table~\ref{tab:natural_double_beta_elements} are expected to have half-lives significantly longer than the usual $0\nu\beta^-\beta^-$ decay and are therefore unlikely to show up in experiments. Despite that let us discuss their potential role in a bit more detail.

\subsubsection{$\mathbf{0\nu \beta^+\beta^+}$}
This process can be treated in a similar way as the usual $0\nu\beta^-\beta^-$ decay, one only needs to consider a negative nuclear charge $Z\rightarrow -Z$ to calculate the positron wave functions. As such, the expected half-life will also be mainly determined by the PSF which goes with $Q^5$. Looking at the second column of Table~\ref{tab:natural_double_beta_elements} we can see that the $Q$-values for naturally occurring isotopes are up to one order of magnitude smaller than usual $0\nu\beta^-\beta^-$ $Q$-values. Additionally, the electromagnetic repulsion of the outgoing positrons deforms the wave functions and decreases the decay rate. Thus, we see that $0\nu \beta^+\beta^+$ will be highly suppressed compared to $0\nu\beta^-\beta^-$. Also, given the similarities of the two decays there does not seem to be a natural way of enhancing the $0\nu \beta^+\beta^+$-decay rate with respect to $0\nu\beta^-\beta^-$. The relevant PSFs for the $0\nu \beta^+\beta^+$-decay of naturally occurring isotopes have been calculated in~\citep{Kotila_2013} and are about 3-5 orders of magnitude smaller than for $0\nu \beta^-\beta^-$ decay.

\subsubsection{$\mathbf{0\nu EC\beta^+}$}
The PSFs for the neutrinoless mode of $\text{EC}\beta^+$ were also calculated with good precision in~\citep{Kotila_2013} and are found to be 3-4 orders of magnitude smaller than those corresponding to $0\nu\beta^-\beta^-$ decay. The reason is the same as in case of $0\nu \beta^+\beta^+$ decay.

\subsubsection{$\mathbf{0\nu ECEC}$}
As this process has no particles other than the daughter isotope in the final state, mass degeneracy between the mother and daughter isotopes is required in order to satisfy conservation of energy and momentum. However, as the daughter isotope is a non-stable state due to the holes left in the electron shell after the electron capture, the corresponding decay width results in a resonance mechanism~\citep{Krivoruchenko_2011,Kotila_2014}. Resonances are often found when considering nuclear excitations in the final state isotope. However, the resonant enhancement strongly depends on the degeneracy between the initial and final state and hence small uncertainties in the mass measurements of these nuclei result in considerable uncertainties of the corresponding half-lives~\citep{Krivoruchenko_2011}. Existing studies tend to show that the resulting half-lives are still considerably longer than for $0\nu\beta^-\beta^-$~\citep{Kotila_2014}. Therefore, we do not consider this process in this work. However, it is fair to note that a close resonance might lead to half-lives comparable or even shorter than for $0\nu\beta^-\beta^-$ decay. Recently, it was shown that further significant enhancement of the ${0\nu \mathrm{ECEC}}$ decay rate can be generated by a non-resonance shake mechanism~\citep{Karpeshin_2020}. In this case, the double electron capture is accompanied by emission of an electron from the shell of the final state isotope, which can carry away energy, thus making the whole process less dependent on the resonant behaviour. A dedicated review of the $0\nu\mathrm{ECEC}$ decay can be found in~\citep{Blaum:2020ogl}.

\subsubsection{Bound-State $\mathbf{0\nu\beta\beta}$}
Bound state \0 decay refers to a decay in which one or both of the two outgoing electrons end up in a bound energy level of the daughter isotope. It is usually referred to as $0\nu\beta \text{EP}$ and $0\nu \text{EPEP}$ for the one and two bound final state electrons, respectively, with EP denoting the electron production or electron placement. Being a reverse process of $0\nu \text{ECEC}$, also $0\nu \text{EPEP}$ requires mass degeneracy of the initial and final nuclei and the decay rate is described by a resonance-like mechanism. The explicit calculations show that the corresponding half-lives are even longer than those of double electron capture~\citep{Krivoruchenko_2011}. The reason is that there are no electron holes in the shell and only the (small) decay width of the nuclear excitation enters into the resonance.

The single bound state double-$\beta$-decay $0\nu\beta \text{EP}$ was investigated in~\citep{Babi__2018} and found to have PSFs 6-7 orders of magnitude smaller than those of $0\nu\beta^-\beta^-$ decay. The decay rates can be significantly enhanced when considering fully ionized nuclei. In that case, the $0\nu\beta \text{EP}$ decay rate can for certain isotopes even exceed the one of $0\nu\beta^-\beta^-$ decay~\citep{Babi__2018}. Although this is an interesting idea, a full ionization of large number of isotopes represents an experimental challenge. Therefore, despite the enhanced decay rate, the number of available ions would be too small to reach the relevant experimental sensitivity.

\subsubsection{Decay to excited final State Nuclei}
Instead of utilizing different isotopes to determine the decay rate ratios one could also compare the ground state decay with the decay into an excited state final nucleus ($0^+,2^+$) using the same initial state isotope~\citep{Tomoda_2000,Duerr_2011_consistency_test}. The potential benefit would be the possibility of studying this interplay within a single experiment. However, the excited state decays can be again expected to be highly suppressed due to the smaller phase-space resulting from the smaller $Q$-value. Additionally, previous studies tend to show that the NMEs for the decay into the excited final state are either of a similar size or smaller than those for the ground state decay~\citep{Duerr_2011_consistency_test,Men_ndez_2009,Barea_2015}, thus the half-lives would be rather further suppressed than enhanced by the nuclear part of the amplitude either.

\subsubsection{Artificial Isotopes}
Although there are 69 naturally occurring double-$\beta$-decaying isotopes, we found about $\sim2700$ possible \0 candidate isotopes when considering the full NIST list of elements~\citep{ElementList}. Some of them have considerably larger Q-values of up to $50\,\rm{MeV}$.\footnote{Considering only isotopes without a single-$\beta$-decay mode already significantly reduces this number down to 86. None of these has, however, a significantly enhanced $Q$-value.} While such a large $Q$-value of $\sim50\,\mathrm{MeV}$ would result in a significant enhancement of the decay rate by $\sim8$ orders of magnitude, there are several fairly obvious experimental problems. Primarily, it is the artificial production of these isotopes, which would strongly limit the scale of the experiment. Again ton, kilogram and even gram scales would usually not be possible. Additionally, many artificial isotopes, especially those with large $Q$-values, come with additional decay modes that strongly dominate and often lead to extremely short half-lives such that storing them to study \0 decay would be impossible.

To sum up the above paragraphs, despite the fact that a variety of $\beta\beta$ processes exist, $0\nu\beta^-\beta^-$ decay is largely the most relevant candidate to study, indeed. Therefore, other possible \0 modes would only become relevant in exotic scenarios leading either to their significant enhancements, or to strong suppression of $0\nu\beta^-\beta^-$ decay. Given the similarity of all the $\beta\beta$ processes, such models would, however, seem to be rather unnatural from a particle physics point of view. 

\section{Distinguishing Specific Models}\label{sec:specific_models}
Following the discussion of possible discrimination among different LEFT operators, let us now have a brief look at complete models. As one would expect, lepton number violating BSM models will typically excite several LEFT operators at a time. While it would be challenging to identify a specific BSM model, as no finite set of BSM models exists and many different scenarios would result in the same low-energy physics, we do expect that, given fixed model parameters, one can at least check whether a model is consistent with the observed data and reject it if it is not. In the following paragraphs, we adopt and briefly discuss three different BSM scenarios that would lead to \0 decay. Each of the models will be compared with the standard mass mechanism predictions.

\subsection{Minimal Left-Right Symmetric Model}

The Standard Model is a chiral theory. That is, parity is explicitly broken due to the gauged $SU(2)_L$ symmetry and the missing right-handed neutrino. This particular choice of symmetries and particle content, additionally, results in vanishing neutrino masses. A simple approach to resolve these phenomena is to extend the Standard Model's gauge group to a left-right symmetric model $SU(3)_C\times SU(2)_L\times SU(2)_R\times U(1)_{B-L}$~\citep{PhysRevD.10.275,PhysRevD.11.2558,PhysRevD.12.1502} which is spontaneously broken to the Standard Model group $SU(3)_C\times SU(2)_L\times U(1)_Y$. A comprehensive review of the minimal left-right symmetric Standard Model (mLRSM) is given in e.g.~\citep{Duka_2000}.

Extending the Standard Model to the left-right symmetric theory requires the existence of additional scalars and fermions. The conventional minimal setting includes two scalar triplets $\Delta_L\in\left(1, 3, 1, 2\right)$ and $\Delta_R\in\left(1, 1, 3, 2\right)$ as well as a scalar bidoublet $\Phi\in\left(1, 2, 2^*, 0\right)$ incorporating the SM Higgs doublet and the right-handed neutrinos $\nu_R$.
The fermions are grouped into left- and right-handed doublets
\begin{align}
    L_L &= \left(\begin{matrix}\nu_L\\e_L\end{matrix}\right)\in\left(1, 2, 1, -1\right),\qquad Q_L= \left(\begin{matrix}u_L\\d_L\end{matrix}\right)\in\left(3, 2, 1, 1/3\right),\\
    L_R &= \left(\begin{matrix}\nu_R\\e_R\end{matrix}\right)\in\left(1, 1, 2, -1\right),\qquad Q_R= \left(\begin{matrix}u_R\\d_R\end{matrix}\right)\in\left(3, 1, 2, 1/3\right),
\end{align}
which under $U\in SU(2)_{L,R}$ transform as
\begin{align}
    \Psi_{L,R}\longrightarrow U_{L,R}\Psi_{L,R}
\end{align}
while the scalar fields transform as
\begin{align}
    \Phi\longrightarrow U_L\Phi U^\dagger_R,\qquad \Delta_L\longrightarrow U_L\Delta_LU_L^\dagger,\qquad \Delta_R\longrightarrow U_R\Delta_RU_R^\dagger.
\end{align}
There are two discrete symmetries that one can impose onto a LR symmetric theory which can relate left- and right-handed fermions. These are parity $P$ and charge conjugation $C$~\citep{Dekens_2014}. Thus, one can define two different discrete symmetry transformations
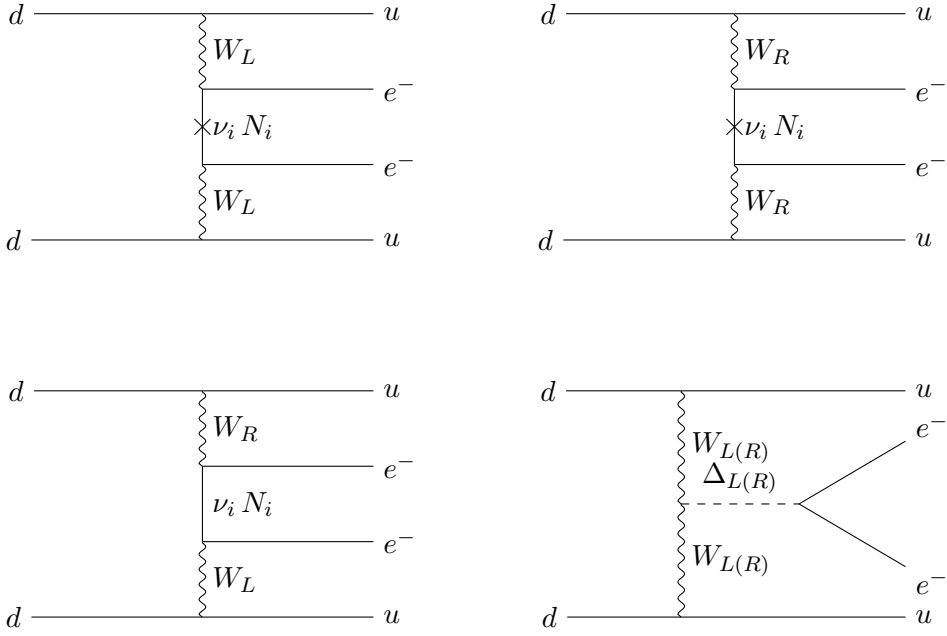
\begin{figure}[t!]
    \centering
    \begin{tikzpicture}
      \begin{feynman}
      \vertex(n){\(d\)};
        \vertex[right=2.45cm of n](neLQ2);
        \vertex[right=2.25cm of neLQ2](e1){\(u\)};
        \vertex[below=1cm of neLQ2](pnu1);
        \vertex[right=2.25cm of pnu1](p1){\(e^-\)};
        \vertex[below=1cm of pnu1](wnue1);
        \vertex[right=2.25cm of wnue1](e2){\(e^-\)};
        \vertex[below=1cm of wnue1](npw1);
        \vertex[left=2.25cm of npw1](n2){\(d\)};
        \vertex[right=2.25cm of npw1](p2){\(u\)};
        \diagram* {
        (n) --  (neLQ2) --  (e1),
        (neLQ2) -- [boson, edge label=\(W_L\)] (pnu1) -- (p1),
        (pnu1) -- [insertion=0.5, edge label=\(\nu_i\,N_i\)] (wnue1) -- (e2),
        (wnue1) -- [boson, edge label=\(W_L\)] (npw1) -- (p2),
        (n2) -- (npw1),
          };
          
        \vertex[right=7cm of n](n3){\(d\)};
        \vertex[right=2.45cm of n3](neLQ);
        \vertex[right=2.25cm of neLQ](e3){\(u\)};
        \vertex[below=1cm of neLQ](pnu);
        \vertex[right=2.25cm of pnu](p3){\(e^-\)};
        \vertex[below=1cm of pnu](wnue2);
        \vertex[right=2.25cm of wnue2](e4){\(e^-\)};
        \vertex[below=1cm of wnue2](npw2);
        \vertex[left=2.25cm of npw2](n4){\(d\)};
        \vertex[right=2.25cm of npw2](p4){\(u\)};
        \diagram* {
        (n3) -- (neLQ) -- (e3),
        (neLQ) -- [boson, edge label=\(W_R\)] (pnu) -- (p3),
        (pnu) -- [insertion=0.5, edge label=\(\nu_i\,N_i\)] (wnue2) -- (e4),
        (wnue2) -- [boson, edge label=\(W_R\)] (npw2) -- (p4),
        (n4) -- (npw2),
          };


      \vertex[below=5cm of n](n1){\(d\)};
        \vertex[right=2.45cm of n1](neLQ21);
        \vertex[right=2.25cm of neLQ21](e11){\(u\)};
        \vertex[below=1cm of neLQ21](pnu11);
        \vertex[right=2.25cm of pnu11](p11){\(e^-\)};
        \vertex[below=1cm of pnu11](wnue11);
        \vertex[right=2.25cm of wnue11](e21){\(e^-\)};
        \vertex[below=1cm of wnue11](npw11);
        \vertex[left=2.25cm of npw11](n21){\(d\)};
        \vertex[right=2.25cm of npw11](p21){\(u\)};
        \diagram* {
        (n1) --  (neLQ21) --  (e11),
        (neLQ21) -- [boson, edge label=\(W_R\)] (pnu11) -- (p11),
        (pnu11) -- [edge label=\(\nu_i\,N_i\)] (wnue11) -- (e21),
        (wnue11) -- [boson, edge label=\(W_L\)] (npw11) -- (p21),
        (n21) -- (npw11),
          };
          
        \vertex[right=7cm of n1](n31){\(d\)};
        \vertex[right=1.75cm of n31](neLQ1);
        \vertex[right=2.95cm of neLQ1](e31){\(u\)};
        \vertex[below=1.5cm of neLQ1](pnu1);
        \vertex[below=1.5cm of pnu1](npw21);
        \vertex[left=1.5cm of npw21](n41){\(d\)};
        \vertex[right=2.95cm of npw21](p41){\(u\)};
        \vertex[right=1.55cm of pnu1](delta1);
        \vertex[right=1.75cm of delta1](mid1);
        \vertex[above=0.75cm of mid1](lepton11){\(e^-\)};
        \vertex[below=0.75cm of mid1](lepton21){\(e^-\)};
        
        \diagram* {
        (n31) -- (neLQ1) -- (e31),
        (neLQ1) -- [boson, edge label=\(W_{L(R)}\)] (pnu1),
        (pnu1) -- [boson, edge label=\(W_{L(R)}\)] (npw21) -- (p41),
        (n41) -- (npw21),
        (pnu1) -- [scalar, edge label=\(\Delta_{L(R)}\)] (delta1) -- (lepton11),
        (delta1) -- (lepton21);
          };
      \end{feynman}
    \end{tikzpicture}
    \caption[Feynman diagrams arising in the mLRSM that contribute to \0.]{Feynman diagrams arising in the mLRSM that contribute to \0. Here, $\nu_i$ and $N_i$ represent the light and heavy neutrino mass eigenstates. It should be noted that, due to mixing of both left- and right-handed neutrinos and gauge bosons, each diagram (except the triplet-exchange diagram) comes with all possible combinations of the outgoing particle's chiralities. However, some diagrams are highly suppressed compared to others.}
    \label{fig:feynman_mLRSM}
\end{figure}
\begin{align}
    &P: \Psi_L\Longleftrightarrow\Psi_R,\qquad\Phi\Longleftrightarrow\Phi^\dagger,\qquad \Delta_{L,R}\Longleftrightarrow\Delta_{R,L}\\
    &C: \Psi_L\Longleftrightarrow\left(\Psi_R\right)^c,\qquad\Phi\Longleftrightarrow\Phi^T,\qquad \Delta_{L,R}\Longleftrightarrow\Delta_{R,L}^*.
\end{align}
Requiring either $P$ or $C$ invariance results in different constraints on the scalar potential as well as the Yukawa coupling matrices~\citep{Dekens_2014}.\footnote{Note that a combination of both does not fit observational constraints~\citep{Dekens_2014}.}

The lepton number violation at low energy stems from the leptonic Yukawa interactions given by
\begin{align}
    \mathcal{L}_{y} = \sum_{ij} \bigg\{
     Y^l_{ij}{\overline{L_L}}_i\Phi L_{R,j} + \Tilde{Y}^l_{ij}\overline{L_L}_{i}\Tilde{\Phi} L_{R,j} + Y^L_{ij}L^T_{L,i}Ci\tau_2\Delta_L L_{L,j} + {Y^R_{ij}}^\dagger L^T_{R,i}Ci\tau_2\Delta_R L_{R,j}
     \bigg\} + \text{h.c.}\label{eq:mLRSMyukawas}
\end{align}
After the neutral components of the scalars have acquired their VEVs
\begin{align}
    \left<\Phi\right> = \frac{1}{\sqrt{2}}\left(\begin{matrix}\kappa & 0 \\ 0 & \kappa'e^{i\alpha}\end{matrix}\right),\qquad \left<\Delta_L\right> = \frac{1}{\sqrt{2}}\left(\begin{matrix}0 & 0 \\ v_Le^{i\theta_L} & 0\end{matrix}\right),\qquad \left<\Delta_R\right> = \frac{1}{\sqrt{2}}\left(\begin{matrix}0 & 0 \\ v_R & 0\end{matrix}\right),
\end{align}
one can infer the neutrino mass matrices from~\eqref{eq:mLRSMyukawas} 
\begin{align}
    \begin{split}
    M^\nu_{D,ij} &= \frac{1}{\sqrt{2}}\left[Y^l_{ij}\kappa + \Tilde{Y}^l_{ij}\kappa'\exp{-i\alpha}\right],\quad\; \\
    {M^\nu_{L,ij}}^\dagger &= \sqrt{2}Y^L_{ij}v_L\exp{i\theta_L},\qquad\qquad\quad\;\;\; M^\nu_{R,ij}=\sqrt{2}Y^R_{ij}v_R\;.
\end{split}
\end{align}
There are several diagrams in the mLRSM setting which can contribute to the $0\nu\beta\beta$ decay at tree level, see Figure~\ref{fig:feynman_mLRSM}. Detailed discussions of $0\nu\beta\beta$ decay within the mLRSM scenario can be found e.g. in~\citep{Deppisch:2012nb,Huang_2014,Li:2020flq}.

The matching of the $C$-symmetric mLRSM onto SMEFT and, subsequently, onto the relevant LEFT operators has been discussed in~\citep{Cirigliano_2018}. Here, we will summarize their findings and study the distinguishability from the usual mass mechanism.

Integrating out the heavy fields with masses proportional to $v_R$ and matching the theory onto SMEFT results in the lepton number violating operators,
\begin{align}
\begin{split}
    \mathcal{L}_{\Delta L}=&C^{(5)}\left(\left(L^TCi\tau_2\Phi_{SM}\right)\left(\Tilde{\Phi}_{SM}^\dagger L\right)\right) \\
    &+ \left(L^T\gamma^\mu e_R\right)i\tau_2\Phi_{SM}\left(C^{(7)}_{Leu\overline{d}\Phi}\overline{d_R}\gamma_\mu u_R +C^{(7)}_{L\Phi De}\Phi_{SM}^Ti\tau_2(D_\mu\Phi_{SM}) \right)\\
    &+\overline{e_R}e_R^c\bigg(C^{(9)}_{eeud}\overline{u_R}\gamma^\mu d_R\overline{u_R}\gamma_\mu d_R + C^{(9)}_{ee\Phi ud}\overline{u_R}\gamma^\mu d_R\left(\left[iD_\mu\Phi_{SM}\right]^\dagger\Tilde{\Phi}_{SM}\right)\\
    &\qquad\qquad+ .C^{(9)}_{ee\Phi D}\left(\left[iD_\mu\Phi_{SM}\right]^\dagger\Tilde{\Phi}_{SM}\right)^2\bigg),
\end{split}
\end{align}
where $\Phi_{SM}$ is the Standard Model Higgs doublet. The matching scale corresponds to $\sim m_{W_R}$ and the Wilson coefficients at SMEFT level are given by
\begin{align}
\begin{split}
    C^{(5)}\;\;\;\;\,&=\frac{1}{v^2}\left({M_D^{\nu}}^T{M_R^\nu}^{-1}M_D^\nu-M_L^\nu\right)\,,\\
    C^{(7)}_{Leu\overline{d}\Phi}&=\frac{\sqrt{2}}{v}\frac{1}{v_R^2}\left(V_R^{ud}\right)^*\left(M_D^{\nu T}{M_R^\nu}^{-1}\right)_{ee}\,,\qquad\qquad\quad
    C^{(7)}_{L\Phi De}\,=\frac{2i\xi\exp{i\alpha}}{\left(1+\xi^2\right){V_R^{ud}}^*}C^{(7)}_{Leu\overline{d}\Phi}\,,\\
    C^{(9)}_{eeud}\;\;\,&=-\frac{1}{2v_R^4}{V_R^{ud}}^2\left[\left({M_R^\nu}^\dagger\right)^{-1}+\frac{2}{m^2_{\Delta_R}}M_R^\nu\right]\,,\qquad
    C^{(9)}_{ee\Phi ud}\,=-4\frac{\xi\exp{-i\alpha}}{\left(1+\xi^2\right)V_R^{ud}}C^{(9)}_{eeud}\,,\\
    C^{(9)}_{ee\Phi D}\;&=4\frac{\xi^2\exp{-2i\alpha}}{\left(1+\xi^2\right)^2 {V_R^{ud}}^2}C^{(9)}_{eeud}\,,
\end{split}
\end{align}
where $v$ is the Standard Model Higgs doublets VEV,
\begin{align}
    v^2 = \kappa^2+\kappa'^2.
\end{align}
Here, $C^{(5)}$ corresponds to the usual seesaw formula. From the matching scale $\sim m_{W_R}$ the above coefficients have to be evolved down to $m_W\sim 80\,\rm{GeV}$, at which one can match onto the relevant LEFT operators by integrating out the remaining heavy particles with masses above $m_W$. By doing so one obtains
\begin{align}
\begin{split}
    &m_{\beta\beta}=-v^2 C^{(5)}_{ee}\\
    &C_{VL}^{(6)}=-iV_L^{ud}\frac{v^3}{\sqrt{2}}{C_{L\Phi De}^{(7)}}^*\,,\qquad\qquad\quad\,
    C_{VR}^{(6)}=\frac{v^3}{\sqrt{2}}{C^{(7)}_{Leu\overline{d}\Phi}}^*\,,\\
    &C_{1R}^{(9)}(m_W)=v^5 {V_L^{ud}}^2 C_{ee\Phi D}^{(9)}(m_W)\,,\qquad
    C_{1R}^{(9)}{'}(m_W)=v^5 C_{eeud}^{(9)}(m_W)\,,\\
    &C_{4R}^{(9)}(m_W)=-v^5V_L^{ud} C_{ee\Phi ud}^{(9)}(m_W).
\end{split}
\end{align}
Evolving the above coefficients down to the $\chi$PT scale of $\sim2\,\rm{GeV}$ also generates a non-zero $C_{5R}^{(9)}$ coefficient since the RGEs of $C_{4,5}^{(9)}$ mix.

The relevant Wilson coefficients are fixed by several physical parameters: the values of the triplet VEVs $v_{L,R}$, the mass of the heavy right-handed triplet $m_{\Delta R}$ as well as the masses of the three heavy neutrinos $\left(m_{\nu_{R1}}, m_{\nu_{R2}}, m_{\nu_{R3}}\right)$ and the lightest neutrino mass $m_{\nu_{\text{min}}}$, the complex phases of the VEVs $\alpha$ and $\theta_L$ and finally the left-right mixing parameter $\xi$. Here, we fix $\xi=\frac{m_b}{m_t}$. The lightest neutrino mass together with the squared mass differences $\Delta m_{ij}^2$ that are known from oscillation data fix the remaining light neutrino masses for a given mass hierarchy. Taking
\begin{align}
    \ket{\nu_{\alpha\; L,R}} = \sum_i U_{\alpha i}^*\ket{\nu_{i\; L,R}}
\end{align}
we obtain
\begin{align}
    \begin{split}
    M_\nu &= v^2C^{(5)} = U_{\text{PMNS}}\,m_\nu \,U_{\text{PMNS}}^T\;,\qquad\;\;\, m_\nu\; = \diag\left(m_{\nu_1}, m_{\nu_2}, m_{\nu_3}\right)\\
    M^\nu_R &=  U_R\, m_{\nu R}\, U_R^T\;,\qquad\qquad\qquad\qquad\qquad m_{\nu R} = \diag\left(m_{{\nu}_{R1}}, m_{{\nu}_{R2}}, m_{{\nu}_{R3}}\right)
    \end{split}
\end{align}
Additionally, the mixing matrix $U$ for the heavy neutrinos must be fixed. Here, we take $U_R=U_{\text{PMNS}}$ for simplicity. In the $C$-symmetric case, one has
\begin{align}
    M_L^\nu = \frac{v_L\exp{i\theta_L}}{v_R}M^R_\nu
\end{align}
and the Dirac mass matrix can be derived as~\citep{Nemevsek:2012iq}
\begin{align}
    M^\nu_D=U_{\text{PMNS}}\;m_{\nu_ R}\sqrt{\frac{v_L\exp{i\theta_L}}{v_R}\mathbb{1}_{(3\times 3)}-m_{\nu R}^{-1}m_\nu}\;U_{\text{PMNS}}^T\;.
\end{align}
\begin{figure}
    \centering
    \includegraphics[width=\textwidth]{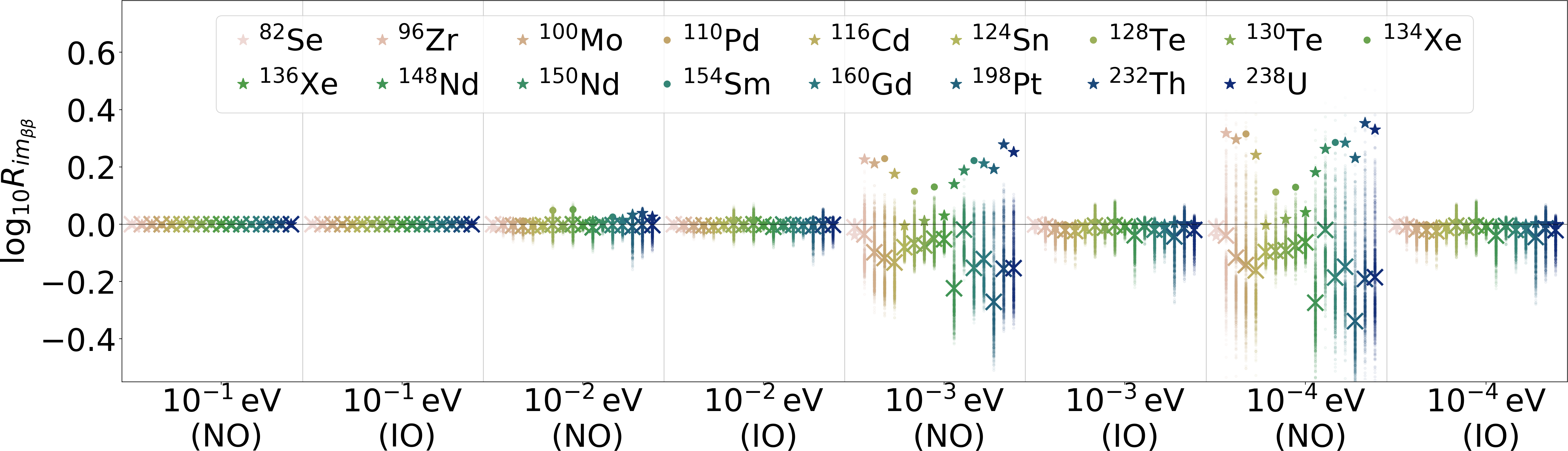}
    \caption[Half-life ratios resulting from different mLRSM settings when taking \textsuperscript{76}Ge as the reference isotope.]{Half-life ratios resulting from different mLRSM settings (different neutrino mass hierarchy and different minimal neutrino mass) when taking \textsuperscript{76}Ge as the reference isotope. The ratios are compared to the standard mass mechanism. We vary both the unknown LECs as well as the unknown phases of the mLRSM model.
    }
    \label{fig:mLRSM_ratios}
\end{figure}
Assuming $V_L^{ud}=V_R^{ud}$ and
\begin{align}
\begin{split}
    &m_{{\nu}_{R1}} = 10\,\textrm{TeV}\;,\quad m_{{\nu}_{R2}}= 12\,\textrm{TeV}\;,\quad m_{{\nu}_{R3}} = 13\,\textrm{TeV}\\
    &m_{\Delta_R} = 4\,\textrm{TeV}\,,\quad v_L=0.1\,\textrm{eV}\quad v_R=10\,\textrm{TeV}
\end{split}
\end{align}
as in~\citep{Cirigliano_2018}, we can derive the LEFT Wilson coefficients in dependence on the minimal light neutrino mass $m_{\text{min}}$, the Majorana phases entering $U_{\text{PMNS}}$ and the VEV phases $\theta_L$ and $\alpha$.

The resulting phase-space observables in this parameter setting of the mLRSM are hardly any different from the standard mechanism. This is because of the specific choice of parameters studied here which results in the scalar short-range contributions dominating over the long-range contributions. Hence, the phase-space is almost indistinguishable from the standard mechanism.

\begin{figure}[t]
    \centering
    \includegraphics[width = 0.7\textwidth]{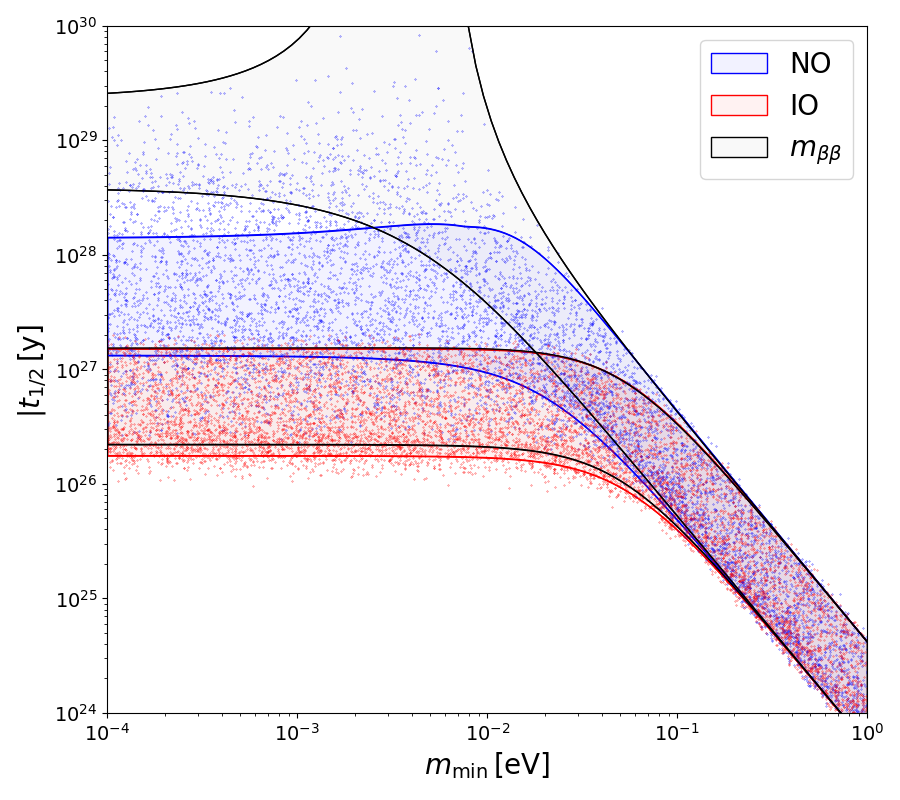}
    \caption{Here we present the half-life of \textsuperscript{76}Ge in the mLRSM model in dependency on the minimal neutrino mass $m_\mathrm{min}$ for both normal (NO) and inverted ordering (IO). The blue and red areas represent the scenario marginalized over the unknown phases with the LECs fixed to their order of magniture estimate while the scattered points show the additional variation of the relevant LECs. The area inside the black borders represents the usual mass-mechanism without any additional contributions.}
    \label{fig:lobster_mLRSM}
\end{figure}

The resulting half-life ratios normalized to the standard mass mechanism are depicted in Figure~\ref{fig:mLRSM_ratios}. Here, additionally to varying the unknown LECs we also marginalized over the unknown phases. We can see that assuming inverted mass ordering there are only minor variations from the standard mechanism. In the case of normal ordering, the non-standard contributions alter the ratios notably only for small $m_\mathrm{min}\leq 10^{-3}\,$eV. In this region, as shown before in Figure~\ref{fig:ratios_ROi}, the central values of the variation differ significantly from the benchmark scenario. A similar behavior is manifested in Figure~\ref{fig:lobster_mLRSM} displaying the half-life in dependence on the minimal neutrino mass $m_\mathrm{min}$ for both orderings and in comparison with the standard mechanism on its own. One can see that in the case of inverted ordering the half-life is almost unaltered from the standard mechanism while for normal ordering the non-standard contributions start to play a substantial role below $\sim 10^{-2}\,$eV decreasing the expected half-life by about one order of magnitude compared to the standard scenario. In the same range of $m_\mathrm{min}$ the uncertainties induced by the unknown LECs start to significantly influence the predicted half-life. On the other hand, the central values of the decay rate ratios alter for $m_\mathrm{min}\lesssim 10^{-3}\,$eV at most by a factor of $R_{im_{\beta\beta}}^{\mathrm{max}}\sim 2.2$ with \textsuperscript{76}Ge as the reference isotope. The reason for this behaviour can be traced back to the dominance of the short-range contributions which (see Section~\ref{sec:distinguishing}) result in relatively small $R_{im_{\beta\beta}}$ despite the appearance of $C_{VR}^{(6)}$. This ratio would translate to a necessary accuracy on the nuclear part of the amplitude of $\frac{\Delta M_\mathrm{eff}}{M_\mathrm{eff}}\lesssim 14\%$.

%
\subsection{Gluino- and Neutralino-exchange in $\cancel{R}_p$ - SUSY}
\label{subsec:rp_susy}
%
Supersymmetric theories contain supermultiplets of fermions and bosons which, under supersymmetry, transform into each other. The most simple constructions are chiral supermultiplets
\begin{align}
    \left(\Psi_{L,R},\Phi_{L,R}^\Psi\right)
\end{align}
which relate two component chiral spinors $\left(\Psi_{L,R}\right)$ and a corresponding complex scalar $\Phi_{L,R}$. To construct a supersymmetric version of the Standard Model, one also needs to consider gauge supermultiplets
\begin{align}
    \left(A_\mu^a,\Psi^a\right)
\end{align}
which relate the Standard Model's gauge bosons $A_\mu^a$ to their superpartner fermions $\Psi^a$. One should note that since gauge bosons have 2 degrees of freedom (d.o.f.) and since this kind of transformation obviously cannot change the number of d.o.f., their superpartners $\Psi^a$ also have 2 degrees of freedom. Therefore, they are Majorana fermions.
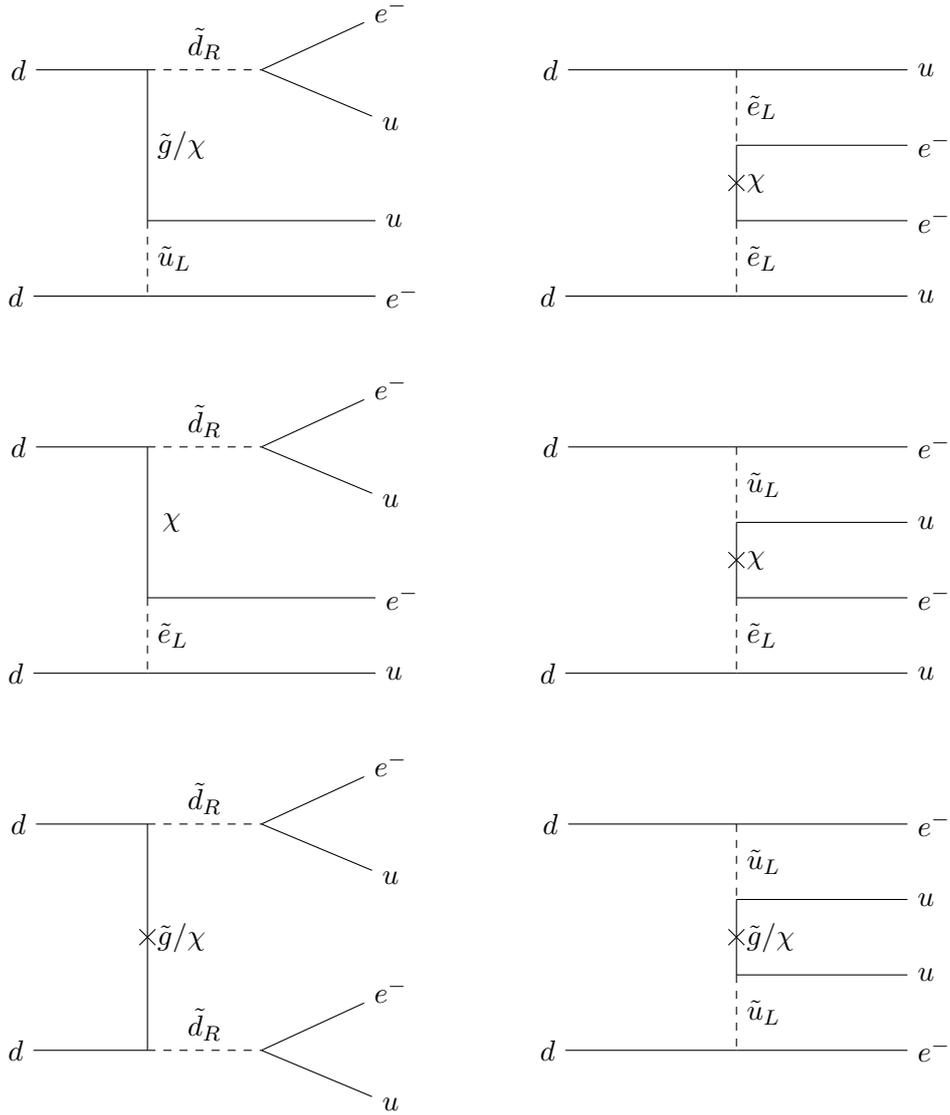
\begin{figure}[t!]
    \centering
    \begin{tikzpicture}
      \begin{feynman}
        \vertex (n){\(d\)};
        \vertex[right=1.7cm of n](nnuLQ);
        \vertex[right=1.5cm of nnuLQ](peLQ);
        \vertex[right=1.7cm of peLQ](pemiddle);
        \vertex[above=0.5cm of pemiddle](p){\(e^-\)};
        \vertex[below=0.5cm of pemiddle](e){\(u\)};
        \vertex[below=2cm of nnuLQ](wnue);
        \vertex[right=3cm of wnue] (e2){\(u\)};
        \vertex[below=1cm of wnue](npw);
        \vertex[left=1.5cm of npw](n2){\(d\)};
        \vertex[right=3cm of npw](p2){\(e^-\)};
        \diagram* {
          (n) --  (nnuLQ) -- [edge label=\(\Tilde{g}/\chi\)] (wnue),
          (n2) --  (npw) --  (p2),
          (nnuLQ) -- [scalar, edge label=\(\Tilde{d}_R\)] (peLQ) -- (p),
          (peLQ) -- (e),
          (wnue) -- [scalar, edge label=\(\Tilde{u}_L\)] (npw),
          (wnue) -- (e2),
          };
          
        \vertex[right=7cm of n](n3){\(d\)};
        \vertex[right=2.45cm of n3](neLQ);
        \vertex[right=2.25cm of neLQ](e3){\(u\)};
        \vertex[below=1cm of neLQ](pnu);
        \vertex[right=2.25cm of pnu](p3){\(e^-\)};
        \vertex[below=1cm of pnu](wnue2);
        \vertex[right=2.25cm of wnue2](e4){\(e^-\)};
        \vertex[below=1cm of wnue2](npw2);
        \vertex[left=2.25cm of npw2](n4){\(d\)};
        \vertex[right=2.25cm of npw2](p4){\(u\)};
        \diagram* {
        (n3) -- (neLQ) -- (e3),
        (neLQ) -- [scalar, edge label=\(\Tilde{e}_L\)] (pnu) -- (p3),
        (pnu) -- [insertion=0.5, edge label=\(\chi\)] (wnue2) -- (e4),
        (wnue2) -- [scalar, edge label=\(\Tilde{e}_L\)] (npw2) -- (p4),
        (n4) -- (npw2),
          };


        \vertex[below=5cm of n] (n1){\(d\)};
        \vertex[right=1.7cm of n1](nnuLQ1);
        \vertex[right=1.5cm of nnuLQ1](peLQ1);
        \vertex[right=1.7cm of peLQ1](pemiddle1);
        \vertex[above=0.5cm of pemiddle1](p1){\(e^-\)};
        \vertex[below=0.5cm of pemiddle1](e1){\(u\)};
        \vertex[below=2cm of nnuLQ1](wnue1);
        \vertex[right=3cm of wnue1] (e21){\(e^-\)};
        \vertex[below=1cm of wnue1](npw1);
        \vertex[left=1.5cm of npw1](n21){\(d\)};
        \vertex[right=3cm of npw1](p21){\(u\)};
        \diagram* {
          (n1) -- (nnuLQ1) -- [edge label=\(\,\chi\)] (wnue1),
          (n21) --  (npw1) -- (p21),
          (nnuLQ1) -- [scalar, edge label=\(\Tilde{d}_R\)] (peLQ1) -- (p1),
          (peLQ1) --  (e1),
          (wnue1) -- [scalar, edge label=\(\Tilde{e}_L\)] (npw1),
          (wnue1) --  (e21),
          };
          
        \vertex[right=7cm of n1](n31){\(d\)};
        \vertex[right=2.45cm of n31](neLQ1);
        \vertex[right=2.25cm of neLQ1](e31){\(e^-\)};
        \vertex[below=1cm of neLQ1](pnu1);
        \vertex[right=2.25cm of pnu1](p31){\(u\)};
        \vertex[below=1cm of pnu1](wnue21);
        \vertex[right=2.25cm of wnue21](e41){\(e^-\)};
        \vertex[below=1cm of wnue21](npw21);
        \vertex[left=2.25cm of npw21](n41){\(d\)};
        \vertex[right=2.25cm of npw21](p41){\(u\)};
        \diagram* {
        (n31) -- (neLQ1) -- (e31),
        (neLQ1) -- [scalar, edge label=\(\Tilde{u}_L\)] (pnu1) -- (p31),
        (pnu1) -- [insertion=0.5, edge label=\(\chi\)] (wnue21) -- (e41),
        (wnue21) -- [scalar, edge label=\(\Tilde{e}_L\)] (npw21) -- (p41),
        (n41) -- (npw21),
          };
          
          
        \vertex[below=5cm of n1] (n2){\(d\)};
        \vertex[right=1.7cm of n2](nnuLQ2);
        \vertex[right=1.5cm of nnuLQ2](peLQ2);
        \vertex[right=1.7cm of peLQ2](pemiddle2);
        \vertex[above=0.5cm of pemiddle2](p2){\(e^-\)};
        \vertex[below=0.5cm of pemiddle2](e2){\(u\)};
        \vertex[below=2cm of nnuLQ2](wnue2);
        \vertex[below=1cm of wnue2](npw2);
        \vertex[left=1.5cm of npw2](n22){\(d\)};
        \vertex[right=1.5cm of npw2](peLQ22);
        \vertex[right=1.7cm of peLQ22](pemiddle22);
        \vertex[above=0.5cm of pemiddle22](p22){\(e^-\)};
        \vertex[below=0.5cm of pemiddle22](e22){\(u\)};
        \diagram* {
          (n2) -- (nnuLQ2) -- [insertion=0.5, edge label=\(\Tilde{g}/\chi\)] (npw2),
          (n22) -- (npw2),
          (nnuLQ2) -- [scalar, edge label=\(\Tilde{d}_R\)] (peLQ2) -- (p2),
          (peLQ2) -- (e2),
          (npw2) -- [scalar, edge label=\(\Tilde{d}_R\)] (peLQ22) -- (p22),
          (peLQ22) -- (e22),
          };
          
        \vertex[right=7cm of n2](n32){\(d\)};
        \vertex[right=2.45cm of n32](neLQ2);
        \vertex[right=2.25cm of neLQ2](e32){\(e^-\)};
        \vertex[below=1cm of neLQ2](pnu2);
        \vertex[right=2.25cm of pnu2](p32){\(u\)};
        \vertex[below=1cm of pnu2](wnue22);
        \vertex[right=2.25cm of wnue22](e42){\(u\)};
        \vertex[below=1cm of wnue22](npw22);
        \vertex[left=2.25cm of npw22](n42){\(d\)};
        \vertex[right=2.25cm of npw22](p42){\(e^-\)};
        \diagram* {
        (n32) -- (neLQ2) -- (e32),
        (neLQ2) -- [scalar, edge label=\(\Tilde{u}_L\)] (pnu2) -- (p32),
        (pnu2) -- [insertion=0.5, edge label=\(\Tilde{g}/\chi\)] (wnue22) -- (e42),
        (wnue22) -- [scalar, edge label=\(\Tilde{u}_L\)] (npw22) -- (p42),
        (n42) -- (npw22),
          };
      \end{feynman}
    \end{tikzpicture}
    \caption[Feynman diagrams contributing to \0 within the $\cancel{R}_p$-MSSM.]{Feynman diagrams of gluino- and neutralino-exchange contributing to \0 within the $\cancel{R}_p$-MSSM~\citep{Hirsch_1996_susy}.}
    \label{fig:feynman_rp-susy}
\end{figure}
As particles within a supermultiplet must share the same mass, quantum numbers (except spin), interactions and couplings, SUSY must be broken at low energies to reproduce the experimentally confirmed SM predictions. Typically, after SUSY breaking there remains a discrete symmetry called $R$-parity ($R_p$) which can be assigned to every field, such that we have $R_p=+1$ for Standard Model fields and $R_p=-1$ for the superpartner fields. One can define $R$-parity as~\citep{MARTIN_1998}
\begin{align}
    R_p = \left(-1\right)^{2s+3(B-L)}
\end{align}
where $s$ is the spin and $B$ and $L$ are the corresponding baryon and lepton numbers of the field, respectively. If $R_p$ is a conserved quantity, it follows that the lightest superpartner cannot decay such that it becomes a candidate for explaining the origin of dark matter.

\begin{figure}[t!]
    \centering
    \includegraphics[width = 0.7\textwidth]{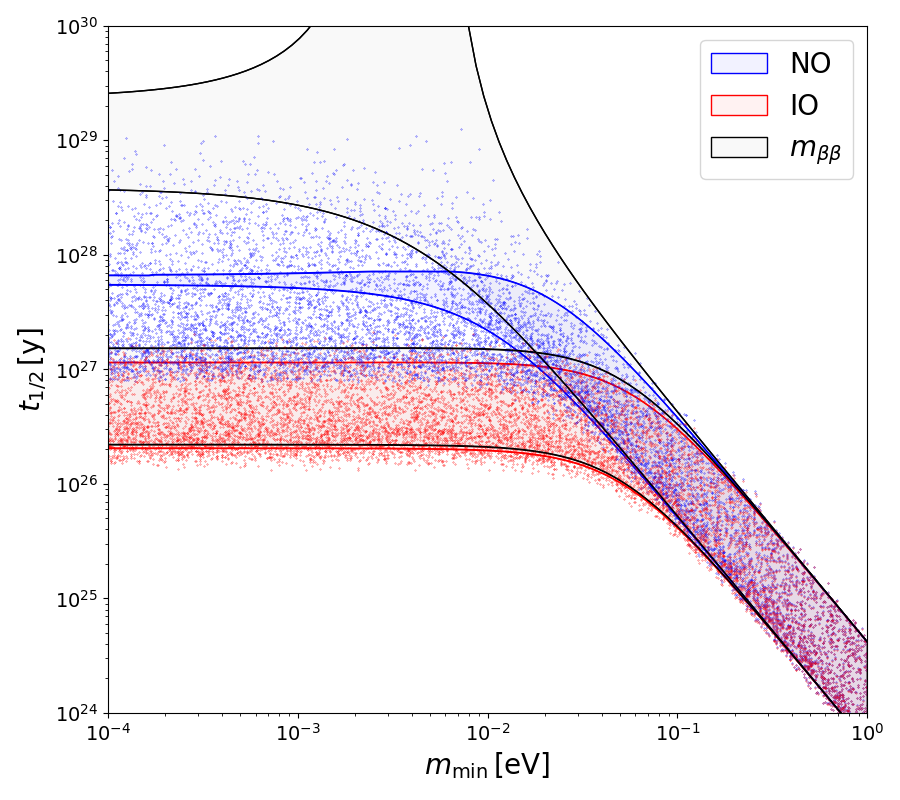}
    \caption{Half-lives in \textsuperscript{76}Ge for the standard mass mechanism accompanied by the exchange of heavy neutralinos and gluinos in the $\cancel{R}_p$-SUSY with $\lambda{'}_{111}=2\times10^{-4}$, $m_{\Tilde{e}_L} = 410\,$GeV, $m_{\Tilde{q}_{L,R}} = 1600\, $GeV,$m_{\Tilde{g}} = 2260\,$GeV, $m_{\chi_1} = \Lambda_\chi=2\,\mathrm{GeV}$ and $m_{\chi_i} \rightarrow \infty$ for $i\neq 1$}
    \label{fig:lobster_RP}
\end{figure}

However, $R_p$ conservation also comes with the conservation of both baryon and lepton number~\citep{MARTIN_1998}. Thus, supersymmetric models aiming to explain the baryon asymmetry of the Universe via explicit violation of either lepton or baryon number need to break $R_p$. This induces new lepton number violating terms~\citep{Hirsch_1996_susy}
\begin{align}
\begin{split}
    \mathcal{L}_{\cancel{R}_p}^{\Delta L=1}=-\lambda{'}_{111}\bigg[\left(\overline{u}_L,\overline{d}_R\right)\left(\begin{matrix}e^c_R\\-\nu^c_R\end{matrix}\right)\Tilde{d}_R
    + \left(\overline{e}_L,\overline{\nu}_L\right)d_R\left(\begin{matrix}\Tilde{u}^*_L\\-\Tilde{d}^*_L\end{matrix}\right)
    + \left(\overline{u}_L,\overline{d}_L\right)d_R\left(\begin{matrix}\Tilde{e}^*_L\\-\Tilde{\nu}^*_L\end{matrix}\right)\bigg] + \text{h.c.},
\end{split}
\end{align}
which can contribute to \0 decay. Contributions to \0 decay from $\cancel{R}_p-$SUSY have been studied first in Refs.~\citep{Hirsch_1996_susy,PhysRevLett.75.17}, the corresponding Feynman diagrams are shown in Figure~\ref{fig:feynman_rp-susy}. The relevant gluino ($\Tilde{g}$) and neutralino ($\chi$) - fermion interactions are given by~\citep{Hirsch_1996_susy, Haber:1984rc, Nilles:1983ge}
 \begin{align}
     \mathcal{L}_{\Tilde{g}} &= -\sqrt{2}g_3\sum_a\frac{\lambda_{\alpha\beta}^{(a)}}{2}\left(\overline{q_L}^\alpha\Tilde{g}\Tilde{q}_L^\beta - \overline{q_R}^\alpha\Tilde{g}\Tilde{q}_R^\beta\right) + \text{h.c.}
    \end{align}
and
    \begin{align}
     \mathcal{L}_\chi &= \sqrt{2}g_2\sum_{i=1}^{4}\left(\epsilon_{Li}\left(\Psi\right)\overline{\Psi_L}\chi_i\Tilde{\Psi}_L + \epsilon_{Ri}\left(\Psi\right)\overline{\Psi_R}\chi_i\Tilde{\Psi}_R + \right) + \text{h.c.}.
 \end{align}
One can obtain the low-energy effective Lagrangian by integrating out the heavy superfields as well as the Standard Model particles with masses $\gtrsim m_W$.
In doing so, one finds the different low-energy effective dimension-9 $\Delta L=2$ operators that contribute to \0 decay~\citep{PhysRevLett.75.17}
\begin{align}
\begin{split}
    \mathcal{L}_{\cancel{R}_p}=\frac{G_F^2}{2m_N}\bigg[&\left(\eta_{\Tilde{g}}+\eta_\chi\right)\left(\left[\overline{u}(1+\gamma^5)d\right]\left[\overline{u}(1+\gamma^5)d\right]-\frac{1}{4}\left[\overline{u}\sigma^{\mu\nu}(1+\gamma^5)d\right]\left[\overline{u}\sigma_{\mu\nu}(1+\gamma^5)d\right]\right) \\
    &+\left(\eta_{\chi\Tilde{e}}+\eta_{\Tilde{g}}{'}-\eta_{\chi\Tilde{f}}\right)\left[\overline{u}(1+\gamma^5)d\right]\left[\overline{u}(1+\gamma^5)d\right]\bigg]\;\left[\overline{e}\left(1+\gamma_5\right)e^c\right].
\end{split}
\end{align}
These can be matched onto the LEFT basis as
\begin{align}
\begin{split}
    C_{2R}^{(9)}{'}&=\frac{2v}{m_N}\left[2\eta_{\Tilde{g}}+2\eta_\chi+\eta_{\chi\Tilde{e}}+\eta_g{'}-\eta_{\chi\Tilde{f}}\right]\\
    C_{3R}^{(9)}{'}&=\frac{4v}{m_N}\left[\eta_{\Tilde{g}}+\eta_\chi\right].
\end{split}
\end{align}
The coupling constants are given in terms of gluino, neutralino and squark masses as~\citep{Hirsch_1996_susy}
\begin{figure}[t!]
    \centering
    \includegraphics[width=1\textwidth]{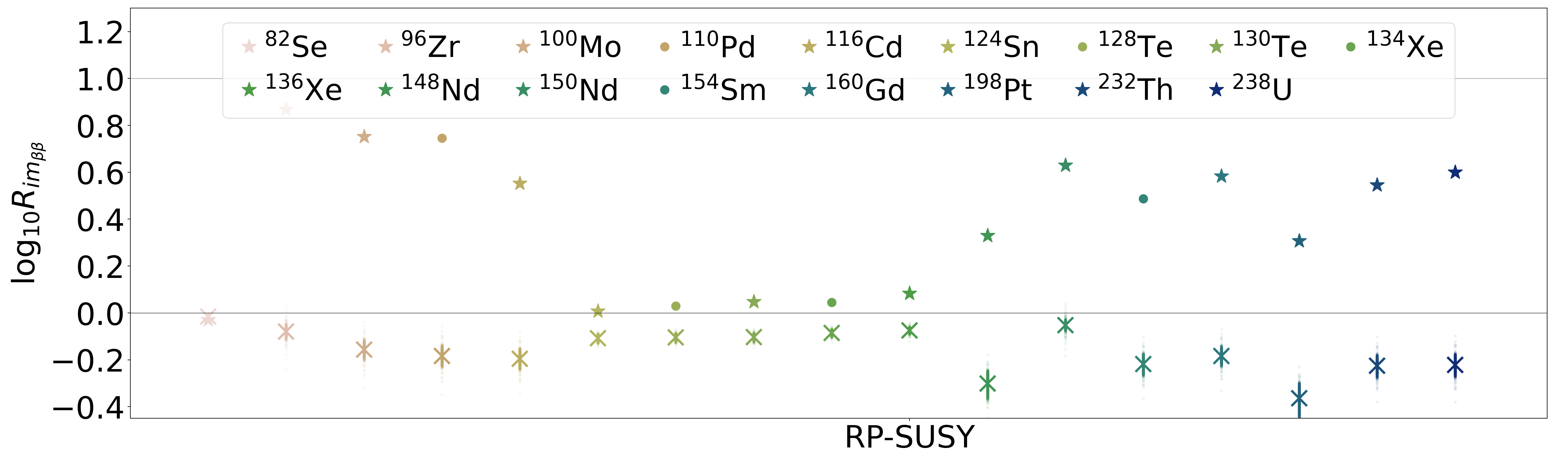}
    \caption{Half-life ratios resulting from the $\cancel{R}_p$-SUSY contributions to \0 normalized to the standard mass mechanism.}
    \label{fig:RP_ratios}
\end{figure}
\begin{align}
    \begin{split}
        \eta_{\Tilde{g}} &= \alpha_s \Lambda^2\frac{m_N}{m_{\Tilde{g}}}\left[1+\left(\frac{m_{\Tilde{d}_R}}{m_{\Tilde{u}_L}}\right)^4\right]\\
        \eta_{\Tilde{g}}{'} &= 2\alpha_s \Lambda^2\frac{m_N}{m_{\Tilde{g}}}\left(\frac{m_{\Tilde{d}_R}}{m_{\Tilde{u}_L}}\right)^2\\
        \eta_{\chi} &= \frac{3\alpha_2}{4} \Lambda^2\sum_{i=1}^4\frac{m_N}{m_{\chi_i}}\left[\epsilon_{Ri}^2(d)+\epsilon_{Li}^2(u)\left(\frac{m_{\Tilde{d}_R}}{m_{\Tilde{u}_L}}\right)^4\right]\\
        \eta_{\chi\Tilde{e}} &= 9\alpha_2\Lambda^2\left(\frac{m_{\Tilde{d}_R}}{m_{\Tilde{e}_L}}\right)^4\sum_{i=1}^4\epsilon_{Li}^2(e)\frac{m_N}{m_{\chi_i}}\\
        \eta_{\chi\Tilde{f}} &= \frac{3\alpha_2}{2}\Lambda^2\left(\frac{m_{\Tilde{d}_R}}{m_{\Tilde{e}_L}}\right)^2\sum_{i=1}^4\frac{m_N}{m_{\chi_i}}\bigg[\epsilon_{Ri}(d)\epsilon_{Li}(e)+\epsilon_{Ri}(u)\epsilon_{Li}(d)\left(\frac{m_{\Tilde{e}_L}}{m_{\Tilde{u}_L}}\right)^2\\&\qquad\qquad\qquad\qquad\qquad\qquad\;\;+\epsilon_{Ri}(u)\epsilon_{Li}(e)\left(\frac{m_{\Tilde{d}_R}}{m_{\Tilde{u}_L}}\right)^2\bigg]
    \end{split}
\end{align}
with
\begin{align}
    \Lambda=\frac{\sqrt{2\pi}}{3}\frac{\lambda{'}_{111}}{G_Fm_{\Tilde{d}_R}^2}.
\end{align}
Both gluino- and neutralino-exchange diagrams contribute to the same low-energy operators. As pointed out in the previous section, distinguishing between the different contributions triggered by $C_{2R}^{(9)}{'}$ and $C_{3R}^{(9)}{'}$ is practically impossible due to the unknown LECs. Given that both operators contribute only to $G_{01}$, the phase-space observables do not provide any additional information either. For completeness, we present the ratios normalized to the mass mechanism in Figure~\ref{fig:RP_ratios}. Clearly, the $\cancel{R}_p$-SUSY model we consider here follows the same pattern as the scalar short-range operators $C_{2R}^{(9)}$ and ${C_{2R}^{(9)}}'$ already discussed in Section~\ref{sec:distinguishing}. In Figure~\ref{fig:lobster_RP} we show the expected half-life assuming the simultaneous existence of the standard mass mechanism and the $\cancel{R}_p$-SUSY induced mechanisms. Here, we assume the masses of the non-neutralino superpartners to be given by the current experimental limits i.e. $m_{\Tilde{e}_L} = 410\,$GeV, $m_{\Tilde{q}_{L,R}} = 1600\,$GeV with $q\in[u,d]$ and $m_{\Tilde{g}} = 2260\,$GeV~\citep{Zyla:2020zbs}. We fix the neutralino masses by requiring that the applied EFT framework holds, which necessitates $m_{\chi_i}\geq \Lambda_\chi\simeq2\,\mathrm{GeV}$. For simplicity we take $m_{\chi_1}= 2\,\mathrm{GeV}$ and $m_{\chi_i}\rightarrow\infty$ for $i\neq 1$. Lighter neutralino masses in connection to \0 have also recently been studied~\citep{Bolton:2021hje}. We set the coupling constant to $\lambda{'}_{111}=2\times10^{-4}$. Similarly to the mLRSM discussed above, the additional non-standard contributions hardly affect the inverted ordering setting. However, in the normal ordering case the non-standard contributions from $\cancel{R}_p$-SUSY model start to significantly influence the expected half-lives decreasing them, again, by about one order of magnitude. While one would naively assume that this should result in significantly enhanced ratios, it is important to bear in mind that any enhancement in the decay rates which is independent of the isotope of interest will drop out when considering the decay rate ratios.

\subsection{Leptoquark Models}
\label{subsec:LQ}

Leptoquarks (LQs) are hypothetical bosons $(3,X,Y)$ with non-zero color charge which couple to both quarks and leptons. They appear in numerous Standard Model extensions such as technicolor and composite models~\citep{DIMOPOULOS1979237,Gripaios_2010} or grand unifications~\citep{FRITZSCH1975193,Dorsner_2005} and can be used to generate neutrino masses at 1-loop level~\citep{Dor_ner_2017}. For a comprehensive review on leptoquarks see e.g.~\citep{Dor_ner_2016}.

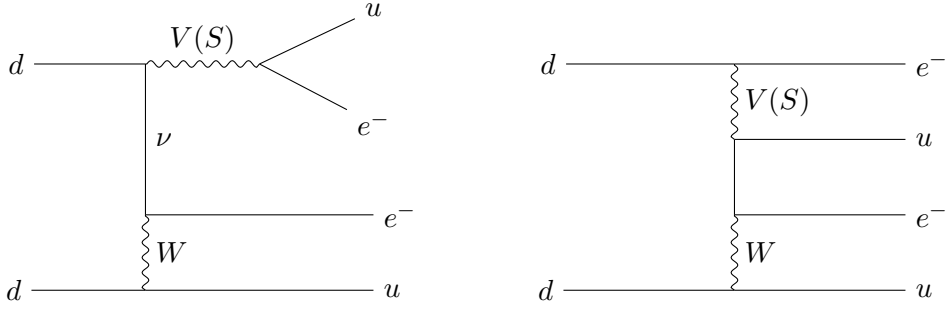
\begin{figure}[t!]
    \centering
    \begin{tikzpicture}
      \begin{feynman}
        \vertex (n){\(d\)};
        \vertex[right=1.7cm of n](nnuLQ);
        \vertex[right=1.5cm of nnuLQ](peLQ);
        \vertex[right=1.5cm of peLQ](pemiddle);
        \vertex[above=0.5cm of pemiddle](p){\(u\)};
        \vertex[below=0.5cm of pemiddle](e){\(e^-\)};
        \vertex[below=2cm of nnuLQ](wnue);
        \vertex[right=3cm of wnue] (e2){\(e^-\)};
        \vertex[below=1cm of wnue](npw);
        \vertex[left=1.5cm of npw](n2){\(d\)};
        \vertex[right=3cm of npw](p2){\(u\)};
        \diagram* {
          (n) -- (nnuLQ) -- [edge label=\(\nu\)] (wnue),
          (n2) --  (npw) -- (p2),
          (nnuLQ) -- [boson, edge label=\(V(S)\)] (peLQ) -- (p),
          (peLQ) -- (e),
          (wnue) -- [boson, edge label=\(W\)] (npw),
          (wnue) -- (e2),
          };
          
        \vertex[right=7cm of n](n3){\(d\)};
        \vertex[right=2.45cm of n3](neLQ);
        \vertex[right=2.25cm of neLQ](e3){\(e^-\)};
        \vertex[below=1cm of neLQ](pnu);
        \vertex[right=2.25cm of pnu](p3){\(u\)};
        \vertex[below=1cm of pnu](wnue2);
        \vertex[right=2.25cm of wnue2](e4){\(e^-\)};
        \vertex[below=1cm of wnue2](npw2);
        \vertex[left=2.25cm of npw2](n4){\(d\)};
        \vertex[right=2.25cm of npw2](p4){\(u\)};
        \diagram* {
        (n3) -- (neLQ) -- (e3),
        (neLQ) -- [boson, edge label=\(V(S)\)] (pnu) -- (p3),
        (pnu) -- (wnue2) -- (e4),
        (wnue2) -- [boson, edge label=\(W\)] (npw2) -- (p4),
        (n4) -- (npw2),
          };
      \end{feynman}
    \end{tikzpicture}
    \caption{Feynman diagrams of the vector $(V)$ and scalar $(S)$ leptoquark interactions contributing to \0.}
    \label{fig:feynman_LQ}
\end{figure}

Ignoring leptoquarks which do not directly couple to the Standard Model's particle content (i.e. without right-handed neutrinos), one can add up to 10 different leptoquarks obeying the Standard Model symmetries~\citep{Hirsch_1996_LQ}.
These are summarized in Table~\ref{tab:leptoquarks}. By looking at the relevant Feynman diagrams in Figure~\ref{fig:feynman_LQ} we can see that the contributions to \0 decay arise from leptoquarks with $Q^{(1)} = \pm 1/3$ (Figure~\ref{fig:feynman_LQ} left) and $Q^{(2)} = \pm 2/3$ (Figure~\ref{fig:feynman_LQ} right).
The full set of renormalizable LQ-fermion interactions is given by~\citep{Hirsch_1996_LQ}
\begin{table}[t!]
    \centering
    \begin{tabular}{l|c c c c}
    \hline\hline
        LQ $(\Omega)$ & $SU(3)_C$ & $SU(2)_L$ & $U(1)_Y$ & Q \\
         \hline
        $S_0$ & 3 & 1 & -2/3 & -1/3 \\
        $\Tilde{S}_0$ & 3 & 1 & -8/3 & -4/3 \\
        $S_{1/2}$ & $\overline{3}$ & 2 & -7/3 & $\left(-2/3, -5/3\right)$ \\
        $\Tilde{S}_{1/2}$ & $\overline{3}$ & 2 & -1/3 & $\left(1/3, -2/3\right)$ \\
        $S_1$ & 3 & 3 & -2/3 & $\left(2/3, -1/3, -4/3\right)$ \\
        $V_0$ & $\overline{3}$ & 1 & -4/3 & -2/3 \\
        $\Tilde{V}_0$ & $\overline{3}$ & 1 & -10/3 & -5/3 \\
        $V_{1/2}$ & 3 & 2 & -5/3 & $\left(-1/3, -4/3\right)$ \\
        $\Tilde{V}_{1/2}$ & 3 & 2 & 1/3 & $\left(2/3, -1/3\right)$ \\
        $V_1$ & $\overline{3}$ & 3 & -4/3 & $\left(1/3, -2/3, -5/3\right)$\\
        \hline\hline
    \end{tabular}
    \caption{List of possible scalar and vector leptoquarks and their transformation properties under the Standard Model symmetries~\citep{Hirsch_1996_LQ}.}
    \label{tab:leptoquarks}
\end{table}
\begin{align}
\begin{split}
    \mathcal{L}_{S,f} =& \left(\lambda_{S_0}^R\right)_{ij}S_0^{R\dagger} \left[\overline{u^c}_iP_Re_j\right]
    + \left(\lambda_{\Tilde{S}_0}^R\right)_{ij}\Tilde{S}_0^\dagger\left[\overline{d^c}_iP_Re_j\right]
    \\+& \left(\lambda_{S_{1/2}}^R\right)_{ij}S_{1/2}^{R\dagger}\left[\overline{u}_iP_LL_j\right]
    + \left(\lambda_{\Tilde{S}_{1/2}}^R\right)_{ij}\Tilde{S}_{1/2}^\dagger\left[\overline{d}_iP_LL_j\right]
    \\+&\left(\lambda_{S_0}^L\right)_{ij}S_0^{L\dagger}\left[\overline{Q^c}_iP_Li\tau_2L_j\right] + \left(\lambda_{S_{1/2}}^L\right)_{ij}S_{1/2}^{L\dagger}\left[\overline{Q^c}_iP_Ri\tau_2e_j\right]
    \\+& \left(\lambda_{S_1}^L\right)_{ij}\left[\overline{Q^c}_iP_Li\tau_2S_1^\dagger L_j\right] + \text{h.c.}
\end{split}
\end{align}
and
\begin{align}
\begin{split}
    \mathcal{L}_{V,f} =& \left(\lambda_{V_0}^R\right)_{ij}V_{0\mu}^{R\dagger}\left[\overline{d}_i\gamma^\mu P_Re_j\right] + \left(\lambda_{\Tilde{V}_0}^R\right)_{ij}\Tilde{V}_{0\mu}^{R\dagger}\left[\overline{u}_i\gamma^\mu P_Re_j\right]
    \\+&\left(\lambda_{V_{1/2}}^R\right)_{ij}V_{1/2\mu}^{R\dagger}\left[\overline{d^c}_i\gamma^\mu P_L L_j\right] + \left(\lambda_{\Tilde{V}_{1/2}}^R\right)_{ij}\Tilde{V}_{1/2}^\dagger\left[\overline{u^c}_i\gamma^\mu P_L L_j\right]
    \\+&\left(\lambda_{V_0}^L\right)_{ij}V_{0\mu}^{L\dagger}\left[\overline{Q}_i\gamma^\mu P_L L_j\right] + \left(\lambda_{V_{1/2}}^L\right)_{ij}V_{1/2\mu}^{L\dagger}\left[\overline{Q^c}_i\gamma^\mu P_R e_j\right]
    \\+&\left(\lambda_{V_1}^L\right)_{ij}\left[\overline{Q}_i\gamma^\mu P_L V_{1\mu}^\dagger L_j\right] + \text{h.c.}
\end{split}
\end{align}
for the scalar (S) and vector (V) leptoquarks, respectively. We follow the notation of~\citep{Hirsch_1996_LQ} distinguishing leptoquarks coupling to left-handed and right-handed quarks. In addition to the LQ-fermion interactions, one can write down the gauge invariant and renormalizable LQ-Higgs interactions,
\begin{align}
\begin{split}
    \mathcal{L}_{LQ,\Phi} =& h_{S_0}^i \Tilde{\Phi}^\dagger \Tilde{S}_{1/2}S_0^i + h_{V_0}^i \Tilde{\Phi}^\dagger \Tilde{V}_{1/2}^\mu V_{0\mu}^i
    \\+&h_{S_1}\Tilde{\Phi}^\dagger S_1\Tilde{S}_{1/2} + h_{V_1}\Tilde{\Phi}^\dagger V_1^\mu\Tilde{V}_{1/2\mu}
    \\+&Y_{S_{1/2}}^i\left(\Tilde{\Phi}^\dagger S_{1/2}^i\right)\left(\Tilde{S}_{1/2}^\dagger\Phi\right) + Y_{V_{1/2}}^i\left(\Tilde{\Phi}^\dagger V_{1/2}^{\mu i}\right)\left(\Tilde{V}_{1/2\mu}^\dagger\Phi\right)
    \\+&Y_{S_{1}}\left(\Tilde{\Phi}^\dagger S_1^\dagger \Phi\right)\Tilde{S}_0 + Y_{V_1}\left(\Tilde{\Phi}^\dagger V_{1\mu}^\dagger\Phi\right)\Tilde{V}_0^\mu
    \\+&\kappa_S^i\left(\Phi^\dagger S_1\Phi\right){S_0^i}^\dagger + \kappa_V^i\left(\Phi^\dagger V_1^\mu\Phi\right){V_{0\mu}^i}^\dagger + \text{h.c.}
    \\-&\sum_{\Omega}\left(\eta_\Omega M^2_\Omega - g_\Omega^{i_1 i_2}\Phi^\dagger \Phi\right){\Omega^{i_1}}^\dagger\Omega^{i_2},
\end{split}
\end{align}
where the leptoquark triplets are defined as
\begin{align}
    V_1 = \sum_i \tau_iV_{1i}\qquad S_1=\sum_i \tau_i S_{1i}.
\end{align}
These LQ-Higgs interactions are essential when considering contributions to \0 decay because they result in non-zero correlation functions for, e.g.,
\begin{align}
    \left<{S_0^i\Tilde{S}_{1/2}}\right>\propto \sum_{\Tilde{I}} \mathcal{N}_{S_0^i\Tilde{I}}\mathcal{N}_{\Tilde{S}_{1/2}\Tilde{I}},
\end{align}
where $\mathcal{N}$ is the mixing matrix which diagonalizes the mass matrix $\mathcal{N}^T\mathcal{M}^2\mathcal{N} = \mathcal{M}^2_{diag}$ and $\Tilde{I}=\mathcal{N}^TI$ are the mass eigenstate fields.
This particular example results in contributions captured by the right diagram in Figure~\ref{fig:feynman_LQ}.

After integrating out the heavy LQ degrees of freedom and rearranging the resulting EFT operators via Fierz transformations one arrives at the effective low-energy four-fermion interactions. The parts of the low-energy Lagrangian relevant for \0 decay are then given by~\citep{Hirsch_1996_LQ}
\begin{align}
\begin{split}
    \mathcal{L}_{LQ} =& \left[\overline{e}P_L\nu^c\right]\left\{\frac{\epsilon_S}{M^2_S}\left[\overline{u}P_Rd\right] + \frac{\epsilon_V}{M^2_V}\left[\overline{u}P_Ld\right]\right\}
    \\-&\left[\overline{e}\gamma^\mu P_L\nu^c\right]\left\{\left(\frac{\alpha_S^R}{M_S^2}+\frac{\alpha_V^R}{M_V^2}\right)\left[\overline{u}\gamma_\mu P_Rd\right] - \sqrt{2}\left(\frac{\alpha_S^L}{M_S^2}+\frac{\alpha_V^L}{M_V^2}\right)\left[\overline{u}\gamma_\mu P_Ld\right]\right\} + \text{h.c.}\label{eq:low_energy_LQ_lagrangian},
\end{split}
\end{align}
\begin{figure}[t!]
    \centering
    \includegraphics[scale=0.75]{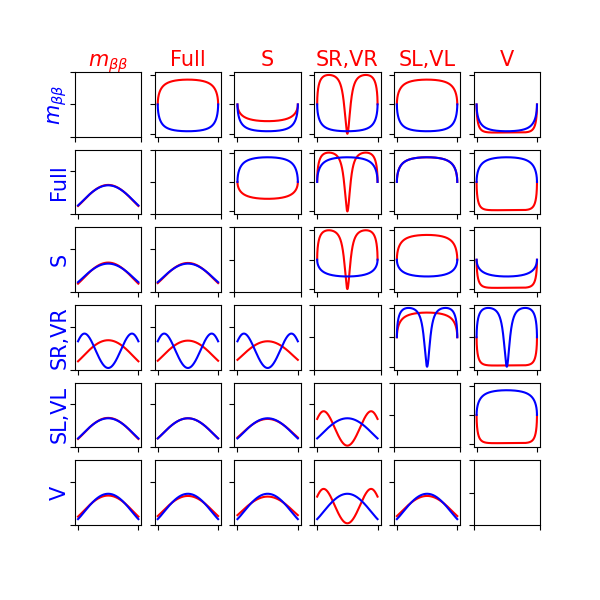}
    \caption[Angular correlations and single electron energy spectra resulting from the different LQ contributions.]{Angular correlations (upper right) and single electron energy spectra (lower left) resulting from the different LQ contributions as well as the standard mass mechanism in \textsuperscript{136}Xe. The unknown LECs are set to their order-of-magnitude estimates. The specific choice of the isotope does slightly influence the shape of the angular correlation.}
    \label{fig:LQ_psf_observables}
\end{figure}
with the low-energy Wilson coefficients
\begin{align}
    \epsilon_I &= 2^{-\eta_I}\left[\lambda_{I_1}^L\lambda_{\Tilde{I}_{1/2}}^R\left(\Tilde{\theta}_{43}^I\left(Q_I^1\right)+\eta_I\sqrt{2}\Tilde{\theta}_{41}^I\left(Q_I^2\right)\right)-\lambda_{I_0}^L\lambda_{\Tilde{I}_{1/2}}^R\Tilde{\theta}_{23}^I\left(Q_I^1\right)\right]\label{eq:LQ_epsilon_from_high_energy}
    \\
    \alpha_I^L &= \frac{2}{3+\eta_I}\lambda_{I_{1/2}}^L\lambda_{I_1}^L\Tilde{\theta}_{24}^I\left(Q_I^2\right),\qquad\qquad\alpha_I^R=\frac{2}{r+\eta_I}\lambda_{I_0}^R\lambda_{\Tilde{I}_{1/2}}^R\Tilde{\theta}_{23}^I\left(Q_I^1\right),\label{eq:LQ_alpha_from_high_energy}
\end{align}
where
\begin{align}
    \Tilde{\theta}^I_{ij} = \sum_k \mathcal{N}_{ik}\mathcal{N}_{jk}\frac{M_I^2}{M_{I_k}^2}.
\end{align}
Here, ``common mass scales'' $M_S$ and $M_V$ have been inserted for convenience. It should be noted that the exact choice of $M_{S,V}$ does not matter as they drop out. However, the exact LQ masses do enter into the calculation such that for leptoquark masses which are about the same order of magnitude one can choose $M_{S,V}$ to represent the suppression factors. Looking at Eq.~\eqref{eq:LQ_epsilon_from_high_energy} and Eq.~\eqref{eq:LQ_alpha_from_high_energy}, there is a priori no reason from, e.g., naturalness arguments why any of the low energy coefficients $\alpha_I$ and $\epsilon_I$ should be suppressed or enhanced compared to the others. However, if the LQ interactions arise from a more complete model or if simply not all possible LQ interactions are realized in nature, hierarchical structures might appear. We will therefore study different settings in which some couplings dominate over the others.
From Eq.~\eqref{eq:low_energy_LQ_lagrangian} we can match the Wilson coefficients in Eq.~\eqref{eq:LQ_epsilon_from_high_energy} and Eq.~\eqref{eq:LQ_alpha_from_high_energy} onto the LEFT basis arriving at
\begin{align}
\begin{split}
    C^{(6)}_{SL} &= \frac{v^2}{M_V^2}\epsilon_V\,,\qquad\qquad\qquad\qquad
    C^{(6)}_{SR} = \frac{v^2}{M_S^2}\epsilon_S
    \\C^{(6)}_{VL} &= \sqrt{2}v^2\left(\frac{\alpha_{S}^L}{M_S^2}+\frac{\alpha_{V}^L}{M_V^2}\right)
    \,,\qquad 
    C^{(6)}_{VR} = -v^2\left(\frac{\alpha_{S}^R}{M_S^2}+\frac{\alpha_{V}^R}{M_V^2}\right).
\end{split}
\end{align}
We study the following 7 different settings of LQ contributions to \0 decay:
\begin{enumerate}
    \item Full LQ Model: \qquad\qquad\qquad\qquad\qquad\qquad\;\;\;\,$\epsilon_S=\epsilon_V=\alpha_S^L=\alpha_S^R=\alpha_V^L=\alpha_V^R=1$
    \item Scalar LQs (S): \qquad\qquad\qquad\qquad\qquad\qquad\;\;\;\,\,$\epsilon_S=\alpha_S^L=\alpha_S^R=1$
    \item Scalar LQs coupling to LH fermions (SL): \quad\,\,$\alpha_S^L=1$
    \item Scalar LQs coupling to RH fermions (SR): \quad$\alpha_S^R=1$
    \item Vector LQs (V): \qquad\qquad\qquad\qquad\qquad\qquad\;\;\,$\epsilon_V=\alpha_V^L=\alpha_V^R=1$
    \item Vector LQs coupling to LH fermions (VL): \;\;\;\,$\alpha_V^L=1$
    \item Vector LQs coupling to RH fermions (VR): \;\;\;$\alpha_V^R=1$
\end{enumerate}

\begin{figure}[t!]
    \centering
    \includegraphics[width=\textwidth]{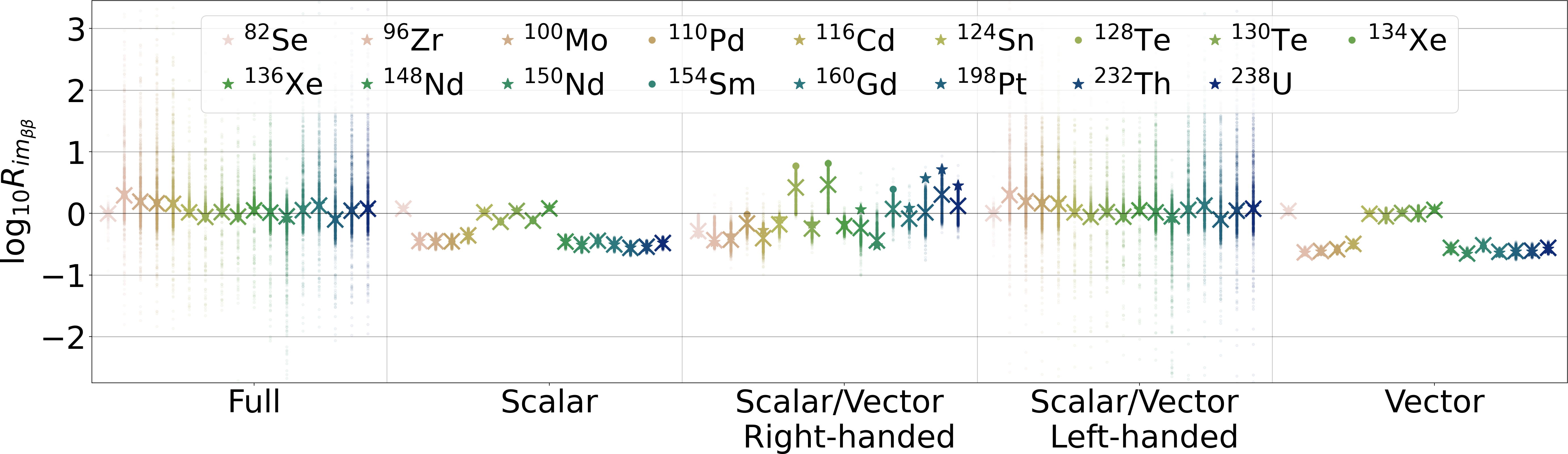}
    \caption[Half-life ratios resulting from different leptoquark settings when taking \textsuperscript{76}Ge as the reference isotope.]{Half-life ratios resulting from different leptoquark settings when taking \textsuperscript{76}Ge as the reference isotope. The ratios are compared to the standard mass mechanism.}
    \label{fig:leptoquark_ratios}
\end{figure}
The left-handed scalar (SL) and left-handed vector (VL) models result in the same low-energy physics because they match onto the same LEFT operator. The same is true for SR and VR. In Figure~\ref{fig:LQ_psf_observables} we show the corresponding single electron energy spectra and angular correlations corresponding to each of the above models and compare them with the standard mechanism scenario. When setting the unknown LECs to their order-of-magnitude estimates we find that except for the vector (V) scenario all other models give shapes distinguishable from the standard mass mechanism for at least one phase-space observable. The resulting half-life ratios normalized to the neutrino mass mechanism for each of the above scenarios are shown in Figure~\ref{fig:leptoquark_ratios}. Except for the SR and VR cases, for which the central values suggest somewhat weaker distinguishability, we find that the central values match fairly well the chosen benchmark scenario. Nonetheless, the spread in $R_{im_{\beta\beta}}$ is still significant for the full model as well as the SL and VL models. Considering the central values, the highest ratio when taking \textsuperscript{76}Ge as the reference isotope is realized in the vector model with $R_{im_{\beta\beta}}^\mathrm{max}\sim 4.5$. Again, assuming that the calculated central values of the half-life ratios represent a reasonable estimate, this would correspond to a necessary theoretical accuracy on the nuclear part of the amplitude to satisfy $\frac{\Delta M_\mathrm{eff}}{M_\mathrm{eff}} \lesssim 19\%.$

\begin{figure}
    \centering
    \includegraphics[width = 0.7\textwidth]{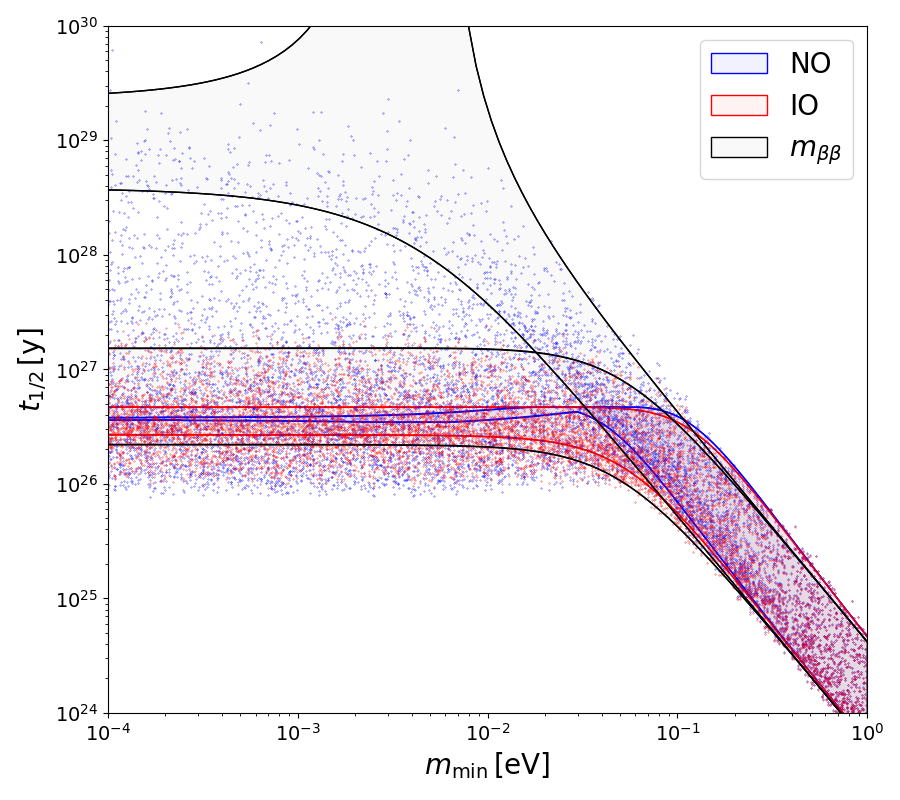}
    \caption{Here we show the expected half-lives for the full LQ model with the parameters fixed to $\epsilon_S=\epsilon_V=\alpha_S^L=\alpha_S^R=\alpha_V^L=\alpha_V^R=1$ and the suppression scales $M_S=M_V = 10^7\,$GeV}
    \label{fig:lobster_LQ}
\end{figure}

In Figure~\ref{fig:lobster_LQ} we show the expected half-lives for the simultaneous realization of the full LQ model and the standard mass mechanism. We assumed the suppression factors to be $M_S=M_V = 10^7\,$GeV. One can see that in this setting the inverted mass ordering case is not altered significantly while the half-life in the normal ordering case is decreased such that the gap between the two mass orderings is closed.

\section{Summary and Conclusions}
\label{sec:summary}
Neutrinoless double beta decay is the best laboratory probe of lepton number violation and as such can naturally shed light on the generation of neutrino masses as well as associated UV physics. The implications of observation of this hypothetical nuclear process would largely depend on the mechanism responsible for the dominant contribution. In this paper we have performed a detailed analysis discussing the possibilities of experimental discrimination among the 32 different LEFT LNV operators of dimension $\leq 9$ triggering \0 decay at low energy.

The main aim of our study is to understand the differences in various \0 decay mechanisms and to investigate the possibilities of identifying the potential exotic contribution in experiments. Assuming only one operator at a time, we found that the 32 different LEFT operators can be split into 12 groups which are distinguishable from each other by comparison of ratios of half-lives in different double-beta-decaying isotopes. We calculated the half-life ratios normalized to the standard mass mechanism $R_{im_{\beta\beta}}$ for each of the operator groups discussing the potential for their identification by experimental observations. Varying the currently unknown low-energy constants (LECs) around their order-of-magnitude estimates obtained using NDA we observed that their impact on the expected half-life ratios can be significant for most operator groups. To quantify this impact and temporarily eliminate it in our conclusions we focused on two different scenarios; namely, we identified the central values of the ratio ranges as well as the worst-case scenario considering the value of the ratios closest to 1 within each ratio range. In Figure~\ref{fig:tile_plot_exp_isotopes} we summarized the potential of distinguishing among different operator groups for both of these scenarios considering all isotopes for which the experimental limit on \0 decay half-life exists. Based on the central-value scenario we estimated the required theoretical accuracy on the nuclear physics calculations, parameterized by an effective nuclear matrix element $M_\mathrm{eff}$, that would allow for identifying non-standard mechanisms in the single operator dominance scenario. We found that identifying all the non-standard mechanisms via half-life ratios would require (at least) a few-percent accuracy. While the required accuracy of the theoretical description is beyond the current status of nuclear uncertainties, advances in \emph{ab initio} calculations of nuclear matrix elements may be able to deliver such precision in the future provided that the currently unknown LECs are fixed with similar accuracy as those that are already under control.

The additional information that can be inferred from the phase-space observables does not allow for distinguishing operators within the 12 operator groups corresponding to distinct half-life ratios. However, the phase-space observables are much less affected by nuclear uncertainties such that they can potentially deliver important insight into the underlying mechanism of \0 decay even if nuclear uncertainties remain significant. For operator groups such as $C_V{(9)}$ or $\Tilde{C}_V{(9)}$ for which the expected half-life ratios do not differ significantly from the standard mass mechanism measurements, tracking the outgoing electrons would be a more promising approach of identification even if nuclear uncertainties are substantially reduced. Therefore, future experiments that would have the required technology such as SuperNEMO seem to be very relevant. The operator groups that could be distinguished by means of phase-space observables are also marked in Figure~\ref{fig:tile_plot_exp_isotopes}.

Besides the effective approach detailing individual operators, we focussed also on \0 decay contributions triggered by three different high-energy models. In each case, we identified and discussed the signatures that could help to distinguish these models in observations of \0 decay. Our approach can be easily extended to any other UV model that can be matched onto the applied EFT framework.

Based on the obtained results and following their discussion it becomes clear that although it might be possible to unravel an exotic contribution to \0 decay, pinpointing the dominant mechanism underlying this hypothetical nuclear process most probably would not be possible without other, complementary experiments. As discussed, the possibilities of employing different double-beta processes seems to be rather unlikely because of their phase-space suppression. On the other hand, the underlying mechanism could be identified by combining the \0 decay data with different experiments searching for lepton number violation, such as meson decays, tau decays, or collider searches, which can, however, more naturally verify a complete UV scenario rather than  a specific effective operator. Useful information will be provided also by measurements aiming to determine the absolute neutrino mass scale, including the CMB data providing a constraint on the sum of neutrino masses, $\sum_i m_{\nu_i}$. A detailed discussion of the interplay with complementary probes of neutrino masses and lepton number (non-)conservation is however beyond the scope of this paper. Similarly, we have not covered the possibility of existence of light sterile neutrinos and its implications for \0 decay. Same methods as employed in this study may allow to unravel additional contributions to \0 decay induced by light sterile neutrinos.

\section*{Acknowledgments}
\noindent
We would like to thank Jordy de Vries for many fruitful discussions and useful comments. We are also grateful to Frank F. Deppisch for valuable comments on the final version of the manuscript. LG acknowledges support from the National Science Foundation, Grant PHY-1630782, and to the Heising-Simons Foundation, Grant 2017-228.
\appendix
\section{Double $\beta$ Modes}\label{app:modes}
\begin{table}[H]
\vspace{-10pt}
    \centering
    \begin{tabularx}{\textwidth}{X X | X X | X X | X X}
    \hline\hline
        \multicolumn{2}{c|}{$2\nu\beta^-\beta^-$} & \multicolumn{2}{c|}{$2\nu\beta^+\beta^+$} & \multicolumn{2}{c|}{$2\nu \mathrm{EC}\beta^+$} & \multicolumn{2}{c}{$2\nu \mathrm{ECEC}$} \\
        $^AZ$ & Q [MeV] & $^AZ$ & Q [MeV] & $^AZ$ & Q [MeV] & $^AZ$ & Q [MeV]  \\ \cline{1-2}\cline{3-4}\cline{5-6}\cline{7-8}
        \textsuperscript{46}Ca & 0.99 & 
        \textsuperscript{78}Kr & 0.80 &
        \textsuperscript{50}Cr & 0.15 &
        \textsuperscript{36}Ar & 0.43\\
        \textsuperscript{48}Ca & 4.27 &
        \textsuperscript{96}Ru & 0.67 &
        \textsuperscript{58}Ni & 0.90 &
        \textsuperscript{40}Ca & 0.19\\
        \textsuperscript{70}Zn & 1.00 &	
        \textsuperscript{106}Cd & 0.73 &
        \textsuperscript{64}Zn & 0.073 &
        \textsuperscript{50}Cr & 1.17\\
        \textsuperscript{76}Ge & 2.04 &
        \textsuperscript{124}Xe & 0.82 &
        \textsuperscript{74}Se & 0.19 &
        \textsuperscript{54}Fe & 0.68\\
        \textsuperscript{80}Se & 0.13 &
        \textsuperscript{130}Ba & 0.57 &
        \textsuperscript{78}Kr 	& 1.82 &
        \textsuperscript{58}Ni & 1.93\\ 
        \textsuperscript{82}Se & 3.00 &
        \textsuperscript{136}Ce & 0.33 &
        \textsuperscript{84}Sr & 0.77 &
        \textsuperscript{64}Zn & 1.09\\
        \textsuperscript{86}Kr & 1.26 &  &	& 
        \textsuperscript{92}Mo & 0.63 &
        \textsuperscript{74}Se & 1.21\\
        \textsuperscript{94}Zr & 1.14 &	& &	
        \textsuperscript{96}Ru & 1.69 &
        \textsuperscript{78}Kr & 2.85\\
        \textsuperscript{96}Zr & 3.35 &	& &	
        \textsuperscript{102}Pd & 0.15 &
        \textsuperscript{84}Sr & 1.79 \\
        \textsuperscript{98}Mo & 0.11 &	& &	
        \textsuperscript{106}Cd & 1.75 &
        \textsuperscript{92}Mo & 1.65 \\
        \textsuperscript{100}Mo & 3.03 & & &	
        \textsuperscript{112}Sn & 0.90 &
        \textsuperscript{96}Ru & 2.71\\
        \textsuperscript{104}Ru & 1.30 & & &	
        \textsuperscript{120}Te & 0.71 &
        \textsuperscript{102}Pd & 1.17\\
        \textsuperscript{110}Pd & 2.02 & & &
        \textsuperscript{124}Xe & 1.84 &
        \textsuperscript{106}Cd & 2.78 \\
        \textsuperscript{114}Cd & 0.54 & & &	
        \textsuperscript{130}Ba & 1.60 & 
        \textsuperscript{108}Cd & 0.27\\
        \textsuperscript{116}Cd & 2.81 	& &	&	
        \textsuperscript{136}Ce & 1.36 & 
        \textsuperscript{112}Sn & 1.92 \\
        \textsuperscript{122}Sn & 0.37 & & &	
        \textsuperscript{144}Sm & 0.76 &
        \textsuperscript{120}Te & 1.73\\
        \textsuperscript{124}Sn & 2.29 & & &
        \textsuperscript{156}Dy & 0.98 &
        \textsuperscript{124}Xe & 2.86\\
        \textsuperscript{128}Te & 0.87 & & &	
        \textsuperscript{162}Er & 0.82 &
        \textsuperscript{126}Xe & 0.92\\
        \textsuperscript{130}Te & 2.53 & & &
        \textsuperscript{168}Yb & 0.39 &
        \textsuperscript{130}Ba & 2.62\\
        \textsuperscript{134}Xe & 0.83 & & &	
        \textsuperscript{174}Hf & 0.077 &
        \textsuperscript{132}Ba & 0.84\\
        \textsuperscript{136}Xe & 2.46 & & &	
        \textsuperscript{184}Os & 0.43 &
        \textsuperscript{136}Ce & 2.38 \\
        \textsuperscript{142}Ce & 1.42 & & &	
        \textsuperscript{190}Pt & 0.36 &
        \textsuperscript{138}Ce & 0.69 \\
        \textsuperscript{146}Nd & 0.070 & & & & &	
        \textsuperscript{144}Sm & 1.78 \\
        \textsuperscript{148}Nd & 1.93 & & & & & 	
        \textsuperscript{152}Gd & 0.056 \\
        \textsuperscript{150}Nd & 3.37 & & & & & 	
        \textsuperscript{156}Dy & 2.01 \\
        \textsuperscript{154}Sm & 1.25 & & & & &	
        \textsuperscript{158}Dy & 0.28 \\
        \textsuperscript{160}Gd & 1.73 & & & & &
        \textsuperscript{162}Er & 1.85\\
        \textsuperscript{170}Er & 0.66 & & & & &
        \textsuperscript{164}Er & 0.025\\
        \textsuperscript{176}Yb & 1.09 & & & & &	
        \textsuperscript{168}Yb & 1.41\\
        \textsuperscript{186}W & 0.49 & & &	& &	
        \textsuperscript{174}Hf & 1.10 \\
        \textsuperscript{192}Os & 0.41 & & & & &
        \textsuperscript{180}W & 0.14 \\
        \textsuperscript{198}Pt & 1.05 & & & & &
        \textsuperscript{184}Os & 1.45\\
        \textsuperscript{204}Hg & 0.42 & & & & &
        \textsuperscript{190}Pt & 1.38\\
        \textsuperscript{232}Th & 0.84 & & & & &
        \textsuperscript{196}Hg & 0.82\\
        \textsuperscript{238}U & 1.14 & & & & & &\\
        \hline\hline
    \end{tabularx}
    \caption[List of natural double-$\beta$ elements and the corresponding $Q$-values]{Complete list of natural double-$\beta$ elements and the corresponding $Q$-values calculated from the NIST list of elements~\citep{ElementList} using the conditions \ref{eq:threshold_bmbm}, \ref{eq:threshold_bpbp}, \ref{eq:threshold_ecbp} and~\ref{eq:threshold_ecec}. Overall there are 69 different natural elements that can decay via at least one double-$\beta$ mode.}
    \label{tab:natural_double_beta_elements}
\end{table}
In Table~\ref{tab:natural_double_beta_elements} we present a list of all naturally occuring isotopes that decay via any of the \0-modes i.e. $0\nu\beta^-\beta^-,0\nu\beta^+\beta^+,0\nu\mathrm{EC}\beta^+$ and $0\nu\mathrm{ECEC}$. The isotopes as well as the corresponding Q-values are taken and calculated from the NIST list of elements~\citep{ElementList}.

%
%
%

\section{Contributions from each Operator}\label{app:all_operators}
Assuming only one non-vanishing operator at a time, the half-life can be written in terms of a single Wilson coefficient, different phase-space factors and the nuclear contributions determined by the different NMEs and LECs. For convenience, we list the explicit decay rate equations for each operator below.
\begin{align}
    \begin{split}
        m_{\beta\beta}:\qquad T_{1/2}^{-1} = & g_A^4\left|m_{\beta\beta}\right|^2G_{01}\left|\frac{V_{ud}^2}{m_e}\left(\frac{1}{g_A^2}M_F-M_{GT}-2\frac{m_\pi^2\mathbf{g_\nu^{NN}}}{g_A^2}M_{F,sd}\right)\right|^2
    \end{split}
\end{align}
\begin{align}
    \begin{split}
        C_{VL}^{(6)}:\qquad T_{1/2}^{-1} = & g_A^4\left|C_{VL}^{(6)}\right|^2\Bigg\{4G_{02}\left|\frac{V_{ud}}{3}\left(\frac{g_V^2}{g_A^2}M_F+\frac{1}{3}\left(2M_{GT}^{AA}+M_T^{AA}\right)+\frac{6\mathbf{g_{VL}^E}}{g_A^2}M_{F,sd}\right)\right|^2\\
        &\qquad\qquad\,+2G_{04}\bigg|\frac{V_{ud}}{6}\bigg(\frac{g_V^2}{g_A^2}M_F-\frac{1}{3}\left(M_{GT}^{AA}-4M_T^{AA}\right)\\&\qquad\qquad\qquad\qquad-3\left(M_{GT}^{AP}+M_{GT}^{PP}+M_T^{AP}+M_T^{PP}\right)-\frac{12\mathbf{g_{VL}^{m_e}}}{g_A^2}M_{F,sd}\bigg)\bigg|^2\\
        &\qquad\qquad\,-4G_{03}\text{Re}\bigg[\frac{\left|V_{ud}\right|^2}{18}\left(\frac{g_V^2}{g_A^2}M_F+\frac{1}{3}\left(2M_{GT}^{AA}+M_T^{AA}\right)+\frac{6\mathbf{g_{VL}^{E}}}{g_A^2}M_{F,sd}\right)^*
        \\
        &\qquad\qquad\qquad\qquad\bigg(\frac{g_V^2}{g_A^2}M_F-\frac{1}{3}\left(M_{GT}^{AA}-4M_T^{AA}\right)
        \\
        &\qquad\qquad\qquad\qquad-3\left(M_{GT}^{AP}+M_{GT}^{PP}+M_T^{AP}+M_T^{PP}\right)-\frac{12\mathbf{g_{VL}^{m_e}}}{g_A^2}M_{F,sd}\bigg)\bigg]
        \\
        &\qquad\qquad\,+G_{09}\bigg|\frac{m_N}{m_e}V_{ud}\bigg[2\frac{g_A}{g_M}\left(M_{GT}^{MM}+M_T^{MM}\right)
        \\
        &\qquad\qquad\qquad\qquad+\frac{m_\pi^2}{m_N^2}\left(-\frac{2}{g_A^2}\mathbf{g_{VL}^{NN}}M_{F,sd}+\frac{1}{2}\mathbf{g_{VL}^{\pi N}}\left(M_{GT,sd}^{AP}+M_{T,sd}^{AP}\right)\right)\bigg]\bigg|^2\Bigg\}
    \end{split}\label{eq:CVL(6)}\\
    \begin{split}
        C_{VR}^{(6)}:\qquad T_{1/2}^{-1} = & g_A^4\left|C_{VR}^{(6)}\right|^2\Bigg\{4G_{02}\bigg|\frac{V_{ud}}{3}\left(\frac{g_V^2}{g_A^2}M_F-\frac{1}{3}\left(2M_{GT}^{AA}+M_T^{AA}\right)+6\frac{\mathbf{g_{VR}^E}}{g_A^2}M_{F,sd}\right)\bigg|^2\\
        &\qquad\qquad\,+2G_{04}\bigg|\frac{V_{ud}}{6}\bigg(\frac{g_V^2}{g_A^2}M_F+\frac{1}{3}\left(M_{GT}^{AA}-4M_T^{AA}\right)\\
        &\qquad\qquad\qquad\qquad+3\left(M_{GT}^{AP}+M_{GT}^{PP}+M_T^{AP}+M_T^{PP}\right)-12\frac{\mathbf{g_{VR}^{me}}}{g_A^2}M_{F,sd}\bigg)\bigg|^2\\
        &\qquad\qquad\,+2G_{03}\text{Re}\bigg[\frac{\left|V_{ud}\right|^2}{18}\left(\frac{g_V^2}{g_A^2}M_F-\frac{1}{3}\left(2M_{GT}^{AA}+M_T^{AA}\right)+6\frac{\mathbf{g_{VR}^E}}{g_A^2}M_{F,sd}\right)^*\\
        &\qquad\qquad\qquad\qquad\quad\bigg(\frac{g_V^2}{g_A^2}M_F+\frac{1}{3}\left(M_{GT}^{AA}-4M_T^{AA}\right)\\
        &\qquad\qquad\qquad\qquad\quad+3\left(M_{GT}^{AP}+M_{GT}^{PP}+M_T^{AP}+M_T^{PP}\right)-12\frac{\mathbf{g_{VR}^{m_e}}}{g_A^2}M_{F,sd}\bigg)\bigg]\Bigg\}\label{eq:CVR(6)}
    \end{split}\\
    \begin{split}
        C_{SL,SR}^{(6)}:\quad T_{1/2}^{-1} = & g_A^4\left|C_{SL,SR}^{(6)}\right|^2G_{01}\left|\frac{B}{m_e}V_{ud}M_{PS}\right|^2
    \end{split}\\
    \begin{split}
        C_{T}^{(6)}:\qquad T_{1/2}^{-1} = & g_A^4\left|C_{T}^{(6)}\right|^2G_{01}\left|V_{ud}\left[2\frac{\mathbf{g_T'}-\mathbf{g_T^{NN}}}{g_A^2}\frac{m_\pi^2}{m_N^2}M_{F,sd}-\frac{8g_T}{g_M}\left(M_{GT}^{MM}+M_T^{MM}\right)\right]\right|^2
    \end{split}\\
    \begin{split}
        C_{VL,VR}^{(7)}:\quad T_{1/2}^{-1} = & g_A^4\left|C_{VL,VR}^{(7)}\right|^2G_{01}\left|\frac{m_\pi^2}{m_ev}V_{ud}M_{PS}\right|^2
    \end{split}\\
    \begin{split}
        C_{1L,1R}^{(9)}{(')}:\quad T_{1/2}^{-1}= & g_A^4\left|C_{1L,1R}^{(9)}{(')}\right|^2G_{01}\bigg|\frac{5g_1^{\pi\pi}}{6m_N^2}\left(\frac{1}{2}M_{GT,sd}^{AP}+M_{GT,sd}^{PP}+\frac{1}{2}M_{T,sd}^{AP}+M_{T,sd}^{PP}\right)
        \\
        &\qquad\qquad\qquad\qquad\quad+\left(\mathbf{g_1^{\pi N}}-\frac{5}{6}g_1^{\pi\pi}\right)\frac{m_\pi^2}{2m_N^2}\left(M_{GT,sd}^{AP}+M_{T,sd}^{AP}\right)\\
        &\qquad\qquad\qquad\qquad\quad-2\frac{\mathbf{g_1^{NN}}}{g_A^2}\frac{m_\pi^2}{m_N^2}M_{F,sd}\bigg|^2
    \end{split}\\
    \begin{split}
        C_{2,3^,L,R}^{(9)}{(')}:\quad T_{1/2}^{-1} = & g_A^4 \left|C_{2,3^,L,R}^{(9)}{(')}\right|^2 G_{01}\bigg|\frac{5g_{2,3}^{\pi\pi}}{6m_N^2}\left(\frac{1}{2}M_{GT,sd}^{AP}+M_{GT,sd}^{PP}+\frac{1}{2}M_{T,sd}^{AP}+M_{T,sd}^{PP}\right)\\
        &\qquad\qquad\qquad\qquad\quad+2\frac{\mathbf{g_{2,3}^{NN}}}{g_A^2}\frac{m_\pi^2}{m_N^2}M_{F,sd}\bigg|
    \end{split}\\
    \begin{split}
        C_{4,5^,L,R}^{(9)}:\quad T_{1/2}^{-1} = & g_A^4 \left|C_{4,5^,L,R}^{(9)}\right|^2 G_{01}\bigg|\frac{5g_{4,5}^{\pi\pi}}{6m_N^2}\left(\frac{1}{2}M_{GT,sd}^{AP}+M_{GT,sd}^{PP}+\frac{1}{2}M_{T,sd}^{AP}+M_{T,sd}^{PP}\right)\\
        &\qquad\qquad\qquad\qquad\quad-2\frac{\mathbf{g_{4,5}^{NN}}}{g_A^2}\frac{m_\pi^2}{m_N^2}M_{F,sd}\bigg|
    \end{split}\\
    \begin{split}
        C_{V}^{(9)}:\qquad T_{1/2}^{-1} = & g_A^4 \left|C_{V}^{(9)}\right|^2G_{09}\left|\frac{m_\pi^2}{m_ev}\left(-\frac{2}{g_A}\mathbf{g_6^{NN}}M_{F,sd}+\frac{1}{2}\mathbf{g_V^{\pi N}}\left(M_{GT,sd}^{AP}+M_{T,sd}^{AP}\right)\right)\right|^2
    \end{split}\\
    \begin{split}
        \Tilde{C}_{V}^{(9)}:\qquad T_{1/2}^{-1} = & g_A^4 \left|\Tilde{C}_{V}^{(9)}\right|^2G_{09}\left|\frac{m_\pi^2}{m_ev}\left(-\frac{2}{g_A}\mathbf{g_7^{NN}}M_{F,sd}+\frac{1}{2}\mathbf{\Tilde{g}_V^{\pi N}}\left(M_{GT,sd}^{AP}+M_{T,sd}^{AP}\right)\right)\right|^2
    \end{split}
\end{align}
\section{Considering all Isotopes}\label{app:all_isotopes}
\begin{figure}
    \centering
    \includegraphics[width=\textwidth]{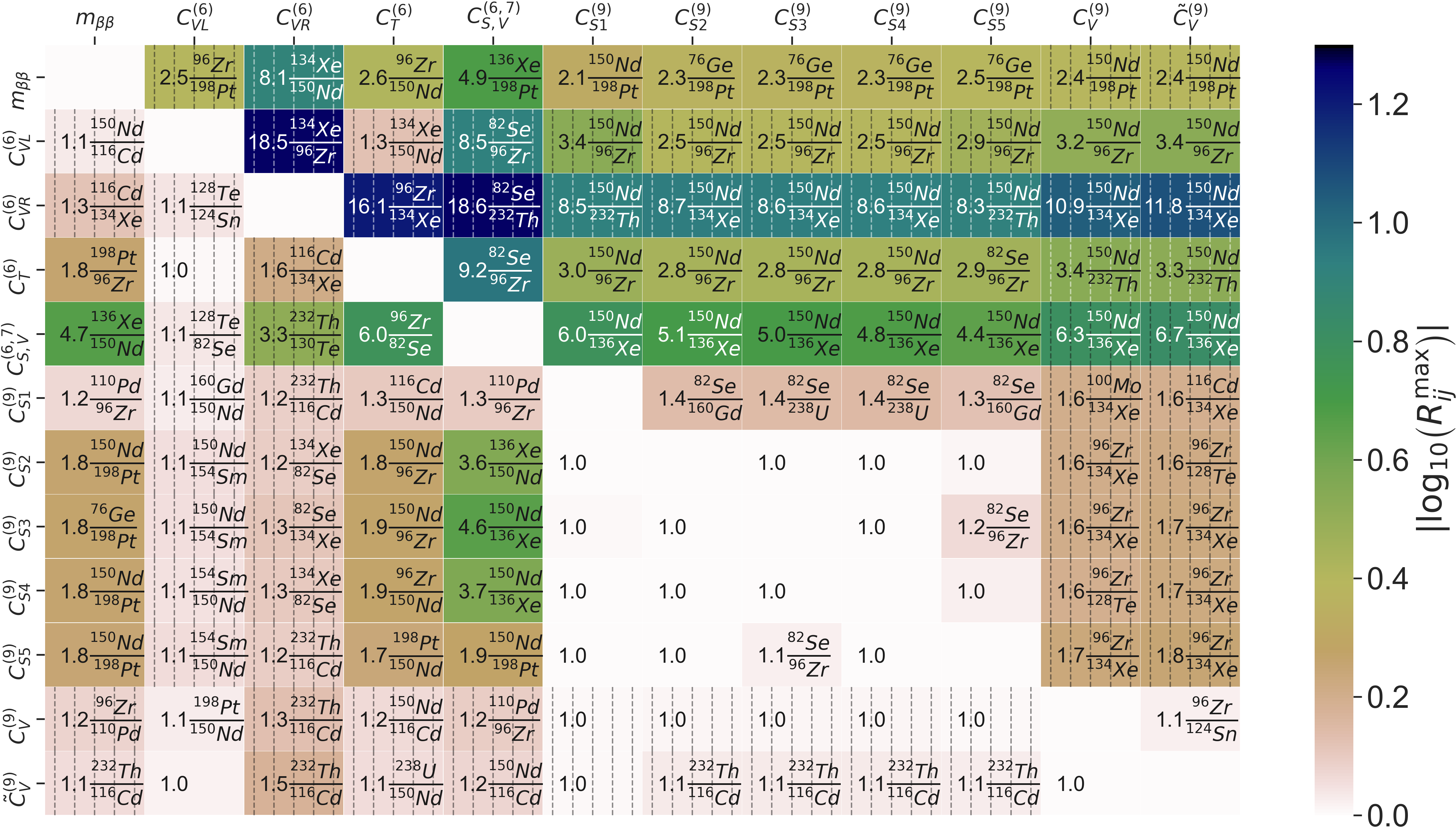}
    \caption{Same as Figure~\ref{fig:tile_plot_exp_isotopes} but now for all isotopes with available NMEs in the IBM2 framework: The maximal ratios $R^\mathrm{max}_{ij}$ for all operator combinations i,j are shown. The exact values and the corresponding isotopes are displayed in each tile. Additionally, operator combinations that result in different phase-space observables are marked by dashed-line shading. In the upper right half of the plot we show the ratios considering the central values from the variation of the LECs. In the lower left half we show the worst-case scenario considering the values of ratios $R_{ij}$ that are closest to 1 within the range obtained by the variation of the LECs.}
    \label{fig:tile_plot_all_isotopes}
\end{figure}

While we have focussed our discussion on isotopes for which experimental limits on the half-lives exist, we want to present our main findings of Figure~\ref{fig:tile_plot_exp_isotopes} here again but now considering all naturally occuring \0 isotopes for which we have nuclear matrix elements available in the IBM2 framework. The corresponding results are presented in Figure~\ref{fig:tile_plot_all_isotopes}. In Figures~\ref{fig:ratios_ROi_all} and~\ref{fig:ratios_ROi_epsilon_all} we show the resulting ratios including variations of the unknown LECs similar to Figures~\ref{fig:ratios_ROi} and~\ref{fig:ratios_ROi_epsilon} when considering the whole set of isotopes available.
\clearpage

\begin{landscape}
\newpage
\begin{figure}
    \includegraphics[width=1.4\textheight]{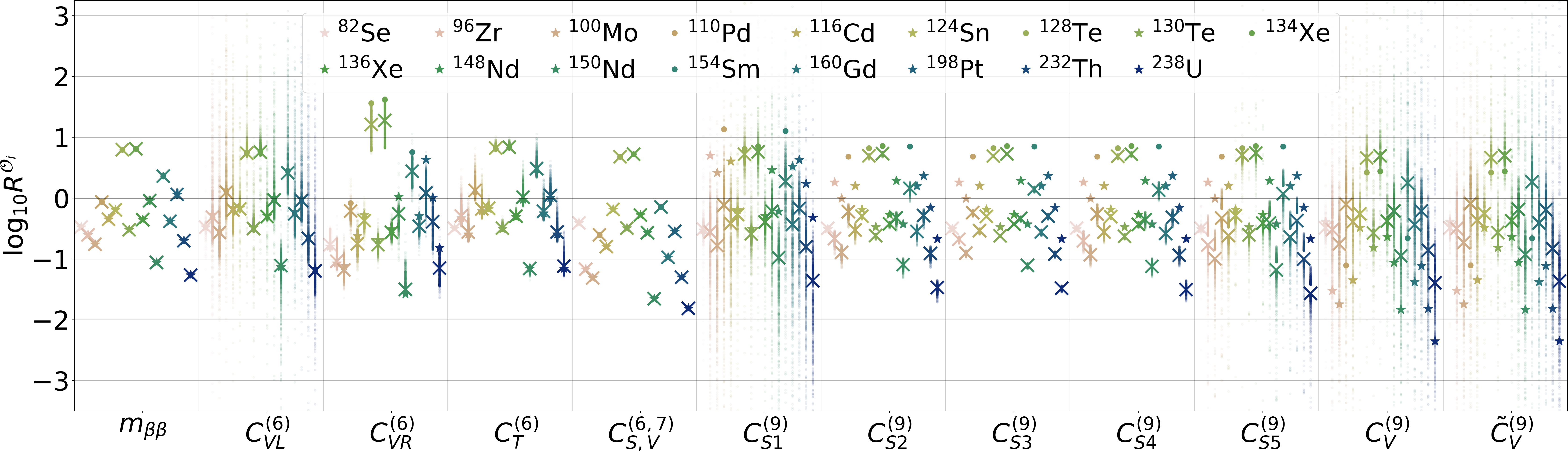}
    \includegraphics[width=1.4\textheight]{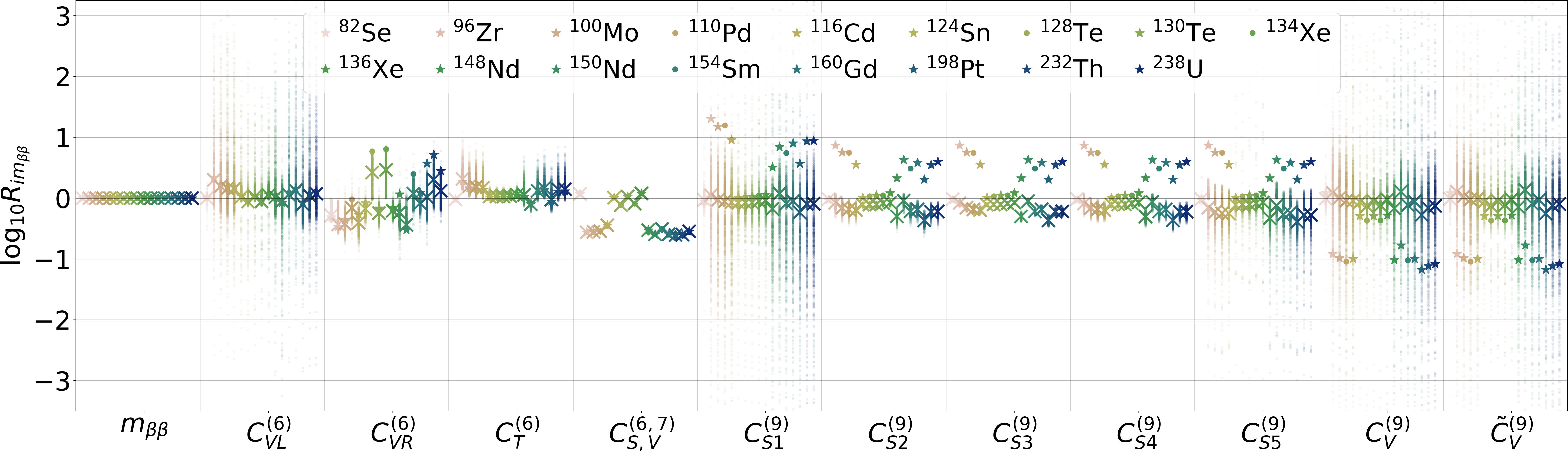}
    \caption{The decay rate ratios $R^{O_i}$ (upper plot) and $R_{i m_{\beta\beta}}$ (lower plot) for the different operator groups are shown. The larger markers represent the choice of vanishing unknown LECs with $g_{6,7}^{NN}=g_V^{\pi N}=\Tilde{g}_V^{\pi N}=1$. Isotopes with a PSF $G_0>10^{-14}\,\mathrm{y^{-1}}$ are represented by stars while isotopes with smaller PSFs are represented by round markers. The additional points represent variations of the different unknown LECs $g_i$ randomly chosen from $\left[-\sqrt{10}, -1/\sqrt{10}\right]\times |g_i|$ and $\left[+1/\sqrt{10}, +\sqrt{10}\right]\times |g_i|$ except for $g_\nu^{NN}$ which is varied in a range of $\pm 50\%$. The crosses represent the central values of the variation i.e. the median values. The reference isotope is chosen to be \textsuperscript{76}Ge.}
    \label{fig:ratios_ROi_all}
\end{figure}
\begin{figure}
    \centering
    \includegraphics[width=1.4 \textheight]{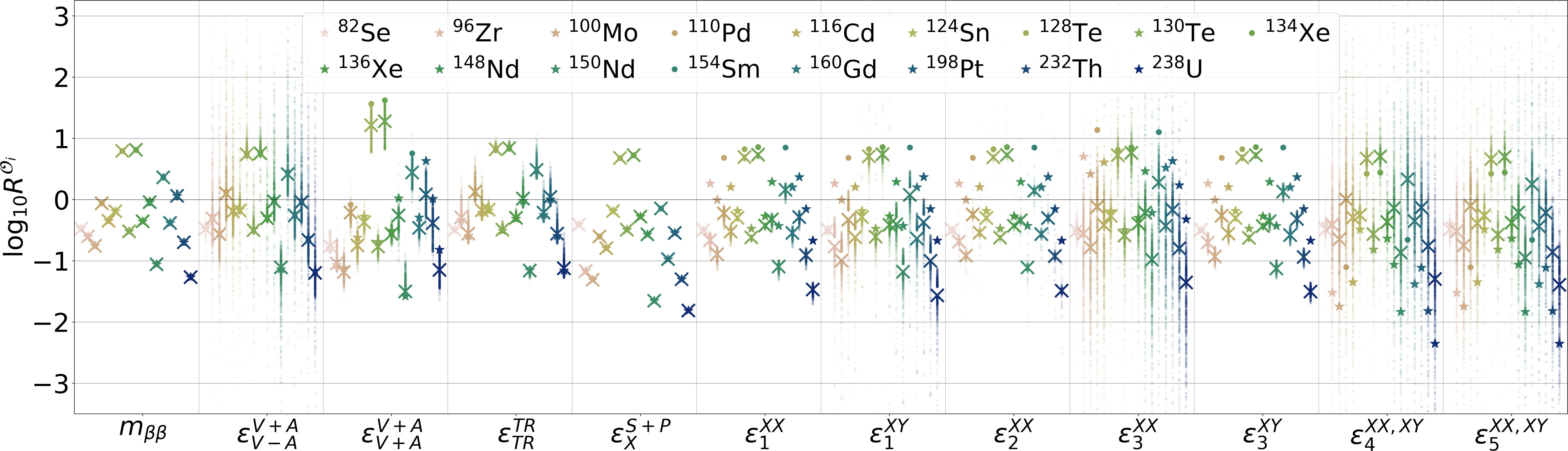}
    \includegraphics[width=1.4\textheight]{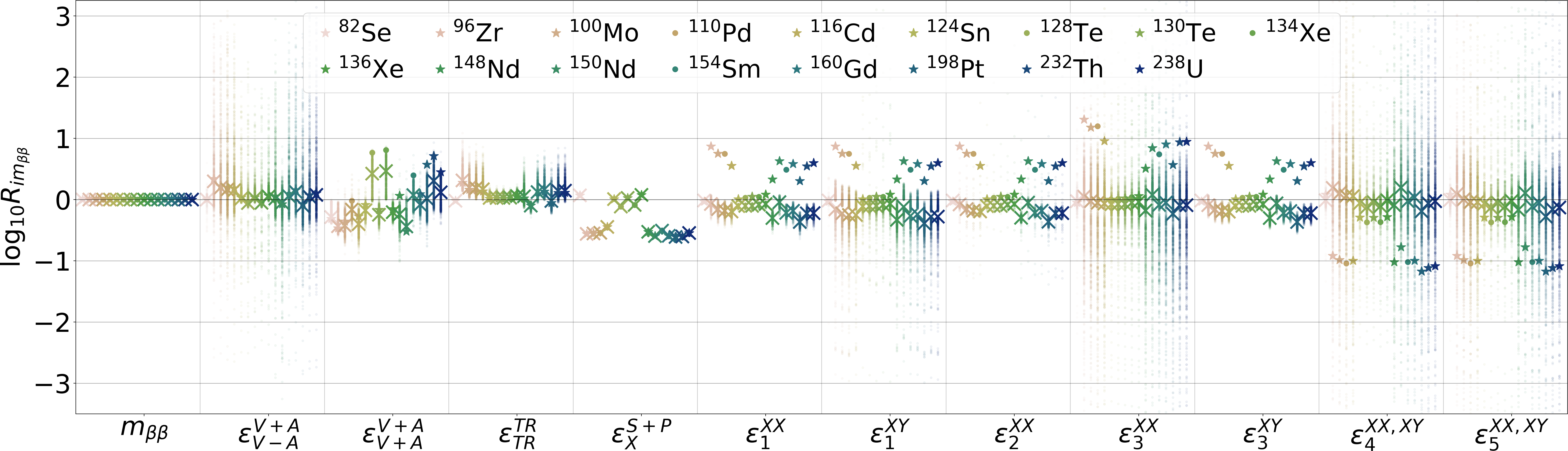}
    \caption{The decay rate ratios $R^{O_i}$ (upper plot) and $R_{i m_{\beta\beta}}$ (lower plot) for the different operator groups in the \textit{$\epsilon$-basis}.}
    \label{fig:ratios_ROi_epsilon_all}
\end{figure}
\clearpage
\global\pdfpageattr\expandafter{\the\pdfpageattr/Rotate 90}
\clearpage
\global\pdfpageattr\expandafter{\the\pdfpageattr/Rotate 0}
\end{landscape}
\newpage
\noindent
\bibliographystyle{utphys}
\bibliography{references}

\providecommand{\href}[2]{#2}\begingroup\raggedright\begin{thebibliography}{100}

\bibitem{Furry:1939qr}
W.~Furry, ``{On transition probabilities in double beta-disintegration},''
  \href{http://dx.doi.org/10.1103/PhysRev.56.1184}{{\em Phys. Rev.} {\bfseries
  56} (1939) 1184--1193}.

\bibitem{PhysRevD.25.2951}
J.~Schechter and J.~W.~F. Valle, ``Neutrinoless double-$\ensuremath{\beta}$
  decay in su(2)$\ifmmode\times\else\texttimes\fi{}$u(1) theories,''
  \href{http://dx.doi.org/10.1103/PhysRevD.25.2951}{{\em Phys. Rev. D}
  {\bfseries 25} (Jun, 1982) 2951--2954}.
  \url{https://link.aps.org/doi/10.1103/PhysRevD.25.2951}.

\bibitem{Pas:1999fc}
H.~Pas, M.~Hirsch, H.~Klapdor-Kleingrothaus, and S.~Kovalenko, ``{Towards a
  superformula for neutrinoless double beta decay},''
  \href{http://dx.doi.org/10.1016/S0370-2693(99)00330-5}{{\em Phys. Lett. B}
  {\bfseries 453} (1999) 194--198}.

\bibitem{Pas:2000vn}
H.~Pas, M.~Hirsch, H.~Klapdor-Kleingrothaus, and S.~Kovalenko, ``{A
  Superformula for neutrinoless double beta decay. 2. The Short range part},''
  \href{http://dx.doi.org/10.1016/S0370-2693(00)01359-9}{{\em Phys. Lett. B}
  {\bfseries 498} (2001) 35--39},
  \href{http://arxiv.org/abs/hep-ph/0008182}{{\ttfamily arXiv:hep-ph/0008182}}.

\bibitem{Deppisch:2012nb}
F.~F. Deppisch, M.~Hirsch, and H.~Päs, ``{Neutrinoless Double Beta Decay and
  Physics Beyond the Standard Model},''
  \href{http://dx.doi.org/10.1088/0954-3899/39/12/124007}{{\em J. Phys. G}
  {\bfseries 39} (2012) 124007},
  \href{http://arxiv.org/abs/1208.0727}{{\ttfamily arXiv:1208.0727 [hep-ph]}}.

\bibitem{Cirigliano_2017}
V.~Cirigliano, W.~Dekens, J.~de~Vries, M.~L. Graesser, and E.~Mereghetti,
  ``Neutrinoless double beta decay in chiral effective field theory: lepton
  number violation at dimension seven,''
  \href{http://dx.doi.org/10.1007/jhep12(2017)082}{{\em Journal of High Energy
  Physics} {\bfseries 2017} no.~12, (Dec, 2017) }.
  \url{http://dx.doi.org/10.1007/JHEP12(2017)082}.

\bibitem{Cirigliano_2018}
V.~Cirigliano, W.~Dekens, J.~de~Vries, M.~L. Graesser, and E.~Mereghetti, ``A
  neutrinoless double beta decay master formula from effective field theory,''
  \href{http://dx.doi.org/10.1007/jhep12(2018)097}{{\em Journal of High Energy
  Physics} {\bfseries 2018} no.~12, (Dec, 2018) }.
  \url{http://dx.doi.org/10.1007/JHEP12(2018)097}.

\bibitem{Graf:2018ozy}
L.~Graf, F.~F. Deppisch, F.~Iachello, and J.~Kotila, ``{Short-Range
  Neutrinoless Double Beta Decay Mechanisms},''
  \href{http://dx.doi.org/10.1103/PhysRevD.98.095023}{{\em Phys. Rev. D}
  {\bfseries 98} no.~9, (2018) 095023},
  \href{http://arxiv.org/abs/1806.06058}{{\ttfamily arXiv:1806.06058
  [hep-ph]}}.

\bibitem{Deppisch:2020ztt}
F.~F. Deppisch, L.~Graf, F.~Iachello, and J.~Kotila, ``{Analysis of light
  neutrino exchange and short-range mechanisms in $0\nu\beta\beta$ decay},''
  \href{http://dx.doi.org/10.1103/PhysRevD.102.095016}{{\em Phys. Rev. D}
  {\bfseries 102} no.~9, (2020) 095016},
  \href{http://arxiv.org/abs/2009.10119}{{\ttfamily arXiv:2009.10119
  [hep-ph]}}.

\bibitem{Pati:1974yy}
J.~C. Pati and A.~Salam, ``{Lepton Number as the Fourth Color},''
  \href{http://dx.doi.org/10.1103/PhysRevD.10.275}{{\em Phys. Rev. D}
  {\bfseries 10} (1974) 275--289}. [Erratum: Phys.Rev.D 11, 703--703 (1975)].

\bibitem{Mohapatra:1974hk}
R.~N. Mohapatra and J.~C. Pati, ``{Left-Right Gauge Symmetry and an
  Isoconjugate Model of CP Violation},''
  \href{http://dx.doi.org/10.1103/PhysRevD.11.566}{{\em Phys. Rev. D}
  {\bfseries 11} (1975) 566--571}.

\bibitem{Mohapatra:1974gc}
R.~N. Mohapatra and J.~C. Pati, ``{A Natural Left-Right Symmetry},''
  \href{http://dx.doi.org/10.1103/PhysRevD.11.2558}{{\em Phys. Rev. D}
  {\bfseries 11} (1975) 2558}.

\bibitem{Senjanovic:1975rk}
G.~Senjanovic and R.~N. Mohapatra, ``{Exact Left-Right Symmetry and Spontaneous
  Violation of Parity},''
  \href{http://dx.doi.org/10.1103/PhysRevD.12.1502}{{\em Phys. Rev. D}
  {\bfseries 12} (1975) 1502}.

\bibitem{Huang_2014}
W.-C. Huang and J.~Lopez-Pavon, ``{On neutrinoless double beta decay in the
  minimal left-right symmetric model},''
  \href{http://dx.doi.org/10.1140/epjc/s10052-014-2853-z}{{\em Eur. Phys. J. C}
  {\bfseries 74} (2014) 2853}, \href{http://arxiv.org/abs/1310.0265}{{\ttfamily
  arXiv:1310.0265 [hep-ph]}}.

\bibitem{Giunti:2010wz}
C.~Giunti and M.~Laveder, ``{Short-Baseline Electron Neutrino Disappearance,
  Tritium Beta Decay and Neutrinoless Double-Beta Decay},''
  \href{http://dx.doi.org/10.1103/PhysRevD.82.053005}{{\em Phys. Rev. D}
  {\bfseries 82} (2010) 053005},
  \href{http://arxiv.org/abs/1005.4599}{{\ttfamily arXiv:1005.4599 [hep-ph]}}.

\bibitem{Barry:2011wb}
J.~Barry, W.~Rodejohann, and H.~Zhang, ``{Light Sterile Neutrinos: Models and
  Phenomenology},'' \href{http://dx.doi.org/10.1007/JHEP07(2011)091}{{\em JHEP}
  {\bfseries 07} (2011) 091}, \href{http://arxiv.org/abs/1105.3911}{{\ttfamily
  arXiv:1105.3911 [hep-ph]}}.

\bibitem{Barea:2015zfa}
J.~Barea, J.~Kotila, and F.~Iachello, ``{Limits on sterile neutrino
  contributions to neutrinoless double beta decay},''
  \href{http://dx.doi.org/10.1103/PhysRevD.92.093001}{{\em Phys. Rev. D}
  {\bfseries 92} (2015) 093001},
  \href{http://arxiv.org/abs/1509.01925}{{\ttfamily arXiv:1509.01925
  [hep-ph]}}.

\bibitem{Dekens:2020ttz}
W.~Dekens, J.~de~Vries, K.~Fuyuto, E.~Mereghetti, and G.~Zhou, ``{Sterile
  neutrinos and neutrinoless double beta decay in effective field theory},''
  \href{http://dx.doi.org/10.1007/JHEP06(2020)097}{{\em JHEP} {\bfseries 06}
  (2020) 097}, \href{http://arxiv.org/abs/2002.07182}{{\ttfamily
  arXiv:2002.07182 [hep-ph]}}.

\bibitem{Li:2020flq}
G.~Li, M.~Ramsey-Musolf, and J.~C. Vasquez, ``{Left-Right Symmetry and Leading
  Contributions to Neutrinoless Double Beta Decay},''
  \href{http://dx.doi.org/10.1103/PhysRevLett.126.151801}{{\em Phys. Rev.
  Lett.} {\bfseries 126} no.~15, (2021) 151801},
  \href{http://arxiv.org/abs/2009.01257}{{\ttfamily arXiv:2009.01257
  [hep-ph]}}.

\bibitem{Bolton:2019pcu}
P.~D. Bolton, F.~F. Deppisch, and P.~S. Bhupal~Dev, ``{Neutrinoless double beta
  decay versus other probes of heavy sterile neutrinos},''
  \href{http://dx.doi.org/10.1007/JHEP03(2020)170}{{\em JHEP} {\bfseries 03}
  (2020) 170}, \href{http://arxiv.org/abs/1912.03058}{{\ttfamily
  arXiv:1912.03058 [hep-ph]}}.

\bibitem{2020GERDA}
M.~Agostini {\em et~al.}, ``Final results of gerda on the search for
  neutrinoless double-$\beta$ decay,''
  \href{http://dx.doi.org/10.1103/physrevlett.125.252502}{{\em Physical Review
  Letters} {\bfseries 125} no.~25, (Dec, 2020) }.
  \url{http://dx.doi.org/10.1103/PhysRevLett.125.252502}.

\bibitem{2019CUPID}
O.~Azzolini {\em et~al.}, ``Final result of cupid-0 phase-i in the search for
  the se82 neutrinoless double-$\beta$ decay,''
  \href{http://dx.doi.org/10.1103/physrevlett.123.032501}{{\em Physical Review
  Letters} {\bfseries 123} no.~3, (Jul, 2019) }.
  \url{http://dx.doi.org/10.1103/PhysRevLett.123.032501}.

\bibitem{2010NEMO3Zr}
J.~Argyriades {\em et~al.}, ``Measurement of the two neutrino double beta decay
  half-life of zr-96 with the nemo-3 detector,''
  \href{http://dx.doi.org/10.1016/j.nuclphysa.2010.07.009}{{\em Nuclear Physics
  A} {\bfseries 847} no.~3-4, (Dec, 2010) 168–179}.
  \url{http://dx.doi.org/10.1016/j.nuclphysa.2010.07.009}.

\bibitem{2021CUPID-Mo}
E.~Armengaud {\em et~al.}, ``New limit for neutrinoless double-beta decay of
  mo100 from the cupid-mo experiment,''
  \href{http://dx.doi.org/10.1103/physrevlett.126.181802}{{\em Physical Review
  Letters} {\bfseries 126} no.~18, (May, 2021) }.
  \url{http://dx.doi.org/10.1103/PhysRevLett.126.181802}.

\bibitem{2016Cd116}
F.~A. Danevich {\em et~al.}, ``Search for double beta decay of116cd with
  enriched116cdwo4crystal scintillators (aurora experiment),''
  \href{http://dx.doi.org/10.1088/1742-6596/718/6/062009}{{\em Journal of
  Physics: Conference Series} {\bfseries 718} (May, 2016) 062009}.
  \url{http://dx.doi.org/10.1088/1742-6596/718/6/062009}.

\bibitem{2003Te128}
C.~Arnaboldi {\em et~al.}, ``A calorimetric search on double beta decay of
  130te,'' \href{http://dx.doi.org/10.1016/s0370-2693(03)00212-0}{{\em Physics
  Letters B} {\bfseries 557} no.~3-4, (Apr, 2003) 167–175}.
  \url{http://dx.doi.org/10.1016/S0370-2693(03)00212-0}.

\bibitem{2020CUORE130Te}
D.~Adams {\em et~al.}, ``Improved limit on neutrinoless double-beta decay in
  te130 with cuore,''
  \href{http://dx.doi.org/10.1103/physrevlett.124.122501}{{\em Physical Review
  Letters} {\bfseries 124} no.~12, (Mar, 2020) }.
  \url{http://dx.doi.org/10.1103/PhysRevLett.124.122501}.

\bibitem{2019EXO}
G.~Anton {\em et~al.}, ``Search for neutrinoless double-$\beta$ decay with the
  complete exo-200 dataset,''
  \href{http://dx.doi.org/10.1103/physrevlett.123.161802}{{\em Physical Review
  Letters} {\bfseries 123} no.~16, (Oct, 2019) }.
  \url{http://dx.doi.org/10.1103/PhysRevLett.123.161802}.

\bibitem{2016KamLandZen}
A.~Gando {\em et~al.}, ``Search for majorana neutrinos near the inverted mass
  hierarchy region with kamland-zen,''
  \href{http://dx.doi.org/10.1103/physrevlett.117.082503}{{\em Physical Review
  Letters} {\bfseries 117} no.~8, (Aug, 2016) }.
  \url{http://dx.doi.org/10.1103/PhysRevLett.117.082503}.

\bibitem{2009NEMO150Nd}
J.~Argyriades {\em et~al.}, ``Measurement of the double-$\beta$ decay half-life
  of nd150 and search for neutrinoless decay modes with the nemo-3 detector,''
  \href{http://dx.doi.org/10.1103/physrevc.80.032501}{{\em Physical Review C}
  {\bfseries 80} no.~3, (Sep, 2009) }.
  \url{http://dx.doi.org/10.1103/PhysRevC.80.032501}.

\bibitem{KamLAND-Zen:2022tow}
{\bfseries KamLAND-Zen} Collaboration, S.~Abe {\em et~al.}, ``{First Search for
  the Majorana Nature of Neutrinos in the Inverted Mass Ordering Region with
  KamLAND-Zen},'' \href{http://arxiv.org/abs/2203.02139}{{\ttfamily
  arXiv:2203.02139 [hep-ex]}}.

\bibitem{2017LEGEND200}
N.~Abgrall {\em et~al.}, ``The large enriched germanium experiment for
  neutrinoless double beta decay (legend),''
  \href{http://dx.doi.org/10.1063/1.5007652}{{\em AIP Conference Proceedings}
  (2017) }. \url{http://dx.doi.org/10.1063/1.5007652}.

\bibitem{legendcollaboration2021legend1000}
{\bfseries LEGEND} Collaboration, N.~Abgrall {\em et~al.}, ``Legend-1000
  preconceptual design report,'' 2021.

\bibitem{2020CupidProspects}
E.~Armengaud {\em et~al.}, ``The cupid-mo experiment for neutrinoless
  double-beta decay: performance and prospects,''
  \href{http://dx.doi.org/10.1140/epjc/s10052-019-7578-6}{{\em The European
  Physical Journal C} {\bfseries 80} no.~1, (Jan, 2020) }.
  \url{http://dx.doi.org/10.1140/epjc/s10052-019-7578-6}.

\bibitem{2021SNOplus}
V.~Albanese {\em et~al.}, ``The sno+ experiment,''
  \href{http://dx.doi.org/10.1088/1748-0221/16/08/p08059}{{\em Journal of
  Instrumentation} {\bfseries 16} no.~08, (Aug, 2021) P08059}.
  \url{http://dx.doi.org/10.1088/1748-0221/16/08/P08059}.

\bibitem{nexocollaboration2021nexo}
{\bfseries nEXO Collaboration} Collaboration, G.~Adhikari {\em et~al.}, ``nexo:
  Neutrinoless double beta decay search beyond $10^{28}$ year half-life
  sensitivity,'' 2021.

\bibitem{Deppisch:2020mxv}
F.~F. Deppisch, L.~Graf, and F.~\v{S}imkovic, ``{Searching for New Physics in
  Two-Neutrino Double Beta Decay},''
  \href{http://dx.doi.org/10.1103/PhysRevLett.125.171801}{{\em Phys. Rev.
  Lett.} {\bfseries 125} no.~17, (2020) 171801},
  \href{http://arxiv.org/abs/2003.11836}{{\ttfamily arXiv:2003.11836
  [hep-ph]}}.

\bibitem{2019review_Werner}
M.~J. Dolinski, A.~W. Poon, and W.~Rodejohann, ``Neutrinoless double-beta
  decay: Status and prospects,''
  \href{http://dx.doi.org/10.1146/annurev-nucl-101918-023407}{{\em Annual
  Review of Nuclear and Particle Science} {\bfseries 69} no.~1, (Oct, 2019)
  219–251}. \url{http://dx.doi.org/10.1146/annurev-nucl-101918-023407}.

\bibitem{Agostini:2022zub}
M.~Agostini, G.~Benato, J.~A. Detwiler, J.~Men\'endez, and F.~Vissani,
  ``{Toward the discovery of matter creation with neutrinoless double-beta
  decay},'' \href{http://arxiv.org/abs/2202.01787}{{\ttfamily arXiv:2202.01787
  [hep-ex]}}.

\bibitem{neutrinotool}
J.~de~Vries, L.~Gr\'{a}f, and O.~Scholer To be published.

\bibitem{Graesser_2017}
M.~L. Graesser, ``An electroweak basis for neutrinoless double $\beta$ decay,''
  \href{http://dx.doi.org/10.1007/jhep08(2017)099}{{\em Journal of High Energy
  Physics} {\bfseries 2017} no.~8, (Aug, 2017) }.
  \url{http://dx.doi.org/10.1007/JHEP08(2017)099}.

\bibitem{Prezeau:2003xn}
G.~Prezeau, M.~Ramsey-Musolf, and P.~Vogel, ``{Neutrinoless double beta decay
  and effective field theory},''
  \href{http://dx.doi.org/10.1103/PhysRevD.68.034016}{{\em Phys. Rev. D}
  {\bfseries 68} (2003) 034016},
  \href{http://arxiv.org/abs/hep-ph/0303205}{{\ttfamily arXiv:hep-ph/0303205}}.

\bibitem{Deppisch_2020}
F.~F. Deppisch, L.~Graf, F.~Iachello, and J.~Kotila, ``Analysis of light
  neutrino exchange and short-range mechanisms in $0\nu\beta\beta$ decay,''
  \href{http://dx.doi.org/10.1103/physrevd.102.095016}{{\em Physical Review D}
  {\bfseries 102} no.~9, (Nov, 2020) }.
  \url{http://dx.doi.org/10.1103/PhysRevD.102.095016}.

\bibitem{Cirigliano:2018hja}
V.~Cirigliano, W.~Dekens, J.~De~Vries, M.~L. Graesser, E.~Mereghetti,
  S.~Pastore, and U.~Van~Kolck, ``{New Leading Contribution to Neutrinoless
  Double-\ensuremath{\beta} Decay},''
  \href{http://dx.doi.org/10.1103/PhysRevLett.120.202001}{{\em Phys. Rev.
  Lett.} {\bfseries 120} no.~20, (2018) 202001},
  \href{http://arxiv.org/abs/1802.10097}{{\ttfamily arXiv:1802.10097
  [hep-ph]}}.

\bibitem{Cirigliano:2019vdj}
V.~Cirigliano, W.~Dekens, J.~De~Vries, M.~L. Graesser, E.~Mereghetti,
  S.~Pastore, M.~Piarulli, U.~Van~Kolck, and R.~B. Wiringa, ``{Renormalized
  approach to neutrinoless double- $\beta$ decay},''
  \href{http://dx.doi.org/10.1103/PhysRevC.100.055504}{{\em Phys. Rev. C}
  {\bfseries 100} no.~5, (2019) 055504},
  \href{http://arxiv.org/abs/1907.11254}{{\ttfamily arXiv:1907.11254
  [nucl-th]}}.

\bibitem{Bhattacharya_2016}
T.~Bhattacharya, V.~Cirigliano, S.~D. Cohen, R.~Gupta, H.-W. Lin, and B.~Yoon,
  ``Axial, scalar, and tensor charges of the nucleon from2+1+1-flavor lattice
  qcd,'' \href{http://dx.doi.org/10.1103/physrevd.94.054508}{{\em Physical
  Review D} {\bfseries 94} no.~5, (Sep, 2016) }.
  \url{http://dx.doi.org/10.1103/PhysRevD.94.054508}.

\bibitem{Nicholson_2018}
A.~Nicholson {\em et~al.}, ``Heavy physics contributions to neutrinoless double
  beta decay from qcd,''
  \href{http://dx.doi.org/10.1103/physrevlett.121.172501}{{\em Physical Review
  Letters} {\bfseries 121} no.~17, (Oct, 2018) }.
  \url{http://dx.doi.org/10.1103/PhysRevLett.121.172501}.

\bibitem{Cirigliano_2021_1}
V.~Cirigliano, W.~Dekens, J.~de~Vries, M.~Hoferichter, and E.~Mereghetti,
  ``Toward complete leading-order predictions for neutrinoless double$\beta$
  decay,'' \href{http://dx.doi.org/10.1103/physrevlett.126.172002}{{\em
  Physical Review Letters} {\bfseries 126} no.~17, (Apr, 2021) }.
  \url{http://dx.doi.org/10.1103/PhysRevLett.126.172002}.

\bibitem{Cirigliano_2021}
V.~Cirigliano, W.~Dekens, J.~de~Vries, M.~Hoferichter, and E.~Mereghetti,
  ``Determining the leading-order contact term in neutrinoless double $\beta$
  decay,'' \href{http://dx.doi.org/10.1007/jhep05(2021)289}{{\em Journal of
  High Energy Physics} {\bfseries 2021} no.~5, (May, 2021) }.
  \url{http://dx.doi.org/10.1007/JHEP05(2021)289}.

\bibitem{Wirth:2021pij}
R.~Wirth, J.~M. Yao, and H.~Hergert, ``{Ab initio calculation of the contact
  operator contribution in the standard mechanism for neutrinoless double beta
  decay},'' \href{http://arxiv.org/abs/2105.05415}{{\ttfamily arXiv:2105.05415
  [nucl-th]}}.

\bibitem{Arnold_2017}
R.~Arnold and others., ``Measurement of the $2\nu\beta\beta$ decay half-life
  and search for the $0\nu\beta\beta$ decay of cd116 with the nemo-3
  detector,'' \href{http://dx.doi.org/10.1103/physrevd.95.012007}{{\em Physical
  Review D} {\bfseries 95} no.~1, (Jan, 2017) }.
  \url{http://dx.doi.org/10.1103/PhysRevD.95.012007}.

\bibitem{Arnold2010}
R.~Arnold {\em et~al.}, ``Probing new physics models of neutrinoless double
  beta decay with supernemo,''
  \href{http://dx.doi.org/10.1140/epjc/s10052-010-1481-5}{{\em The European
  Physical Journal C} {\bfseries 70} no.~4, (Dec, 2010) 927--943}.
  \url{https://doi.org/10.1140/epjc/s10052-010-1481-5}.

\bibitem{Kotila_2012}
J.~Kotila and F.~Iachello, ``Phase-space factors for double-$\beta$ decay,''
  \href{http://dx.doi.org/10.1103/physrevc.85.034316}{{\em Physical Review C}
  {\bfseries 85} no.~3, (Mar, 2012) }.
  \url{http://dx.doi.org/10.1103/PhysRevC.85.034316}.

\bibitem{_tef_nik_2015}
D.~Štefánik, R.~Dvornický, F.~Šimkovic, and P.~Vogel, ``Reexamining the
  light neutrino exchange mechanism of the $0\nu\beta\beta$ decay with left-
  and right-handed leptonic and hadronic currents,''
  \href{http://dx.doi.org/10.1103/physrevc.92.055502}{{\em Physical Review C}
  {\bfseries 92} no.~5, (Nov, 2015) }.
  \url{http://dx.doi.org/10.1103/PhysRevC.92.055502}.

\bibitem{Ali:2007ec}
A.~Ali, A.~V. Borisov, and D.~V. Zhuridov, ``{Probing new physics in the
  neutrinoless double beta decay using electron angular correlation},''
  \href{http://dx.doi.org/10.1103/PhysRevD.76.093009}{{\em Phys. Rev. D}
  {\bfseries 76} (2007) 093009},
  \href{http://arxiv.org/abs/0706.4165}{{\ttfamily arXiv:0706.4165 [hep-ph]}}.

\bibitem{Deppisch_2007}
F.~Deppisch and H.~Päs, ``Pinning down the mechanism of neutrinoless
  double$\beta$ decay with measurements in different nuclei,''
  \href{http://dx.doi.org/10.1103/physrevlett.98.232501}{{\em Physical Review
  Letters} {\bfseries 98} no.~23, (Jun, 2007) }.
  \url{http://dx.doi.org/10.1103/PhysRevLett.98.232501}.

\bibitem{Gehman:2007qg}
V.~M. Gehman and S.~R. Elliott, ``{Multiple-Isotope Comparison for Determining
  0 nu beta beta Mechanisms},''
  \href{http://dx.doi.org/10.1088/0954-3899/34/4/006}{{\em J. Phys. G}
  {\bfseries 34} (2007) 667--678},
  \href{http://arxiv.org/abs/hep-ph/0701099}{{\ttfamily arXiv:hep-ph/0701099}}.
  [Erratum: J.Phys.G 35, 029701 (2008)].

\bibitem{Lisi:2015yma}
E.~Lisi, A.~Rotunno, and F.~Simkovic, ``{Degeneracies of particle and nuclear
  physics uncertainties in neutrinoless $\beta \beta$ decay},''
  \href{http://dx.doi.org/10.1103/PhysRevD.92.093004}{{\em Phys. Rev. D}
  {\bfseries 92} no.~9, (2015) 093004},
  \href{http://arxiv.org/abs/1506.04058}{{\ttfamily arXiv:1506.04058
  [hep-ph]}}.

\bibitem{Fogli:2009py}
G.~L. Fogli, E.~Lisi, and A.~M. Rotunno, ``{Probing particle and nuclear
  physics models of neutrinoless double beta decay with different nuclei},''
  \href{http://dx.doi.org/10.1103/PhysRevD.80.015024}{{\em Phys. Rev. D}
  {\bfseries 80} (2009) 015024},
  \href{http://arxiv.org/abs/0905.1832}{{\ttfamily arXiv:0905.1832 [hep-ph]}}.

\bibitem{Simkovic:2010ka}
F.~Simkovic, J.~Vergados, and A.~Faessler, ``{Few active mechanisms of the
  neutrinoless double beta-decay and effective mass of Majorana neutrinos},''
  \href{http://dx.doi.org/10.1103/PhysRevD.82.113015}{{\em Phys. Rev. D}
  {\bfseries 82} (2010) 113015},
  \href{http://arxiv.org/abs/1006.0571}{{\ttfamily arXiv:1006.0571 [hep-ph]}}.

\bibitem{Meroni:2012qf}
A.~Meroni, S.~T. Petcov, and F.~Simkovic, ``{Multiple CP non-conserving
  mechanisms of $(\beta\beta)_{0\nu}$-decay and nuclei with largely different
  nuclear matrix elements},''
  \href{http://dx.doi.org/10.1007/JHEP02(2013)025}{{\em JHEP} {\bfseries 02}
  (2013) 025}, \href{http://arxiv.org/abs/1212.1331}{{\ttfamily arXiv:1212.1331
  [hep-ph]}}.

\bibitem{Horoi:2018fls}
M.~Horoi and A.~Neacsu, ``{Shell model study of using an effective field theory
  for disentangling several contributions to neutrinoless double-$\beta$
  decay},'' \href{http://dx.doi.org/10.1103/PhysRevC.98.035502}{{\em Phys. Rev.
  C} {\bfseries 98} no.~3, (2018) 035502},
  \href{http://arxiv.org/abs/1801.04496}{{\ttfamily arXiv:1801.04496
  [nucl-th]}}.

\bibitem{Nicholson:2018mwc}
A.~Nicholson {\em et~al.}, ``{Heavy physics contributions to neutrinoless
  double beta decay from QCD},''
  \href{http://dx.doi.org/10.1103/PhysRevLett.121.172501}{{\em Phys. Rev.
  Lett.} {\bfseries 121} no.~17, (2018) 172501},
  \href{http://arxiv.org/abs/1805.02634}{{\ttfamily arXiv:1805.02634
  [nucl-th]}}.

\bibitem{Detmold:2020jqv}
{\bfseries NPLQCD} Collaboration, W.~Detmold and D.~J. Murphy, ``{Neutrinoless
  Double Beta Decay from Lattice QCD: The Long-Distance $\pi^{-} \rightarrow
  \pi^{+} e^{-} e^{-}$ Amplitude},''
  \href{http://arxiv.org/abs/2004.07404}{{\ttfamily arXiv:2004.07404
  [hep-lat]}}.

\bibitem{Tuo:2019bue}
X.-Y. Tuo, X.~Feng, and L.-C. Jin, ``{Long-distance contributions to
  neutrinoless double beta decay $\pi^- \to\pi^+ e e$},''
  \href{http://dx.doi.org/10.1103/PhysRevD.100.094511}{{\em Phys. Rev. D}
  {\bfseries 100} no.~9, (2019) 094511},
  \href{http://arxiv.org/abs/1909.13525}{{\ttfamily arXiv:1909.13525
  [hep-lat]}}.

\bibitem{2017EXO}
J.~Albert {\em et~al.}, ``Searches for double beta decay of xe134 with
  exo-200,'' \href{http://dx.doi.org/10.1103/physrevd.96.092001}{{\em Physical
  Review D} {\bfseries 96} no.~9, (Nov, 2017) }.
  \url{http://dx.doi.org/10.1103/PhysRevD.96.092001}.

\bibitem{menendez2017neutrinoless}
J.~Men{\'e}ndez, ``Neutrinoless $\beta$$\beta$ decay mediated by the exchange
  of light and heavy neutrinos: The role of nuclear structure correlations,''
  {\em Journal of Physics G: Nuclear and Particle Physics} {\bfseries 45}
  no.~1, (2017) 014003.

\bibitem{2021tpSM_NMEs}
Y.~K. Wang, P.~W. Zhao, and J.~Meng, ``Nuclear matrix elements of neutrinoless
  double-$\beta$ decay in the triaxial projected shell model,''
  \href{http://dx.doi.org/10.1103/physrevc.104.014320}{{\em Physical Review C}
  {\bfseries 104} no.~1, (Jul, 2021) }.
  \url{http://dx.doi.org/10.1103/PhysRevC.104.014320}.

\bibitem{2020rSM_NMEs}
L.~Coraggio, A.~Gargano, N.~Itaco, R.~Mancino, and F.~Nowacki, ``Calculation of
  the neutrinoless double-$\beta$ decay matrix element within the realistic
  shell model,'' \href{http://dx.doi.org/10.1103/physrevc.101.044315}{{\em
  Physical Review C} {\bfseries 101} no.~4, (Apr, 2020) }.
  \url{http://dx.doi.org/10.1103/PhysRevC.101.044315}.

\bibitem{PhysRevC.91.024613}
J.~Hyv\"arinen and J.~Suhonen, ``Nuclear matrix elements for
  $0\ensuremath{\nu}\ensuremath{\beta}\ensuremath{\beta}$ decays with light or
  heavy majorana-neutrino exchange,''
  \href{http://dx.doi.org/10.1103/PhysRevC.91.024613}{{\em Phys. Rev. C}
  {\bfseries 91} (Feb, 2015) 024613}.
  \url{https://link.aps.org/doi/10.1103/PhysRevC.91.024613}.

\bibitem{2018dQRPAFang}
D.-L. Fang, A.~Faessler, and F.~Šimkovic, ``$0\nu\beta\beta$ -decay nuclear
  matrix element for light and heavy neutrino mass mechanisms from deformed
  quasiparticle random-phase approximation calculations for
  ge76, se82, te130, xe136 , and nd150 with isospin restoration,''
  \href{http://dx.doi.org/10.1103/physrevc.97.045503}{{\em Physical Review C}
  {\bfseries 97} no.~4, (Apr, 2018) }.
  \url{http://dx.doi.org/10.1103/PhysRevC.97.045503}.

\bibitem{PhysRevC.87.064302}
M.~T. Mustonen and J.~Engel, ``Large-scale calculations of the
  double-$\ensuremath{\beta}$ decay of
  ${}^{76}\mathbf{Ge},\phantom{\rule{0.28em}{0ex}}{}^{130}\mathbf{Te},\phantom{\rule{0.28em}{0ex}}{}^{136}\mathbf{Xe}$,
  and ${}^{150}\mathbf{Nd}$ in the deformed self-consistent skyrme
  quasiparticle random-phase approximation,''
  \href{http://dx.doi.org/10.1103/PhysRevC.87.064302}{{\em Phys. Rev. C}
  {\bfseries 87} (Jun, 2013) 064302}.
  \url{https://link.aps.org/doi/10.1103/PhysRevC.87.064302}.

\bibitem{PhysRevC.91.024316}
J.~M. Yao, L.~S. Song, K.~Hagino, P.~Ring, and J.~Meng, ``Systematic study of
  nuclear matrix elements in neutrinoless double-$\ensuremath{\beta}$ decay
  with a beyond-mean-field covariant density functional theory,''
  \href{http://dx.doi.org/10.1103/PhysRevC.91.024316}{{\em Phys. Rev. C}
  {\bfseries 91} (Feb, 2015) 024316}.
  \url{https://link.aps.org/doi/10.1103/PhysRevC.91.024316}.

\bibitem{2017_rEDF_Song}
L.~S. Song, J.~M. Yao, P.~Ring, and J.~Meng, ``Nuclear matrix element of
  neutrinoless double-$\beta$ decay: Relativity and short-range correlations,''
  \href{http://dx.doi.org/10.1103/physrevc.95.024305}{{\em Physical Review C}
  {\bfseries 95} no.~2, (Feb, 2017) }.
  \url{http://dx.doi.org/10.1103/PhysRevC.95.024305}.

\bibitem{PhysRevLett.111.142501}
N.~L. Vaquero, T.~R. Rodr\'{\i}guez, and J.~L. Egido, ``Shape and pairing
  fluctuation effects on neutrinoless double beta decay nuclear matrix
  elements,'' \href{http://dx.doi.org/10.1103/PhysRevLett.111.142501}{{\em
  Phys. Rev. Lett.} {\bfseries 111} (Sep, 2013) 142501}.
  \url{https://link.aps.org/doi/10.1103/PhysRevLett.111.142501}.

\bibitem{2020_ab_initio}
J.~Yao, B.~Bally, J.~Engel, R.~Wirth, T.~Rodríguez, and H.~Hergert,
  ``Ab initio treatment of collective correlations and the neutrinoless double
  beta decay of ca48,''
  \href{http://dx.doi.org/10.1103/physrevlett.124.232501}{{\em Physical Review
  Letters} {\bfseries 124} no.~23, (Jun, 2020) }.
  \url{http://dx.doi.org/10.1103/PhysRevLett.124.232501}.

\bibitem{2021_ab_initio}
A.~Belley, C.~Payne, S.~Stroberg, T.~Miyagi, and J.~Holt, ``Ab initio
  neutrinoless double-beta decay matrix elements for ca48 , ge76 , and se82,''
  \href{http://dx.doi.org/10.1103/physrevlett.126.042502}{{\em Physical Review
  Letters} {\bfseries 126} no.~4, (Jan, 2021) }.
  \url{http://dx.doi.org/10.1103/PhysRevLett.126.042502}.

\bibitem{Kotila_2013}
J.~Kotila and F.~Iachello, ``Phase space factors for$\beta^+\beta^+$ decay and
  competing modes of double-$\beta$ decay,''
  \href{http://dx.doi.org/10.1103/physrevc.87.024313}{{\em Physical Review C}
  {\bfseries 87} no.~2, (Feb, 2013) }.
  \url{http://dx.doi.org/10.1103/PhysRevC.87.024313}.

\bibitem{Krivoruchenko_2011}
M.~Krivoruchenko, F.~Šimkovic, D.~Frekers, and A.~Faessler, ``Resonance
  enhancement of neutrinoless double electron capture,''
  \href{http://dx.doi.org/10.1016/j.nuclphysa.2011.04.009}{{\em Nuclear Physics
  A} {\bfseries 859} no.~1, (Jun, 2011) 140–171}.
  \url{http://dx.doi.org/10.1016/j.nuclphysa.2011.04.009}.

\bibitem{Kotila_2014}
J.~Kotila, J.~Barea, and F.~Iachello, ``Neutrinoless double-electron capture,''
  \href{http://dx.doi.org/10.1103/physrevc.89.064319}{{\em Physical Review C}
  {\bfseries 89} no.~6, (Jun, 2014) }.
  \url{http://dx.doi.org/10.1103/PhysRevC.89.064319}.

\bibitem{Karpeshin_2020}
F.~F. Karpeshin, M.~B. Trzhaskovskaya, and L.~F. Vitushkin, ``Nonresonance
  shake mechanism in neutrinoless double electron capture,''
  \href{http://dx.doi.org/10.1134/s1063778820030126}{{\em Physics of Atomic
  Nuclei} {\bfseries 83} no.~4, (Jul, 2020) 608–612}.
  \url{http://dx.doi.org/10.1134/S1063778820030126}.

\bibitem{Blaum:2020ogl}
K.~Blaum, S.~Eliseev, F.~A. Danevich, V.~I. Tretyak, S.~Kovalenko, M.~I.
  Krivoruchenko, Y.~N. Novikov, and J.~Suhonen, ``{Neutrinoless Double-Electron
  Capture},'' \href{http://dx.doi.org/10.1103/RevModPhys.92.045007}{{\em Rev.
  Mod. Phys.} {\bfseries 92} (2020) 045007},
  \href{http://arxiv.org/abs/2007.14908}{{\ttfamily arXiv:2007.14908
  [hep-ph]}}.

\bibitem{Babi__2018}
A.~Babič, D.~Štefánik, M.~I. Krivoruchenko, and F.~Šimkovic, ``Bound-state
  double-$\beta$ decay,''
  \href{http://dx.doi.org/10.1103/physrevc.98.065501}{{\em Physical Review C}
  {\bfseries 98} no.~6, (Dec, 2018) }.
  \url{http://dx.doi.org/10.1103/PhysRevC.98.065501}.

\bibitem{Tomoda_2000}
T.~Tomoda, ``0+$\rightarrow$2+ $0\nu\beta\beta$ decay triggered directly by the
  majorana neutrino mass,''
  \href{http://dx.doi.org/10.1016/s0370-2693(00)00025-3}{{\em Physics Letters
  B} {\bfseries 474} no.~3-4, (Feb, 2000) 245–250}.
  \url{http://dx.doi.org/10.1016/S0370-2693(00)00025-3}.

\bibitem{Duerr_2011_consistency_test}
M.~Duerr, M.~Lindner, and K.~Zuber, ``Consistency test of neutrinoless double
  beta decay with one isotope,''
  \href{http://dx.doi.org/10.1103/physrevd.84.093004}{{\em Physical Review D}
  {\bfseries 84} no.~9, (Nov, 2011) }.
  \url{http://dx.doi.org/10.1103/PhysRevD.84.093004}.

\bibitem{Men_ndez_2009}
J.~Menéndez, A.~Poves, E.~Caurier, and F.~Nowacki, ``Disassembling the nuclear
  matrix elements of the neutrinoless $\beta\beta$ decay,''
  \href{http://dx.doi.org/10.1016/j.nuclphysa.2008.12.005}{{\em Nuclear Physics
  A} {\bfseries 818} no.~3-4, (Mar, 2009) 139–151}.
  \url{http://dx.doi.org/10.1016/j.nuclphysa.2008.12.005}.

\bibitem{Barea_2015}
J.~Barea, J.~Kotila, and F.~Iachello, ``$0\nu\beta\beta$ and $2\nu\beta\beta$
  nuclear matrix elements in the interacting boson model with isospin
  restoration,'' \href{http://dx.doi.org/10.1103/physrevc.91.034304}{{\em
  Physical Review C} {\bfseries 91} no.~3, (Mar, 2015) }.
  \url{http://dx.doi.org/10.1103/PhysRevC.91.034304}.

\bibitem{ElementList}
J.~Coursey, D.~Schwab, J.~Tsai, and R.~Dragoset, ``Atomic weights and isotopic
  compositions (version 4.1).,'' 2015.
\newblock [Online] Available: \url{http://physics.nist.gov/Comp} [2019, 12, 12]
  National Institute of Standards and Technology, Gaithersburg, MD.

\bibitem{PhysRevD.10.275}
J.~C. Pati and A.~Salam, ``Lepton number as the fourth "color",''
  \href{http://dx.doi.org/10.1103/PhysRevD.10.275}{{\em Phys. Rev. D}
  {\bfseries 10} (Jul, 1974) 275--289}.
  \url{https://link.aps.org/doi/10.1103/PhysRevD.10.275}.

\bibitem{PhysRevD.11.2558}
R.~N. Mohapatra and J.~C. Pati, ``"natural" left-right symmetry,''
  \href{http://dx.doi.org/10.1103/PhysRevD.11.2558}{{\em Phys. Rev. D}
  {\bfseries 11} (May, 1975) 2558--2561}.
  \url{https://link.aps.org/doi/10.1103/PhysRevD.11.2558}.

\bibitem{PhysRevD.12.1502}
G.~Senjanovic and R.~N. Mohapatra, ``Exact left-right symmetry and spontaneous
  violation of parity,'' \href{http://dx.doi.org/10.1103/PhysRevD.12.1502}{{\em
  Phys. Rev. D} {\bfseries 12} (Sep, 1975) 1502--1505}.
  \url{https://link.aps.org/doi/10.1103/PhysRevD.12.1502}.

\bibitem{Duka_2000}
P.~Duka, J.~Gluza, and M.~Zrałek, ``Quantization and renormalization of the
  manifest left–right symmetric model of electroweak interactions,''
  \href{http://dx.doi.org/10.1006/aphy.1999.5988}{{\em Annals of Physics}
  {\bfseries 280} no.~2, (Mar, 2000) 336–408}.
  \url{http://dx.doi.org/10.1006/aphy.1999.5988}.

\bibitem{Dekens_2014}
W.~Dekens and D.~Boer, ``Viability of minimal left–right models with discrete
  symmetries,'' \href{http://dx.doi.org/10.1016/j.nuclphysb.2014.10.025}{{\em
  Nuclear Physics B} {\bfseries 889} (Dec, 2014) 727–756}.
  \url{http://dx.doi.org/10.1016/j.nuclphysb.2014.10.025}.

\bibitem{Nemevsek:2012iq}
M.~Nemevsek, G.~Senjanovic, and V.~Tello, ``{Connecting Dirac and Majorana
  Neutrino Mass Matrices in the Minimal Left-Right Symmetric Model},''
  \href{http://dx.doi.org/10.1103/PhysRevLett.110.151802}{{\em Phys. Rev.
  Lett.} {\bfseries 110} no.~15, (2013) 151802},
  \href{http://arxiv.org/abs/1211.2837}{{\ttfamily arXiv:1211.2837 [hep-ph]}}.

\bibitem{Hirsch_1996_susy}
M.~Hirsch, H.~V. Klapdor-Kleingrothaus, and S.~G. Kovalenko, ``Supersymmetry
  and neutrinoless double $\beta$ decay,''
  \href{http://dx.doi.org/10.1103/physrevd.53.1329}{{\em Physical Review D}
  {\bfseries 53} no.~3, (Feb, 1996) 1329–1348}.
  \url{http://dx.doi.org/10.1103/PhysRevD.53.1329}.

\bibitem{MARTIN_1998}
S.~P. MARTIN, ``A supersymmetry primer,''
  \href{http://dx.doi.org/10.1142/9789812839657_0001}{{\em Advanced Series on
  Directions in High Energy Physics} (Jul, 1998) 1–98}.
  \url{http://dx.doi.org/10.1142/9789812839657_0001}.

\bibitem{PhysRevLett.75.17}
M.~Hirsch, H.~V. Klapdor-Kleingrothaus, and S.~G. Kovalenko, ``New constraints
  on $\mathit{R}$-parity-broken supersymmetry from neutrinoless double beta
  decay,'' \href{http://dx.doi.org/10.1103/PhysRevLett.75.17}{{\em Phys. Rev.
  Lett.} {\bfseries 75} (Jul, 1995) 17--20}.
  \url{https://link.aps.org/doi/10.1103/PhysRevLett.75.17}.

\bibitem{Haber:1984rc}
H.~E. Haber and G.~L. Kane, ``{The Search for Supersymmetry: Probing Physics
  Beyond the Standard Model},''
  \href{http://dx.doi.org/10.1016/0370-1573(85)90051-1}{{\em Phys. Rept.}
  {\bfseries 117} (1985) 75--263}.

\bibitem{Nilles:1983ge}
H.~P. Nilles, ``{Supersymmetry, Supergravity and Particle Physics},''
  \href{http://dx.doi.org/10.1016/0370-1573(84)90008-5}{{\em Phys. Rept.}
  {\bfseries 110} (1984) 1--162}.

\bibitem{Zyla:2020zbs}
{\bfseries Particle Data Group} Collaboration, P.~Zyla {\em et~al.}, ``{Review
  of Particle Physics},'' \href{http://dx.doi.org/10.1093/ptep/ptaa104}{{\em
  PTEP} {\bfseries 2020} no.~8, (2020) 083C01}.

\bibitem{Bolton:2021hje}
P.~D. Bolton, F.~F. Deppisch, and P.~S.~B. Dev, ``{Neutrinoless double beta
  decay via light neutralinos in R-parity violating supersymmetry},''
  \href{http://dx.doi.org/10.1007/JHEP03(2022)152}{{\em JHEP} {\bfseries 03}
  (2022) 152}, \href{http://arxiv.org/abs/2112.12658}{{\ttfamily
  arXiv:2112.12658 [hep-ph]}}.

\bibitem{DIMOPOULOS1979237}
S.~Dimopoulos and L.~Susskind, ``Mass without scalars,''
  \href{http://dx.doi.org/https://doi.org/10.1016/0550-3213(79)90364-X}{{\em
  Nuclear Physics B} {\bfseries 155} no.~1, (1979) 237 -- 252}.
  \url{http://www.sciencedirect.com/science/article/pii/055032137990364X}.

\bibitem{Gripaios_2010}
B.~Gripaios, ``Composite leptoquarks at the lhc,''
  \href{http://dx.doi.org/10.1007/jhep02(2010)045}{{\em Journal of High Energy
  Physics} {\bfseries 2010} no.~2, (Feb, 2010) }.
  \url{http://dx.doi.org/10.1007/JHEP02(2010)045}.

\bibitem{FRITZSCH1975193}
H.~Fritsch. and P.~Minkowski, ``Unified interactions of leptons and hadrons,''
  \href{http://dx.doi.org/https://doi.org/10.1016/0003-4916(75)90211-0}{{\em
  Annals of Physics} {\bfseries 93} no.~1, (1975) 193 -- 266}.
  \url{http://www.sciencedirect.com/science/article/pii/0003491675902110}.

\bibitem{Dorsner_2005}
I.~Dorsner and P.~Fileviez~Pérez, ``Unification without supersymmetry:
  Neutrino mass, proton decay and light leptoquarks,''
  \href{http://dx.doi.org/10.1016/j.nuclphysb.2005.06.016}{{\em Nuclear Physics
  B} {\bfseries 723} no.~1-2, (Sep, 2005) 53–76}.
  \url{http://dx.doi.org/10.1016/j.nuclphysb.2005.06.016}.

\bibitem{Dor_ner_2017}
I.~Doršner, S.~Fajfer, and N.~Košnik, ``Leptoquark mechanism of neutrino
  masses within the grand unification framework,''
  \href{http://dx.doi.org/10.1140/epjc/s10052-017-4987-2}{{\em The European
  Physical Journal C} {\bfseries 77} no.~6, (Jun, 2017) }.
  \url{http://dx.doi.org/10.1140/epjc/s10052-017-4987-2}.

\bibitem{Dor_ner_2016}
I.~Doršner, S.~Fajfer, A.~Greljo, J.~Kamenik, and N.~Košnik, ``Physics of
  leptoquarks in precision experiments and at particle colliders,''
  \href{http://dx.doi.org/10.1016/j.physrep.2016.06.001}{{\em Physics Reports}
  {\bfseries 641} (Jun, 2016) 1–68}.
  \url{http://dx.doi.org/10.1016/j.physrep.2016.06.001}.

\bibitem{Hirsch_1996_LQ}
M.~Hirsch, H.~V. Klapdor-Kleingrothaus, and S.~G. Kovalenko, ``New leptoquark
  mechanism of neutrinoless double$\beta$ decay,''
  \href{http://dx.doi.org/10.1103/physrevd.54.r4207}{{\em Physical Review D}
  {\bfseries 54} no.~7, (Oct, 1996) R4207–R4210}.
  \url{http://dx.doi.org/10.1103/PhysRevD.54.R4207}.

\end{thebibliography}\endgroup

\end{document}